\documentclass[12pt]{article}

\usepackage{natbib}
\usepackage[nottoc,notlof,notlot]{tocbibind} 

\bibpunct{(}{)}{;}{a}{}{;}
\usepackage{breakcites}
\usepackage{graphicx,xcolor}
\graphicspath{ {images/} }
\usepackage[algoruled]{algorithm2e}
\usepackage[pagebackref=false,breaklinks=true]{hyperref} 
\hypersetup{
    colorlinks=true,
    citecolor=black,
    filecolor=black,
    linkcolor=black,
    urlcolor=black,
    bookmarksopen=true,
    pdfstartview=FitH
}
\usepackage{caption,threeparttable}

\usepackage{amsthm,amsmath}
\usepackage{bbm}
\usepackage{amsfonts}%
\usepackage{amssymb}%

\usepackage{geometry} 
\usepackage{graphicx} 
\usepackage{framed}
\usepackage{caption}
\usepackage{subcaption}

\usepackage{booktabs} 
\usepackage{array} 
\usepackage{paralist} 
\usepackage{verbatim} 

\usepackage{lipsum}
\usepackage{setspace}
\renewcommand{\baselinestretch}{1.5}
\usepackage[singlelinecheck=false]{caption}
\usepackage{tabularx}
\usepackage{tikz}

\usepackage{multicol}
\usepackage{multirow}

\newtheorem{assump*}{Assumption}[section]

\theoremstyle{definition}
\newtheorem{definition}{Definition}

\newtheorem{example}{Example}

\newtheorem{remark}{Remark}

\def\equationautorefname~#1\null{(#1)\null} 

\newtheorem{myexp}{Example}

\newenvironment{myexpcont}
{\addtocounter{myexp}{-1}\begin{myexp}{\textit{\textbf{{(continued)}}}}}
  {\end{myexp}}

\newcommand{\argmax}{\operatornamewithlimits{argmax}}
\newcommand{\argmin}{\operatornamewithlimits{argmin}}

\title{Identifying Socially Disruptive Policies} 
\author{Eric Auerbach\footnote{Department of Economics, Northwestern University. E-mail: eric.auerbach@northwestern.edu.} \and Yong Cai\footnote{Department of Economics, University of Wisconsin-Madison. E-mail: yong.cai@wisc.edu.\newline We thank Hossein Alidaee, Jon Auerbach, Lori Beaman, Vivek Bhattacharya, Stephane Bonhomme, Federico Bugni, Ivan Canay, Matias Cattaneo, Max Cytrynbaum, Ben Golub, Bryan Graham, Joel Horowitz, Gaston Illanes, Guido Imbens, Chuck Manski, Francesca Molinari, Roger Moon, Rob Porter, Chris Udry, Takuya Ura, and Martin Weidner  for helpful feedback. Research supported by NSF grant SES-2149422. }}
 \date{\parbox{\linewidth}{\centering%
  \today\endgraf}} 

\begin{document}
\maketitle

\begin{abstract} \setstretch{1}\noindent
Social disruption occurs when a policy creates or destroys network connections between agents. It is a costly side effect of many interventions and so a growing empirical literature recommends measuring and accounting for social disruption when evaluating the welfare impact of a policy. However, there is currently little work characterizing what can actually be learned about social disruption from data. In this paper, we consider the problem of identifying social disruption in an experimental setting. We show that social disruption is not generally point identified, but informative bounds can be constructed by rearranging the eigenvalues of the conditional distribution of network connections between pairs of agents identified from the experiment. We apply our bounds to the setting of \cite{banerjee2021changes} and find large disruptive effects that the authors miss by only considering regression estimates.   \looseness=-1 
\end{abstract}

\section{Introduction}

Many policies are socially disruptive in that they alter a substantial fraction of agents' social or economic connections. Since networks determine a wide range of economic activities, disrupting them can lead to unintended consequences. For example, \cite{carrell2013natural} study a change in classroom composition that was supposed to improve academic performance but instead segregated students which exacerbated inequality. \cite{barnhardt2017moving} analyze an antipoverty program that was intended to provide economic opportunity but instead isolated participants which led to financial insecurity. Both policies were well intentioned but, because they were socially disruptive, ultimately hurt the agents that they were designed to help.\looseness=-1 

In light of these and other examples, a growing literature recommends measuring and accounting for social disruption when evaluating the welfare impact of a policy \citep[see, for instance,][]{banerjee2021changes,jackson2021inequality}. But identifying social disruption from data is not always straightforward in practice. Economists typically characterize the disruptive impact of a policy by running dyadic regressions, which amounts to comparing the average number of connections between agents with and without the policy. While easy to implement, these regressions generally understate the total amount of social disruption. The reason for this is that economic policies usually have heterogeneous effects: they create some connections and destroy others. If the amount of created connections is roughly the same as the amount of destroyed connections, then the average difference will be small, even when the total number of connections affected by the policy is not.\footnote{This disruption is policy relevant. It represents actual relationships that are upended, requiring time and resources to replace. The literature shows that new connections may be of lower quality, associated with less trust, communication, peer influence, etc. As a result, this disruption impacts welfare, even when the average number of connections with and without the policy is similar.} \looseness=-1

In this paper, we go beyond comparing averages and consider the problem of separately identifying the amount of connections created and the amount of connections destroyed by a policy. We focus on a research design that is popular in the literature. Agents are first randomly (or as good as randomly) assigned to one of two groups. The policy is implemented in one of the groups but not the other. The agents in each group then interact and form network connections. Versions of this design are considered by \cite{carrell2013natural,feigenberg2013economic,graham2014complementarity,cai2015social,bajari2021multiple,banerjee2021changes,comola2021treatment,comola2023interplay, hess2021development,johari2022experimental}. \looseness=-1 

Our first contribution is to propose a new framework to characterize the impact of a policy on the structure of a network in a randomized experiment. We use the classical implication of random assignment, that the agents in the group subjected to the policy form connections that are representative of what the agents in the group not subjected to the policy would have realized had they been subjected to the policy. Formalizing this condition in the context of network data is not standard in econometrics, however, and our framework builds on ideas from the graph theory and operations research literatures \cite[see, generally,][]{lovasz2012large,cela2013quadratic}.  \looseness=-1 

Our second contribution is to derive new identification results. We find that the amount of connections created or destroyed by a policy is partially identified. Sharp bounds on the identified set are given by a quadratic assignment problem (QAP), but these bounds are analytically and computationally intractable. Instead, we propose conservative outer bounds based on intersecting several relaxations of the QAP. These bounds are formed by simple rearrangements of the eigenvalues of the conditional distribution of network connections between pairs of agents identified from the experiment, which are relatively straightforward to analyze and compute. Though not the focus of our paper, we also show how to consistently estimate the bounds and construct valid confidence intervals in Online Appendix Section D.3, focusing on a class of network formation models called dyadic regression models that are popular in the network economics literature, including the motivating examples referenced above \citep[for a textbook treatment, see Section 4 of][]{graham2020network}. R code for implementation can be found at \url{https://github.com/yong-cai/MatrixHTE}.  \looseness=-1 


We demonstrate our bounds with an empirical illustration using data from \cite{banerjee2021changes}. In the illustration, villages participate in a microfinance program and the network connections are informal risk sharing links between households. The authors compare the average number of connections between villages that do and do not participate with various dyadic regression models. They find that participation is associated with a roughly one percentage point decrease in connections between households. We find disruptive effects that are sixteen to twenty-three times larger using our bounds. We conclude that the microfinance program is substantially more disruptive than what is indicated by the authors' regression estimates.\footnote{The reason why we find a much larger effect size than \cite{banerjee2021changes} is that their regressions only approximate the total amount of connections impacted by the microfinance program if the program's effect is monotonic: e.g. it only creates connections or it only destroys connections. See Section 6.4.3 below. Our results indicate that such a monotonicity assumption does not hold in this setting.   }\looseness=-1 

Our paper relates to two relatively new literatures on endogenous network formation and partial identification with network data \cite[see, broadly, reviews by][]{de2016econometrics,bramoulle2020peer,graham2020network,molinari2020microeconometrics}. Most of this work focuses on recovering the structural parameters of a social interaction or network formation model. Two exceptions we know of are \cite{chandrasekhar2011econometrics,thirkettle2019identification}. While these authors focus on identifying centrality measures from sampled networks, our interest is in social disruption from an experiment.  \looseness=-1 

Our paper also relates to an older literature on Fr\'echet-Hoeffding-Makarov bounds \citep{hoeffding1940masstabinvariante,frechet1951tableaux,makarov1982estimates} and quantile treatment effects \citep{doksum1974empirical,lehmann1975nonparametrics,whitt1976bivariate}. See, for instance, \cite{manski1997mixing,manski2003partial,heckman1997making,bitler2006mean,firpo2007efficient,fan2010sharp,tamer2010partial,abadie2018econometric,masten2018identification,masten2020inference,firpo2019partial,frandsen2021partial} for work in econometric program evaluation. However, the structure of our identification problem is fundamentally different, introducing challenges not present in this literature. Intuitively, what distinguishes our framework is that while agents are individually assigned to policies, connections are measured between pairs of agents. It turns out that this distinction substantially alters the identification problem. Standard results are not generally valid and standard tools when naively applied often fail to identify any social disruption. We provide intuition as to how our problem is different and why our methodology is appropriate in Section 2. Our formal framework is in Section 3 and identification results are in Sections 4 and 5. The empirical illustration is in Section 6. Proofs are in the appendix. Additional details and results are in the online appendix.  \looseness=-1

\section{An illustration of the main identification problem}
In this section, we provide a simplified illustration of the main identification problem, deferring the formal framework and identification results to Sections 3-5. We focus on identifying the magnitude of connections destroyed by a change in policy. The identification problem is similar in spirit to that of bounding the joint distribution of two random variables using their marginals originally considered by \cite{hoeffding1940masstabinvariante,frechet1951tableaux}. However, our problem has a fundamentally different structure. We give an example where naively applying the bounds from this literature fails to identify any social disruption, but our methodology does. \looseness=-1  

\subsection{A simplified setup}
To learn about the disruptive impact of a new policy, we consider an experiment where we randomly assign $N$ agents to a treatment group and $N$ agents to a control group. We implement the new policy in the treatment group and maintain the status quo policy in the control group.  \looseness=-1 

The mechanics of the identification problem do not depend on what the two policies actually do. It only matters that they somehow determine the agents' network connections. So, for example, a policy could be that every agent in the group participates in an antipoverty program, no agent participates, a random subset of agents participate, a nonrandom subset of agents are encouraged but not required to participate, etc. Once the new and status quo policies are chosen and implemented,  the $N$ agents in the treatment group interact and form one network. The $N$ agents in the control group interact and form another network. We take the size of the two groups to be the same to simplify the illustration, but this is straightforward to relax. \looseness=-1  

We use potential outcome notation to describe the identification problem. Let policy $1$ refer to the new policy, policy $0$ refer to the status quo, group $1$ refer to the treatment group, and group $0$ refer to the control group. Then $Y_{ij,t}(s)$ is the potential connection between agents $i$ and $j$ in group $t$ under policy $s$. That is, $Y_{ij,t}(s)$ describes what the connection between agents $i$ and $j$ in group $t$ would be if policy $s$ were implemented in that group. Since, in the experiment, policy $s$ is implemented in group $s$, $Y_{ij,t}(s)$ is observed if $s = t$ and is unobserved if $s \neq t$. To simplify our illustration, we assume that the networks are unweighted and undirected so that the matrix $Y_{s}(t) = \{Y_{ij,s}(t)\}_{i,j = 1}^{N}$ is symmetric with $\{0,1\}$-valued entries. Weighted networks are accommodated in Section 3 by thresholding. Directed networks are accommodated in Online Appendix Section D.1 by symmetrization.   \looseness=-1  

In this simplified setup, we assume that the potential outcomes are not random. In practice, researchers often specify stochastic models of network formation where a random connection between pairs of agents reflects the fact that network formation is the result of human decision making which is naturally indeterminate. We ignore the issue of random networks to simplify the presentation of the main identification problem in this section, but incorporate it into our formal framework in Section 3.\looseness=-1  

To illustrate the main identification problem, we take as the parameter of interest the number of network connections between the $N$ agents assigned to the control group that would be destroyed by implementing the new policy in that group. That is, 
\begin{align}\label{param}
\frac{1}{2}\sum_{i,j = 1}^{N}(1-Y_{ij,0}(1))Y_{ij,0}(0). 
\end{align}
$Y_{ij,0}(0)$ is observed but $Y_{ij,0}(1)$ is unobserved and  without additional assumptions could take any value in $\{0,1\}$. It follows that the identified set for (\ref{param}) is 
\begin{align}\label{paramset}
\left\{\theta \in \mathbb{N} : \theta = \frac{1}{2}\sum_{i,j = 1}^{N}M_{ij}Y_{ij,0}(0) \text{ for any } M_{ij} = M_{ji} \in \{0,1\}\right\}.
\end{align}
This set is typically too large to be informative in practice, and so we refine it below by using the assumption that the agents are randomly assigned to the treatment and control groups.  \looseness=-1  

\begin{myexp}
A toy example with $N = 6$ is illustrated in Figure 1. The six agents assigned to the treatment group form a line. The six agents assigned to the control group form a star.  There are five connections between the agents in the control group out of a possible total of fifteen. Without additional assumptions,  (\ref{paramset}) says that the number of connections that would be destroyed by implementing the policy in the control group is between $0$ and $5$. \looseness=-1  
\end{myexp}

\begin{figure}
\centering
\tikzset{every picture/.style={line width=0.75pt}} 

\begin{tikzpicture}[x=0.75pt,y=0.75pt,yscale=-1,xscale=1]
	
	\draw [color={rgb, 255:red, 235; green, 124; blue, 29 }  ,draw opacity=1 ][line width=2.25]    (285,155) -- (325,155) ;
	\draw [color={rgb, 255:red, 235; green, 124; blue, 29 }  ,draw opacity=1 ][line width=2.25]    (285,75) -- (325,75) ;
	\draw [line width=2.25]    (105,65) -- (105,125) ;
	\draw [line width=2.25]    (65,165) -- (105,125) ;
	\draw [line width=2.25]    (105,125) -- (145,165) ;
	\draw [color={rgb, 255:red, 126; green, 211; blue, 33 }  ,draw opacity=1 ][line width=2.25]    (245,75) -- (285,75) ;
	\draw [color={rgb, 255:red, 126; green, 211; blue, 33 }  ,draw opacity=1 ][line width=2.25]    (245,155) -- (285,155) ;
	\draw [color={rgb, 255:red, 80; green, 181; blue, 227 }  ,draw opacity=1 ][line width=2.25]    (245,75) -- (245,155) ;
	\draw  [fill={rgb, 255:red, 0; green, 0; blue, 0 }  ,fill opacity=1 ] (100,65) .. controls (100,62.24) and (102.24,60) .. (105,60) .. controls (107.76,60) and (110,62.24) .. (110,65) .. controls (110,67.76) and (107.76,70) .. (105,70) .. controls (102.24,70) and (100,67.76) .. (100,65) -- cycle ;
	\draw  [fill={rgb, 255:red, 0; green, 0; blue, 0 }  ,fill opacity=1 ] (100,125) .. controls (100,122.24) and (102.24,120) .. (105,120) .. controls (107.76,120) and (110,122.24) .. (110,125) .. controls (110,127.76) and (107.76,130) .. (105,130) .. controls (102.24,130) and (100,127.76) .. (100,125) -- cycle ;
	\draw  [fill={rgb, 255:red, 0; green, 0; blue, 0 }  ,fill opacity=1 ] (60,165) .. controls (60,162.24) and (62.24,160) .. (65,160) .. controls (67.76,160) and (70,162.24) .. (70,165) .. controls (70,167.76) and (67.76,170) .. (65,170) .. controls (62.24,170) and (60,167.76) .. (60,165) -- cycle ;
	\draw  [fill={rgb, 255:red, 0; green, 0; blue, 0 }  ,fill opacity=1 ] (140,165) .. controls (140,162.24) and (142.24,160) .. (145,160) .. controls (147.76,160) and (150,162.24) .. (150,165) .. controls (150,167.76) and (147.76,170) .. (145,170) .. controls (142.24,170) and (140,167.76) .. (140,165) -- cycle ;
	\draw  [fill={rgb, 255:red, 0; green, 0; blue, 0 }  ,fill opacity=1 ] (240,75) .. controls (240,72.24) and (242.24,70) .. (245,70) .. controls (247.76,70) and (250,72.24) .. (250,75) .. controls (250,77.76) and (247.76,80) .. (245,80) .. controls (242.24,80) and (240,77.76) .. (240,75) -- cycle ;
	\draw  [fill={rgb, 255:red, 0; green, 0; blue, 0 }  ,fill opacity=1 ] (320,75) .. controls (320,72.24) and (322.24,70) .. (325,70) .. controls (327.76,70) and (330,72.24) .. (330,75) .. controls (330,77.76) and (327.76,80) .. (325,80) .. controls (322.24,80) and (320,77.76) .. (320,75) -- cycle ;
	\draw  [fill={rgb, 255:red, 0; green, 0; blue, 0 }  ,fill opacity=1 ] (320,155) .. controls (320,152.24) and (322.24,150) .. (325,150) .. controls (327.76,150) and (330,152.24) .. (330,155) .. controls (330,157.76) and (327.76,160) .. (325,160) .. controls (322.24,160) and (320,157.76) .. (320,155) -- cycle ;
	\draw  [fill={rgb, 255:red, 0; green, 0; blue, 0 }  ,fill opacity=1 ] (240,155) .. controls (240,152.24) and (242.24,150) .. (245,150) .. controls (247.76,150) and (250,152.24) .. (250,155) .. controls (250,157.76) and (247.76,160) .. (245,160) .. controls (242.24,160) and (240,157.76) .. (240,155) -- cycle ;
	\draw [line width=2.25]    (105,125) -- (155,95) ;
	\draw  [fill={rgb, 255:red, 0; green, 0; blue, 0 }  ,fill opacity=1 ] (150,95) .. controls (150,92.24) and (152.24,90) .. (155,90) .. controls (157.76,90) and (160,92.24) .. (160,95) .. controls (160,97.76) and (157.76,100) .. (155,100) .. controls (152.24,100) and (150,97.76) .. (150,95) -- cycle ;
	\draw  [fill={rgb, 255:red, 0; green, 0; blue, 0 }  ,fill opacity=1 ] (50,95) .. controls (50,92.24) and (52.24,90) .. (55,90) .. controls (57.76,90) and (60,92.24) .. (60,95) .. controls (60,97.76) and (57.76,100) .. (55,100) .. controls (52.24,100) and (50,97.76) .. (50,95) -- cycle ;
	\draw [line width=2.25]    (55,95) -- (105,125) ;
	\draw  [fill={rgb, 255:red, 0; green, 0; blue, 0 }  ,fill opacity=1 ] (280,75) .. controls (280,72.24) and (282.24,70) .. (285,70) .. controls (287.76,70) and (290,72.24) .. (290,75) .. controls (290,77.76) and (287.76,80) .. (285,80) .. controls (282.24,80) and (280,77.76) .. (280,75) -- cycle ;
	\draw  [fill={rgb, 255:red, 0; green, 0; blue, 0 }  ,fill opacity=1 ] (280,155) .. controls (280,152.24) and (282.24,150) .. (285,150) .. controls (287.76,150) and (290,152.24) .. (290,155) .. controls (290,157.76) and (287.76,160) .. (285,160) .. controls (282.24,160) and (280,157.76) .. (280,155) -- cycle ;
	
	\draw (106,140.5) node   [align=left] {$\displaystyle a$};
	\draw (90.38,54) node   [align=left] {$\displaystyle b$};
	\draw (50.38,174) node   [align=left] {$\displaystyle b$};
	\draw (160.38,174) node   [align=left] {$\displaystyle b$};
	\draw (234.38,59) node   [align=left] {$\displaystyle 3$};
	\draw (234,169.5) node   [align=left] {$\displaystyle 3$};
	\draw (338.38,169) node   [align=left] {$\displaystyle 1$};
	\draw (338.38,59) node   [align=left] {$\displaystyle 1$};
	\draw (286.5,200) node   [align=left] {Treatment Group};
	\draw (105.5,200) node   [align=left] {Control Group};
	\draw (286,59.5) node   [align=left] {$\displaystyle 2$};
	\draw (286,169.5) node   [align=left] {$\displaystyle 2$};
	\draw (166,80.5) node   [align=left] {$\displaystyle b$};
	\draw (46.5,81) node   [align=left] {$\displaystyle b$};

\end{tikzpicture}
\vspace{-6mm}
\caption{A toy example of an experiment. The six agents assigned to the treatment group form a line under the new policy. The six agents assigned to the control group form a star under the status quo.}
\label{fig:toyexample_problem}
\end{figure}
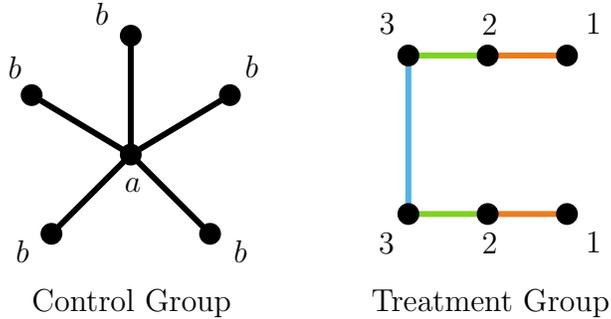

\subsection{The main identification assumption}
Our first main contribution is to propose a condition that formalizes how the random assignment of agents to groups refines the identified set. Our main identification assumption is that $Y_{0}(1)$ and $Y_{1}(1)$ are \emph{weakly isomorphic}. We defer a formal definition of this assumption to Section 4.1, but, intuitively, it says that the configuration of network connections between agents assigned to the treatment group describe how the agents assigned to the control group would be connected if that group were assigned the new policy instead of the status quo. This assumption is a network analog of the conventional assumption that the entries of $Y_{0}(1)$ and $Y_{1}(1)$ have the same empirical distribution, which is a strong but ubiquitous implication of random assignment, see Chapter 7.3 of \cite{manski2009identification} for a textbook discussion. \looseness=-1

As in a conventional experiment, randomization only generally implies that $Y_{0}(1)$ and $Y_{1}(1)$ are weakly isomorphic in expectation. However, to illustrate the main identification problem of our paper, we will, for this simplified illustration, make the strong and unrealistic assumption that it holds exactly in the realized experiment.  \looseness=-1  

Formally, the main identification assumption is that there exists an $N\times N$ permutation matrix $\Pi$,\footnote{A permutation matrix is a square matrix with $\{0,1\}$-valued entries. Every row and column sums to $1$.} unknown to the researcher, such that  \looseness=-1 
\begin{align}\label{simAss}
Y_{ij,0}(1) = \sum_{k,l=1}^{N}Y_{kl,1}(1)\Pi_{ik}\Pi_{jl}.
\end{align}
In words, condition (\ref{simAss}) says that there exists a one-to-one match between agents in the two groups such that the counterfactual outcomes of the agents in group $0$ are given by their matches in group $1$. It could be rationalized by a specific experimental design conducted on $N$ pairs or ``clones'' of agents. One member of each pair is randomly assigned to each group and the counterfactual connection between agents in the control group is given by their clones in the treatment group. If the researcher knows which pairs of agents are clones, then they can  compute (\ref{param}) by simply substituting $Y_{ij,0}(1)$ with $Y_{c_{i}c_{j},1}(1)$ where $c_{i}$ is the identity of $i$'s clone. But the researcher has forgotten this information so that, in principle, any matching between the agents of the treatment and control groups (as represented by some permutation matrix) could be the correct one.  \looseness=-1  

If the researcher conducts such a clone experiment, then  (\ref{simAss}) holds exactly in a finite sample. However, outside this specific setting, randomization does not generally imply this condition.  Instead, it is an approximation to what randomization does in large samples. The idea that randomization can be characterized by an approximate matching is not original to our paper: an analogous condition plays a key role in the identification arguments of the Fr\'echet-Hoeffding-Makarov bounds and quantile treatment effects literature. See \cite{whitt1976bivariate,heckman1997making} for detailed discussions. What is new in our setting is that the quadratic structure of (\ref{simAss}) makes the problem of identifying social disruption with network data fundamentally different. We discuss this complication in Section 2.2.2 below.   \looseness=-1

\subsubsection{The identified set under the main identification assumption}
Though condition (\ref{simAss}) is intended to be an approximation, to illustrate its identifying content in this section we suppose that it holds exactly. Plugging (\ref{simAss}) into (\ref{param}) implies that the number of links destroyed by the policy is \looseness=-1  
\begin{align*}
\frac{1}{2}\sum_{i,j = 1}^{N}(1-Y_{ij,0}(1))Y_{ij,0}(0) = \frac{1}{2}\sum_{i,j,k,l = 1}^{N}(1-Y_{kl,1}(1))Y_{ij,0}(0)\Pi_{ik}\Pi_{jl}.
\end{align*}
The substitution solves the initial problem that $Y_{0}(1)$ is not known because both $Y_{1}(1)$ and $Y_{0}(0)$ on the right-hand side are observed. However, the right-hand side now depends on the unknown $\Pi$. Since, under (\ref{simAss}), any permutation matrix suggests a number of destroyed links that is consistent with the observed network connections, the identified set is \looseness=-1  
 \begin{align}\label{paramset2}
 \left\{\theta \in \mathbb{N}: \theta = \frac{1}{2}\sum_{i,j,k,l = 1}^{N}(1-Y_{kl,1}(1))Y_{ij,0}(0)P_{ik}P_{jl} \text{ for any permutation matrix } P \right\}.
 \end{align}

\begin{myexpcont}
In the toy example, condition (\ref{simAss}) implies that if the new policy were implemented in the control group, the agents would change their social connections to form a line. Under this assumption, the identified set for the number of connections destroyed by the policy is $\{3,4\}$. The logic behind this result is illustrated in Figure 2. There are three ways up to symmetry to match the six agents in the control group to the six agents in the treatment group. Matching  $a$ in the control group to position $1$ in the treatment group destroys four connections. This is because all five connections are adjacent to $a$ in the control group and $1$ has only one connection in the treatment group. Similarly, matching $a$ to positions $2$ or $3$ destroys three connections. Since these are the only unique matches up to a relabeling of the agents, the policy must destroy 3 or 4 out of 5 connections ($60$ or $80$ percent). It follows that, under  (\ref{simAss}), this example necessarily has a large amount of social disruption. \looseness=-1  
\end{myexpcont}

\begin{figure}
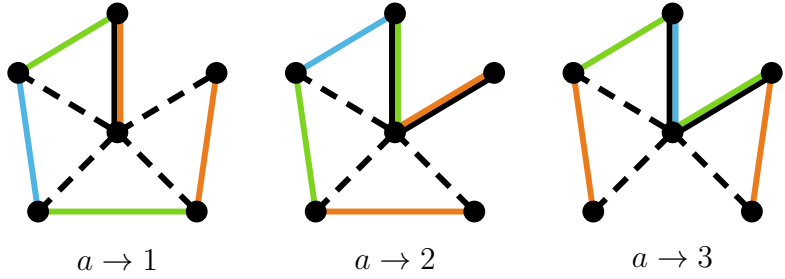

\centering
\include{fig_toyexample_matrix}
\vspace{-6mm}
\caption{There are three ways to match agents in the control group to the treatment group up to symmetry. A black dashed line indicates a destroyed connection. It exists under the status quo but not the new policy. A colored and a black solid line indicates a maintained connection. It exists under both the status quo and the new policy. A colored line only indicates a created connection. It exists under the new policy but not the status quo.}
\label{fig:toyexample_matrix}
\end{figure}

\subsubsection{The identified set is typically uncomputable}
It is straightforward to compute the identified set (\ref{paramset2}) in our toy example because $N$ is small. However, computing the identified set is not possible in many cases of interest. This is because the problem of finding the largest or smallest element of (\ref{paramset2}) is equivalent to solving a quadratic assignment problem which is strongly NP hard in theory and practically uncomputable for instances with more than a few dozen agents. See Section 1.5 of \cite{cela2013quadratic}.  \looseness=-1  

Our second main contribution is to instead propose tractable outer bounds that are both informative about social disruption and computationally feasible even for large networks. Intuitively, it is hard to compute sharp bounds on  (\ref{paramset2}) because searching over permutation matrices is difficult. Our bounds instead search over orthogonal matrices.\footnote{An orthogonal matrix is a square matrix where the inner product of any row or column with itself is $1$ and the inner product of any two distinct rows or two distinct columns is $0$.} To illustrate this idea, we replace (\ref{paramset2}) with 
 \begin{align}\label{paramset3}
 \left\{\theta \in \mathbb{N}: \theta = \frac{1}{2}\sum_{i,j,k,l = 1}^{N}(1-Y_{kl,1}(1))Y_{ij,0}(0)O_{ik}O_{jl} \text{ for any orthogonal matrix } O \right\}.
 \end{align}
There are two reasons for this substitution. First, because all permutation matrices are orthogonal, (\ref{paramset3}) contains (\ref{paramset2}) and so any bounds on (\ref{paramset3})  will also be valid for (\ref{paramset2}). Second, solving for the smallest and largest element of (\ref{paramset3})  is relatively straightforward: the minimum is $\sum_{r=1}^{N}\lambda_{r}(1)\lambda_{N-r}(0)$ and the maximum is  $\sum_{r=1}^{N}\lambda_{r}(1)\lambda_{r}(0)$ where $\lambda_{r}(s)$ is the $r$th largest eigenvalue of $(1-Y_{s}(s))^{s}Y_{s}(s)^{1-s}$. See Lemma 2 in Appendix Section A.3 for a proof.\footnote{The idea of bounding a QAP by searching over orthogonal matrices was originally proposed by \cite{finke1987quadratic}, however applying this logic to our general setting is not straightforward and requires arguments not typical of the QAP literature. See Section 4.1 for a discussion. } \looseness=-1

\subsubsection{Our bounds can be more informative than conventional methods}
Conventional methods such as computing the difference in the number of connections or the Fr\'echet-Hoeffding bounds are not generally effective at identifying social disruption.\footnote{Typically the literature runs dyadic regressions. They regress the magnitude of the connection between a pair of agents on an indicator for whether the agents are subjected to the new policy and additional covariates. The coefficient on the policy indicator is then used to measure social disruption. Without covariates, this is equivalent to computing the difference in the number of connections. Including covariates does not generally make these regressions informative about social disruption. See our empirical illustration in Section 6.} The difference in the number of connections, $\frac{1}{2}\sum_{i,j = 1}^{N}Y_{ij,0}(0) - \frac{1}{2}\sum_{i,j = 1}^{N}Y_{ij,1}(1)$, is also the number of connections destroyed by the new policy minus the number of connections created. It is only a good approximation of the number of destroyed connections if the number of created connections is close to zero, which is rare in practice.  \looseness=-1 

The Fr\'echet-Hoeffding  lower bound on (\ref{param}) is $\max(\frac{1}{2}\sum_{i,j=1}^{N}Y_{ij,0}(0)- \frac{1}{2}\sum_{i,j=1}^{N}Y_{ij,1}(1),0)$. The upper bound is $\min(\frac{1}{2}\sum_{i,j=1}^{N}(1-Y_{ij,1}(1)),\frac{1}{2}\sum_{i,j=1}^{N}Y_{ij,0}(0))$. These bounds are valid in that the number of destroyed connections is necessarily between them, but they are not generally informative. The lower bound is essentially the difference in the number of connections and is zero if the policy creates at least as many connections as it destroys. The upper bound is large if there are many pairs of agents that are connected in the control group and many pairs of agents that are not connected in the treatment group. Both are common in practice.  \looseness=-1 

The problem with these methods is that they do not use all of the information available in condition (\ref{simAss}). They only use the relatively weak implication that $Y_{0}(1)$ and $Y_{1}(1)$ have the same number of connections. Our bounds often perform better because there can be important identifying information in the former restriction that is not in the latter. \looseness=-1

\begin{myexpcont}
Neither conventional method described in Section 2.2.3 is informative about social disruption in the toy example. The difference in the number of connections between treatment groups is $5 - 5 = 0$, which does not identify any social disruption. The upper Fr\'echet-Hoeffding bound is $\min(5,10) = 5$ and the lower bound is $5 - 5 = 0$, which are equivalent to the trivial bounds derived in Section 2.1.1 that do not use condition (\ref{simAss}). \looseness=-1 

Our bounds, in contrast, give an upper bound of 4.17 and a lower bound of 1.6.\footnote{These bounds are in Proposition 2 of Section 5.2 and use the adjustment in Online Appendix Section  D.2. They are implemented in an R package available at \url{https://github.com/yong-cai/MatrixHTE}.} They imply that the number of destroyed links belongs to $\{2,3,4\}$ which is close to the identified set of $\{3,4\}$ derived in Section 2.2.2. In particular, our bounds imply that the fraction of destroyed links is between $40$ and $80$ percent, which is not quite the sharp bounds of $60$ and $80$ percent, but is a substantial improvement on the conventional/trivial bounds of $0$ and $100$ percent. \looseness=-1 \end{myexpcont}

The performance of our methodology is not limited to the toy example. Section 6 provides empirical evidence showing that our bounds can be much more effective at identifying social disruption than conventional methods in a real world setting.\looseness=-1

\section{Main framework}
In this section, we describe our framework. We first define the parameters of interest and then specify a model for the data. We focus on undirected networks, i.e. networks that can be represented by a symmetric real-valued adjacency matrix. The framework and results immediately extended to the directed case by a symmetrization argument described in Online Appendix Section D.1. This use of symmetrization to extend results from the symmetric case to the asymmetric case is standard and comes from the literature on U-statistics. See, for instance, Section 5.1.1 of \cite{serfling2009approximation} for a textbook presentation.  \looseness=-1 

\subsection{Potential outcomes and the parameters of interest}
\subsubsection{Potential outcomes}
A community of $N$ agents may be subjected to one of two policies indexed by $s \in \{0,1\}$. In our framework, we use the word ``policy'' to refer to any community-level intervention that influences the network connections between agents. One policy could, for instance, assign an individualized treatment to every agent in the community. Another policy could assign the treatment to a subpopulation of agents. Still another policy could be the introduction of a public good that is available to all members of the community. All of these examples fall within our framework.  \looseness=-1 

Once the community is subjected to one of the two policies, the agents interact and form connections. For two arbitrary agents $i$ and $j$ in the community, the variable $Y_{ij}(s)$ describes the magnitude of the connection between agents $i$ and $j$ when the community is assigned policy $s$. We refer to these quantities as potential outcomes or potential connections. \looseness=-1 

\begin{remark}
When the policy concerns the assignment of an individualized treatment to agents in the community, our framework does not rule out treatment spillovers, but it does not explicitly model them either. Similarly, our framework neither rules out nor explicitly models strategic interactions between agents in the community. We discuss the sense in which treatment spillovers and strategic interactions are accommodated after we introduce our parameters of interest (Section 3.1.2) and model (Sections 3.2.2) below.\looseness=-1 
\end{remark}

\subsubsection{Parameters of interest}
We focus on the joint distribution of potential outcomes (DPO): \looseness=-1 
\begin{align}\label{joint}
F(y_{1},y_{0}) := \mathbb{P}\left(Y_{ij}(1)  \leq y_1, Y_{ij}(0)  \leq y_{0}\right)
\end{align}
where $y_{1},y_{0} \in \mathbb{R}$ and the measure $\mathbbm{P}$ refers to the joint distribution of potential outcomes, to be specified in Section 3.2 below. In words, $F(y_{1},y_{0}) $ is the mass of agent pairs with potential outcomes less than $y_{1}$ under policy $1$ and less than $y_{0}$ under policy $0$. Many measures of social disruption, including the fraction of network connections created or destroyed by a change in policy, can be written as simple functions of the DPO. For example, the fraction of binary $(\{0,1\}$-valued) network connections destroyed by a change in policy is $F(0,1) - F(0,0)$. The fraction created is $F(1,0) - F(0,0)$. \looseness=-1 

A closely related parameter to the DPO is the distribution of treatment effects (DTE): \looseness=-1 
\begin{align}\label{te}
\Delta(y) := \mathbb{P}\left(Y_{ij}(1) - Y_{ij}(0) \leq y \right)
\end{align}
where $y \in \mathbb{R}$ is arbitrary. In words,  $Y_{ij}(1) - Y_{ij}(0)$ is the change in potential outcomes for agents $i$ and $j$ caused by switching from policy $0$ to policy $1$. $\Delta(y)$ is the mass of individual treatment effects that are less than $y$. While we focus our identification results on the DPO, we show how our proposed bounds for the DPO translate to bounds on the DTE in Proposition 3 of Section 5.2 below. \looseness=-1   

\begin{remark}
When the policy concerns the assignment of an individualized treatment to agents, the effect of treating one agent may spillover and affect the magnitude of connections between other agents in the community. In these settings, we interpret the DPO and DTE as capturing the ``total effect'' of the treatment assignment in the sense that they incorporate both the direct effect of the treatment assigned to an individual and the indirect effect of the treatment assigned to other agents in the community. We conjecture that it is possible to extend our framework to separately identify the distribution of direct and indirect treatment effects, but we leave this extension to future work.  \looseness=-1 
\end{remark}

\begin{remark}
In some settings, the network connections may be the result of strategic interactions in a network formation game. In these settings, we assume that there is a deterministic equilibrium selection process and interpret the DPO and DTE as describing the effect of the policy on the within-equilibrium distribution of outcomes. That is, comparing the distribution of connections for whatever equilibrium was reached by the community under policy $0$ to the distribution of connections under whatever equilibrium was reached by the community under policy $1$. We provide a concrete example in Section 3.2.2 below.  \looseness=-1 
\end{remark}

\subsection{Model of the data observed from the experiment}
The researcher observes network data for two communities, each subjected to a different policy. It is assumed that the observed data are distributed as though a randomized experiment were conducted. Randomized experiments are common in the network economics literature, see, for examples, the literature cited in the third paragraph of the introduction. The assumption is also commonly employed in settings where randomization did not actually occur, but the researcher has reason to believe that the communities that make up the data are balanced in the sense that they contain a similar composition of agent types. \looseness=-1 

In this subsection, we first specify a model of an experiment where agents are randomly assigned to communities. We then specify a stochastic model of link formation. Finally, we describe our main identifying assumption. This assumption is used to characterize the identified set for the DPO in Section 4.  \looseness=-1 

\subsubsection{Experiment}
We model a completely randomized balanced experiment conducted with $2N$ agents. In the experiment, the researcher defines two communities indexed by $\{0,1\}$. $N$ agents are then chosen uniformly at random and assigned to community $1$. The remaining $N$ agents are assigned to community $0$. We focus on the setting where the two communities are the same size because it greatly simplifies the exposition. It is straightforward to extend the framework to allow for differently-sized communities. For $s \in \{0,1\}$, policy $s$ is implemented in community $s$. We use $D_i \in \{0,1\}$ to indicate the policy/community assignment of agent $i$ and collect the community assignments for the $2N$ agents into the vector $D := \{D_i\}_{i=1}^{N}$.  \looseness=-1 

After the community assignments are made, the agents interact and form network connections, as described in Section 3.2.2 below. For two arbitrary agents $i$ and $j$ assigned to the same community $s \in \{0,1\}$ the researcher observes the potential outcome $Y_{ij}(s)$. That is, if $D_i = D_j$ then the observed magnitude of the connection between $i$ and $j$ is given by \looseness=-1 
\begin{align*}
Y_{ij} = Y_{ij}(1)D_iD_j + Y_{ij}(0)(1-D_i)(1-D_j).
\end{align*}
If $D_i \neq D_j$ then no connection is observed. We collect the $N(N-1)$ observed network connections into the vector $Y := \{Y_{ij}\}_{ij : D_i = D_j}$. A key assumption of our framework is that the potential outcomes and community assignments are unrelated, i.e. the entries of $Y(1)$ and $Y(0)$ are mutually independent of the entries of $D$. We describe this condition in Section 3.3 below. \looseness=-1 

At the conclusion of the experiment, the researcher observes the vectors $\{Y,D\}$ as data. \looseness=-1

\begin{remark}
The completely randomized balanced experiment design presumes that the community assignments are homogenous in the sense that every collection of $N$ agents is equally likely to be assigned to community $1$. However, the assumption is not required for our identification results, which continue to hold if the community assignments are heterogeneous, so long as the probability that any $N$ agents is assigned to community $1$ is nonzero. See Remark 15 Appendix Section A.2 for a discussion. \looseness=-1 
\end{remark}

\subsubsection{Network Formation Model}
For a fixed $s \in \{0,1\}$, we assume that the connections between the $N$ agents in community $s$ are given by the model \looseness=-1 
\begin{align}\label{model}
Y_{ij}(s) = g_{N,s}(w_{i,s},w_{j,s},\eta_{ij,s})
\end{align}
where $w_{i,s} \in [0,1]$ is a random agent effect, $\eta_{ij,s} \in [0,1]$ is a random idiosyncratic error, and $g_{N,s} : [0,1]^{3}\to \mathbb{R}$ is a measurable link function that may depend on the community size $N$. The entries of $\{w_{i,s}\}_{i=1}^{N}$ are independent across agents, the entries of  $\{\eta_{ij,s}\}_{i,j = 1}^{N}$ are independent across pairs of agents, and the entries of $\{w_{i,s}\}_{i=1}^{N}$ and $\{\eta_{ij,s}\}_{i,j = 1}^{N}$ are mutually independent. The agent effects, idiosyncratic errors, and link function may all depend on the policy $s$. For our identification results, we assume that these three determinants of link formation are all unobserved. That is, the researcher only observes data on the network connections $Y$. \looseness=-1 



A key feature of this network formation model is that for any two pairs of agents $\{i,j\}$ and $\{k,l\}$ in the community, the connections $Y_{ij}(s)$ and $Y_{kl}(s)$ are independent if $\{i,j\}$ and $\{k,l\}$ do not share an index. Though strong, this dependence structure describes a number of specifications that are popular in the economics literature. We provide three broad examples. \looseness=-1 

\begin{example}
The first example is the literature on dyadic regression, which is commonly used in network economics to model trade between countries, risk sharing relationships between households, peer relationships between students, and more. Nearly all of the motivating examples we reference in our introduction specify a dyadic regression model. See Section 4 of \cite{graham2020network} for a review. In this example, the researcher typically makes a  parametric functional form restriction on the link function. A concrete example is \cite{fafchamps2007risk}, who specify the linear model of risk sharing \looseness=-1 
\begin{equation}\label{dyadicreg}
	Y_{ij} = \alpha + \beta_1 (z_i - z_j) + \beta_2 (z_i + z_j) + \gamma d(x_i,x_j) + u_{ij}
\end{equation}
where $z_i$ is an attribute of agent $i$ such as household education, $x_i$ is the geographic position of agent $i$, $d$ is a measure of distance, and $u_{ij}$ is an idiosyncratic error where $u_{ij}$ and $u_{kl}$ are independent if $\{i,j\}$ and $\{k,l\}$ do not share an index. Taking $(z_i,x_i)$ to be random, we write $z_i = z(w_i)$, $x_i = x(w_i)$, $u_{ij} = u(w_i,w_j,\eta_{ij})$ for measurable functions $z$, $x$, and $u$. Following \cite{graham2020network}, the model can be cast as a special case of  (\ref{model}) by defining the link function to be $g_{N}(u,v,w) = \alpha + \beta_1(z(u)-z(v)) + \beta_2 (z(u) +z(v)) + \gamma d(x(u),x(v)) + u(w)$. A related dyadic regression model is used by \cite{banerjee2021changes}, which is the setting of our empirical illustration in Section 6.  \looseness=-1 
\end{example}

\begin{example}
The second example is the literature on exchangeable arrays. In this literature, the model (\ref{model}) is often motivated as a consequence of the assumption that the connections are exchangeable due to a result called the Aldous-Hoover-Kallenberg representation theorem. See Section 3.4 of \cite{graham2020network} for a review. A concrete example is \cite{bickel2009nonparametric}, who consider unweighted networks where the connections are binary ($\{0,1\}$-valued) and specify the model\looseness=-1 
\begin{equation*}
Y_{ij} = \mathbbm{1}\{\rho_{N}f(w_i,w_j) - \eta_{ij} \geq 0\}
\end{equation*}
where $\rho_N$ is a deterministic sequence of constants that vanishes with the community size to model  sparse networks. The function $f$ is measureable and normalized so that $\int\int f(u,v)dudv = 1$. In this case the model can be cast as a special case of (\ref{model}) by defining the link function to be $g_{N}(u,v,w) = \mathbbm{1}\{\rho_Nf(u,v) - w \geq 0\}$. For applications of this model in the network econometrics literature see, for instance, \cite{menzel2017bootstrap,zeleneev2020identification,auerbach2022identification}.\looseness=-1 
\end{example}

\begin{example}
The third example is the literature on strategic models of network formation. In this literature, the model (\ref{model}) can be thought of as a reduced-form description of equilibrium linking behavior in a strategic game. See Section 8 of \cite{graham2020network} for a review.  For example, \cite{leung2015two} specifies a model of link formation where the marginal utility that agent $i$ forms from forming a connection with agent $j$ may depend on the other connections between agents in the community. He assumes that the connections are in a Bayes-Nash equilibrium where the connections are binary and formed according to 
\begin{equation*}
Y_{ij} = \mathbbm{1}\{\alpha_0 + \beta_0 P_{ij} + \gamma_0 \sum_{k=1}^{N} P_{ki}P_{kj} + t(x_i,x_j)'\delta_0 - u_{ij} \geq 0\}
\end{equation*}
where $x_i$ are iid covariates that are common knowledge to the agents, $u_{ij}$ is an iid idiosyncratic error that is private knowledge to agent $i$, and $P_{ij}$ is the common prior held by all agents other than $i$ regarding the probability that agent $i$ links with agent $j$. \cite{leung2015two} goes on to assume that the network connections are exchangeable and he focuses on symmetric equilibria in that agents with the same covariates have the same ex ante linking probabilities. Under these assumptions, his model of equilibrium link formation can be cast as a special case of  (\ref{model}) by writing $x_i = x(w_i)$, $u_{ij} = u(\eta_{ij})$, and defining the link function to be $g_{N}(u,v,w) = \mathbbm{1}\{\alpha_0 + \beta_0 P(x(u),x(v)) + \gamma_0 \sum_{k=1}^{N} P(x(u),x_k)P(x_k,x(v)) + t(x(u),x(v))'\delta_0 - u(w) \geq 0\}$, where $P(a,b)$ is the common prior that agents with covariate values $a$ and $b$ form a connection. In our setting, we take the function $P$ to be a fixed parameter and absorb it into the link function $g_N$. As a result, our DPO refers specifically to the distribution of connections under whatever collection of common priors is chosen by the community, and any change in these priors is interpreted as a policy effect. Related models of strategic network formation are also considered by \cite{ridder2015estimation,menzel2015strategic}. \looseness=-1  
\end{example}

\begin{remark}
Following the second example literature on exchangeable arrays, we normalize the marginal distributions of the entries of $\{w_{i,s}\}_{i=1}^{N}$ and $\{\eta_{ij,s}\}_{i,j = 1}^{N}$ to be standard uniform. This normalization is without loss of generality since the marginal distributions of $w_{i,s}$ and $\eta_{ij,s}$ can be absorbed into the link function $g_{N,s}$. It is done to simplify the notation. \looseness=-1 
\end{remark}

\begin{remark}
Following the second example literature on exchangeable arrays, allowing the function $g_{N,s}$ to vary with $N$ allows for sparse network asymptotics. For the identification results of this paper, the fact that $g_{N,s}$ varies with $N$ is not important, and so we suppress this dependence in our notation, simply writing $g_s$. In particular, the main identification results in Sections 4 and 5 are valid regardless of the level of sparsity. We give an intuition for why this is the case in Remark 9 below. Our estimation and inference results in Online Appendix Section D.3, however, rely on a class of dyadic regression models whose link function does not depend on $N$, and so does not accommodate sparse network asymptotics. We think it is possible to extend those results to the sparse regime, but leave the details to future work. \looseness=-1 
\end{remark}

\begin{remark}
As highlighted by the third example literature, the model (\ref{model}) can accommodate settings with strategic interactions between agents. When the policy consists of assigning an individualized treatment to agents, the model can, in a similar way, allow for treatment spillovers. For example, if the magnitude of a connection between agents $i$ and $j$ depends on the total number of agents in the community that have been assigned to treatment under the policy, this variable can also be absorbed into the link function $g_{N,s}$. If the connection depends on the number of treated agents that are nearby agents $i$ and $j$ as measured by some distance, this can be incorporated as a function of the agent effects $w_{i}$ and $w_{j}$. \looseness=-1 
\end{remark}

\subsubsection{Relating the network connections across policies}
In the model (\ref{model}), the agent effects $w_{i,s}$, idiosyncratic errors $\eta_{ij,s}$, and link function $g_{s}$ (where the dependence on $N$ has been suppressed- see Remark 7 above) may all vary with the policy $s$. We do not place any additional restrictions on the agent effects or the link functions. That is, the two link functions $g_1$ and $g_2$ can be any measurable functions and the vector $\{w_{i,1},w_{i,0}\}$ is allowed to have an arbitrary joint distribution subject to the normalizing of the marginal distributions of $w_{i,1}$ and $w_{i,0}$ made in Remark 6. We represent the distribution of $\{w_{i,1},w_{i,0}\}$ in a particular way. Specifically, by Lemma 2.7 of Whitt (1976), there exists a policy-invariant $w_i$ and unknown measurable functions $\varphi_{1}, \varphi_{0}: [0,1] \to [0,1]$ such that $w_{i,s} = \varphi_{s}\left(w_i\right)$ for $s \in \{0,1\}$ where the entries of $\{w_i\}_{i=1}^{N}$ can be taken to be independent with standard uniform marginal distributions. We emphasize that this representation is without loss of generality.  Since the marginal distributions of $w_{i,1}$, $w_{i,0}$ and $w_i$ are all standard uniform, the functions $\varphi_1$ and $\varphi_0$ are measure-preserving. This property plays a key role in our analysis, and so we display it as \looseness=-1 
\begin{definition}
A function $\psi: [0,1] \to [0,1]$ is a measure-preserving transformation if for any Lebesgue-measurable $A \subseteq [0,1]$ we have that $|\psi^{-1}(A)| = |A|$ where $|\cdot|$ refers to the Lebesgue measure. We denote the set of all measure-preserving transformations with $\mathcal{M}$.  \looseness=-1 
\end{definition}
Intuitively, the measure-preserving transformation is an infinite-dimensional analog of the permutation that we used to characterize the identified set in the simplified setting of Section 2. It follows from this representation that the magnitude of the connection between agents $i$ and $j$ under policy $s$ is represented by the model  \looseness=-1 
\begin{align}\label{model2}
Y_{ij}(s) =  g_{s}(\varphi_s(w_{i}),\varphi_s(w_{j}),\eta_{ij,s})~.
\end{align}
  \looseness=-1

Finally, we restrict the idiosyncratic errors to be unrelated across the two policies. That is, we assume that the entries of $\{\eta_{ij,1}\}_{i,j = 1}^{N}$ and  $\{\eta_{ij,0}\}_{i,j = 1}^{N}$ are mutually independent. We interpret this restriction as saying that a change in policy does not alter the idiosyncratic errors in any systematic way.   \looseness=-1 

\subsubsection{Main identifying assumption}
We close our specification of the model by highlighting our main identifying assumption. It is that the community assignments are unrelated to the factors that determine the network connections under each policy. That is,  \looseness=-1 
\begin{flushleft}
\textbf{Assumption 1:}  $\{D_i\}_{i=1}^{N}$ and $\{w_{i,1},w_{i,0},\eta_{ij,1},\eta_{ij,0}\}_{i,j=1}^{N}$ have mutually independent entries. 
\end{flushleft}
Since $Y_{ij}(s)$ is determined by $\{w_{i,s},w_{j,s},\eta_{ij,s}\}$, Assumption 1 implies that the entries of $\{D_i\}_{i=1}^{N}$ and $\{Y_{ij}(1), Y_{ij}(0)\}_{i,j=1}^{N}$ are mutually independent. \looseness=-1 

Researchers often justify Assumption 1 in practice by conducting a randomized experiment. The assumption is also common in non-experimental settings when the researcher believes that the communities in the data are balanced so that the agent interactions in one community under the policy implemented in that community describe how the agents in the other community would interact under that same policy. \looseness=-1 

It follows from Assumption 1 that the distribution of the observed data $\{Y,D\}$ is parametrized by the functions $\{g_s,\varphi_s\}_{s \in \{0,1\}}$. Specifically, for any constants $\{y_{ij}, d_i\}_{i,j=1}^{2N} \in \mathbbm{R}^{2N\times 2N}\times \{0,1\}^{2N}$ the distribution function $\mathbbm{P}\left(\{Y_{ij} \leq y_{ij}, D_i = d_i\}_{i,j=1}^{2N}\right)$ can be written as \looseness=-1 
\begin{align}\label{df}
&\int_{u_1,\ldots,u_{2N}}\prod_{i,j=1}^{2N}\left[ \int_{w}\mathbbm{1}\left\{\sum_{s \in \{0,1\}}g_{s}\left(\varphi_{s}(u_i),\varphi_s(u_j),w\right)d_i^sd_j^s(1-d_i)^{1-s}(1-d_j)^{1-s} \leq y_{ij}\right\}dw\right] du_1,\ldots,du_{2N} \nonumber \\
&\hspace{50mm}\times {2N\choose N}^{-1}\mathbbm{1}\left\{\sum_{i=1}^{2N}d_i = N\right\}.
\end{align}
In words, the first line describes the conditional distribution of the network connections given the community assignments, $\mathbbm{P}\left(\{Y_{ij} \leq y_{ij}\}_{i,j=1}^{2N}|\left\{D_i = d_i\right\}_{i=1}^{2N}\right)$. The second line describes the marginal distribution of the community assignments $\mathbbm{P}\left(\{D_i = d_i\}_{i=1}^{2N}\right)$. They are both determined by $N$, the constants $\{y_{ij}, d_i\}_{i,j=1}^{2N}$, and parameters $\{g_s,\varphi_s\}_{s \in \{0,1\}}$.\looseness=-1


\section{Identification}
In this section, we characterize the identified set for the DPO given the distribution of the data $\{Y,D\}$ under the model of Section 3 as parametrized by $\{g_s,\varphi_s\}_{s \in \{0,1\}}$. Our characterization uses a (slight variation on a) construction from the graph theory literature called a graph function. See Chapter 7.1 of \cite{lovasz2012large} for a textbook introduction. We first define this construction and then use it to characterize the identified set. \looseness=-1

\subsection{Graph function}
For any $s \in \{0,1\}$ and $y_s \in \mathbb{R}$, we define the graph function $h_{s}: [0,1]^2\times \mathbbm{R} \to [0,1]$ to be
  \looseness=-1
\begin{align*}
h_{s}(a,b; y_s) := \mathbbm{P}\left(Y_{ij}(s) \leq y_s| w_{i} = a, w_{j} = b\right) = \int \mathbbm{1}\{g_{s}(\varphi_s(a),\varphi_s(b),w) \leq y_s\}dw.
\end{align*}
In words, $h_{s}(a,b;y_s)$ is the conditional probability that the magnitude of a connection between agent $i$ with $w_i = a$ and agent $j$ with  $w_j = b$ is less than $y_s$ under policy $s$. The function depends on a fixed choice of  $y_s$, which we suppress to simplify our notation, simply writing $h_s(a,b,;y_s) = h_s(a,b)$. It also implicitly depends on the community size $N$ through $g_s$, which we also suppress.   \looseness=-1

We do not distinguish between graph functions that are equivalent up to a measure-preserving transformation. That is,
\begin{definition} Two graph functions $h$ and $h'$ are equivalent up to a measure-preserving transformation at a fixed $y \in \mathbb{R}$ if there exists a $\psi, \psi' \in \mathcal{M}$ such that  \looseness=-1
\begin{align}\label{mpt}
h(\psi(a),\psi(b);y) = h'(\psi'(a),\psi'(b);y) \text{ for almost every } (a,b) \in [0,1]^2
\end{align} 
where $\mathcal{M}$ is the set of measure preserving transformations from Definition 1 of Section 3.2.3 and ``almost every'' refers to the Lebesgue measure. When (\ref{mpt}) holds for every $y \in \mathbb{R}$, we write that $h_{s} \sim h'_{s}$. Otherwise $h_{s} \not\sim h'_{s}$.  \looseness=-1
\end{definition}
The reason why we only consider graph functions defined up to a measure-preserving transformation is because it is the limit of what is identified from the data. That is, \looseness=-1

\begin{flushleft}
\textbf{Lemma 1:} The graph functions $h_{1}$ and $h_0$ are identified from the distribution of $\{Y,D\}$ up to a measure-preserving transformation. \end{flushleft}

The proof of Lemma 1 can be found in Appendix Section A.2. When we write that $h_{s}$ is ``identified up to a measure-preserving transformation,'' we make two claims. The first claim is that the distribution of the data can distinguish between two graph functions that are not equivalent up to a measure-preserving transformation. That is, formally, suppose that $\{g_s,\varphi_s\}_{s \in \{0,1\}}$  and $\{g'_s,\varphi'_s\}_{s \in \{0,1\}}$ are two sets of model parameters that generate the data $\{Y,D\}$ and $\{Y',D'\}$. Let $h_{s} =  \int \mathbbm{1}\{g_{s}(\varphi_s(a),\varphi_s(b),w) \leq y_s\}dw$  and $h'_{s} =  \int \mathbbm{1}\{g'_{s}(\varphi'_s(a),\varphi'_s(b),w) \leq y_s\}dw$ be the graph functions associated with these two models. Then the first claim is that $h_{1} \not\sim h'_{1}$ or $h_{0} \not\sim h'_{0}$ implies that $\{Y,D\}$ and $\{Y',D'\}$ do not have the same distribution (i.e. they have different distribution functions (\ref{df})). \looseness=-1

The second claim in the statement of Lemma 1 is that the distribution of the data cannot distinguish between two graph functions that are equivalent up to a measure-preserving transformation. That is, in the above setting, if $h_{1} \sim h'_{1}$ and $h_0 \sim h'_0$ then $\{Y,D\}$ and $\{Y',D'\}$ have the same distribution function  (\ref{df}).\looseness=-1

\subsection{Identified set}
To derive the identified set for the DPO, we represent it as the inner product of $h_{1}$ and $h_{0}$.
\begin{align*}
DPO = \mathbbm{P}\left(Y_{ij}(1) \leq y_1, Y_{ij}(0) \leq y_0\right)
&= \mathbbm{E}\left[\mathbbm{P}\left(Y_{ij}(1) \leq y_1| w_i, w_j\right)\mathbbm{P}\left(Y_{ij}(0) \leq y_0| w_i, w_j\right)\right] \\
&= \int\int h_{1}(u,v)h_{0}(u,v)dudv
\end{align*}
where the first equality is due to the definition of the DPO in Section 3.1.2, the second equality follows from the law of iterated expectations, and the third equality is due to the definition of the graph function in Section 4.1. The identified set for the DPO then follows immediately from the fact that $h_1$ and $h_0$ are identified up to a measure-preserving transformation (Lemma 1 of Section 4.1). That is,  \looseness=-1

\begin{flushleft}
\textbf{Proposition 1:} The identified set for the DPO is 
\begin{align}\label{identified set}
\left\{ \int\int \prod_{s \in \{0,1\}} h_{s}(\psi_s(u),\psi_s(v))dudv \in \mathbb{R} : \psi_1, \psi_0 \in \mathcal{M}\right\}. 
\end{align}
\end{flushleft}
In words, (\ref{identified set}) is the set of all values that the DPO can potentially take given the distribution of $\{Y,D\}$ as determined by the model described in Section 3.  \looseness=-1

\section{Bounds on the identified set}
We first describe sharp but infeasible bounds on the DPO. We then propose tractable outer bounds on the DPO and DTE. Our proposed bounds are based on rearrangements of the eigenvalues of the graph functions $h_1$ and $h_0$. Eigenvalues of functions are defined a bit differently than their matrix counterparts, see Appendix Section A.1 for details. Proof of claims are in Appendix Sections A.3 and A.4. \looseness=-1

\subsection{Sharp bounds are intractable}
Proposition 1 implies that (pointwise) sharp bounds on the identified set for the DPO are \looseness=-1
\begin{align}\label{qap}
\min_{\psi_{0},\psi_{1} \in \mathcal{M}}\int\int \prod_{s \in \{0,1\}} h_{s}(\psi_s(u),\psi_s(v))dudv 
\leq F(y_{1},y_{0}) \nonumber \\ \leq  \max_{\psi_{0},\psi_{1} \in \mathcal{M}}\int\int \prod_{s \in \{0,1\}} h_{s}(\psi_s(u),\psi_s(v))dudv.
\end{align} 
These bounds are sharp because $h_1$ and $h_0$ are only identified up to a measure preserving transformation by Lemma 1, and so, as far as the distribution of the data is concerned, any choice of $\psi_1, \psi_0 \in \mathcal{M}$ could be the one that defines the true DPO. They suggest a natural, though infeasible, two-step estimation strategy. In the first step, the researcher constructs an estimator  $\hat{h}_{s}$ for the graph function $h_{s}$. In the second step, the researcher computes the empirical analog of (\ref{qap}), \looseness=-1
\begin{align}\label{naive estimator}
\left\{\min_{\psi_{0},\psi_{1} \in \mathcal{M}}\int\int \prod_{s \in \{0,1\}} \hat{h}_{s}(\psi_s(u),\psi_s(v))dudv ,  \max_{\psi_{0},\psi_{1} \in \mathcal{M}}\int\int \prod_{s \in \{0,1\}} \hat{h}_{s}(\psi_s(u),\psi_s(v))dudv\right\}.
\end{align}

The first step of this strategy is feasible. That is, even though both the graph function $\{h_1,h_0\}$ and the agent effects $\{w_i\}_{i=1}^{N}$ are unobserved, it is possible to consistently estimate the function $\hat{h}_s$ (up to a measure-preserving transformation) under relatively mild conditions. See, for instance, \cite{olhede2014network,chatterjee2015matrix,zhang2015estimating}. \looseness=-1

It is the second step of the strategy that is infeasible. In particular, the optimization problem described in this step is both analytically and computationally intractable, even for relatively small community sizes, because the function $\psi_{s}$ appears twice in the optimization problems on the right and left-hand sides of (\ref{qap}). As a result, is generalizes the quadratic assignment problem described in Section 2. Solutions to this problem are only known for stylized examples of $Y_{1}$ and $Y_{0}$ that are not relevant for the kinds of network data typically observed by economists. See Online Appendix Section C.3 or \cite{cela2013quadratic} Section 1.5 for examples. \looseness=-1 

\begin{remark}
The bounds in Proposition 1 are pointwise sharp, but not sharp as a function of $(y_1,y_0) \in \mathbb{R}^{2}$. See Section 2.1 of \cite{molinari2020microeconometrics} for a discussion.
\end{remark}

\subsection{Proposed outer bounds}
Our proposed bounds are based on the intersection of several relaxations of (\ref{qap}). Our bounds are not sharp, but they are tractable and use enough information from Lemma 1 to outperform conventional methods in many settings. Our main idea is to rearrange the eigenvalues of the graph functions associated with each policy. Specifically, let $\lambda_{1s}(y_{s}) \geq  \lambda_{2s}(y_{s}) \geq ... \geq \lambda_{Rs}(y_{s})$ be the $R$ largest in magnitude eigenvalues of $h_{s}$ ordered to be decreasing and $\rho_{R}(r) = R - r + 1$. For $s, s' \in \{0,1\}$, let $\sum_{r}\lambda_{rs}\lambda_{rs'} := \lim_{R \to \infty}\sum_{r=1}^{R}\lambda_{rs}(y_{s})\lambda_{rs'}(y_{s'})$, $\sum_{r}\lambda_{rs}\lambda_{\rho(r)s'} :=    \lim_{R \to \infty}\sum_{r=1}^{R}\lambda_{rs}(y_{s})\lambda_{\rho_{R}(r)s'}(y_{t'})$ and $\sum_{r}\lambda_{rs}^{2} := \sum_{r}\lambda_{rs}\lambda_{rs}$.\footnote{These limits exist because the series $\{\lambda_{rs}\lambda_{rs'}\}_{r\in\mathbb{N}}$ and $\{\lambda_{rs}\lambda_{\rho_{R}(r)s'}\}_{r\in\mathbb{N}}$ are absolutely summable by the Cauchy-Schwarz inequality and the fact that $\sum_r \lambda_{rs}^2 = \int\int h_s(u,v)^2dudv \leq 1$ for $s,s' \in \{0,1\}$.}  We note that the eigenvalues of $h_s$ are identified because $h_s$ is identified up to a measure preserving transformation by Lemma 1 of Section 4.1, and any two graph functions that are equivalent up to a measure preserving transformation necessarily have the same eigenvalues.   \looseness=-1 

Our proposed bounds on the DPO are  \looseness=-1 
\begin{flushleft}
\textbf{Proposition 2:} Suppose Assumption 1. Then for any $(y_{1},y_{0}) \in \mathbb{R}^{2}$
\begin{align}\label{SFH}
\max\left(\sum_{r}\left(\lambda_{r1}^{2} + \lambda_{r0}^{2}\right) - 1,\sum_{r}\lambda_{r1}\lambda_{s(r)0},0\right)
\leq F(y_{1},y_{0})  \nonumber \\
\leq \min\left(\sum_{r}\lambda_{r1}^{2},\sum_{r}\lambda_{r0}^{2},\sum_{r}\lambda_{r1}\lambda_{r0}\right).
\end{align} 
\end{flushleft}
where the eigenvalue $\lambda_{rs}$ is implicitly a function of $y_{s}$ through the definition of $h_{s}$.

The proof of Proposition 2 can be found in Appendix Section A.4. The result is similar in spirit to the conventional Fr\'echet-Hoeffding bounds, but builds on a relaxation of (\ref{qap}), and so the arguments behind the proofs are fundamentally different. Intuitively, a common way to derive the Fr\'echet-Hoeffding bounds is to rearrange the quantiles of the distribution of outcomes associated with each policy. See, for instance, the second proof of Theorem 2.1 in \cite{whitt1976bivariate}. Our bounds instead work by rearranging the eigenvalues of the graph functions building on a proposal by \cite{finke1987quadratic} for the finite dimensional QAP described in Section 2.2.2. That their logic extends to graph functions and so applies to the nonparametric model described in Section 3.1 is  not obvious, requires arguments from functional analysis that are not typical of the QAP literature, and is,  to our knowledge, original to our paper. Intuitively, the \cite{finke1987quadratic} bounds work in the finite dimensional case because the rows and columns of the matrix of eigenvectors are orthogonal.  An analogous property does not hold for eigenfunctions and so we instead bound the DPO on a sequence of histogram-like approximations. See our Lemmas 1 and 3 in Appendix Section A.3. Unlike the bounds in  (\ref{qap}), those in  (\ref{SFH}) are tractable because they only depend on the eigenvalues of $h_{s}$. The bounds are also straightforward to estimate, see Sections 5.3 and Online Appendix Section D.3 below. \looseness=-1 

The bounds on the DPO can be used to bound the DTE. Our proposed bounds are
\begin{flushleft}
\textbf{Proposition 3:} Suppose Assumption 1. Then for any $y \in \mathbb{R}$
\begin{align}\label{SM}
\sup_{\substack{(y_{1},y_{0}) \in \mathbb{R}^{2}:\\ y_{1}-y_{0} = y}}\max\left(\sum_{r}\left(\lambda_{r1}^{2} - \lambda_{r0}^{2}\right),\sum_{r}\left(\lambda_{r1}^{2}-\lambda_{r1}\lambda_{r0}\right),0\right)
\leq \Delta(y) \nonumber \\
\leq 1 + \inf_{\substack{(y_{1},y_{0}) \in \mathbb{R}^{2}:\\ y_{1}-y_{0} = y}}\min\left(\sum_{r}\left(\lambda_{r1}^{2} - \lambda_{r0}^{2}\right),\sum_{r}\left(\lambda_{r1}\lambda_{r0}- \lambda_{r0}^{2}\right),0\right)
\end{align}
\end{flushleft}
where the eigenvalue $\lambda_{rs}$ is implicitly a function of $y_{s}$ through the definition of $h_{s}$. The proof of Proposition 3 can be found in Appendix Section A.5. The result is similar in spirit to the conventional Makarov bounds, but uses our Proposition 2 instead of Fr\'echet-Hoeffding. 

\begin{remark}
Some readers may be surprised to see that the validity of Propositions 1-3 do not require any restrictions on the sparsity of the network. That is, these results hold for any link functions $f_1$ and $f_0$, which are allowed to vary with $N$ in an arbitrary way, which includes taking values that are arbitrarily close to $0$. Intuitively, the reason why Propositions 1-3 do not depend on the level of network sparsity is because the DPO is just a measure of the mass of connections between agents. If the network is sparse, the mass of connections will concentrate at $0$, and the DPO will concentrate at either $0$ or $1$ (depending on the signs of $y_1$ and $y_0$). But the data contains the same amount of information about the DPO regardless of how much concentration occurs. In fact, in the extreme case that the network is so sparse that $Y_{ij} = 0$ for every pair of agents, the DPO is point identified and equal to $\mathbbm{1}\{\min(y_1,y_0) \geq 0\}$. In this case, both the lower and upper bound in Proposition 2 will also collapse to this point. \looseness=-1 
\end{remark}

\begin{remark}
In our empirical work, we have found that the bounds described in (\ref{SFH}) and (\ref{SM}) are typically improved by adjusting for row and column heterogeneity.  We describe this procedure in Online Appendix Section D.2 and use it in our empirical work in Section 6 below. \looseness=-1 
\end{remark}

\subsection{Estimating the bounds}
We propose estimating the bounds in Propositions 2 and 3 using a feasible version of the plug-in strategy described in Section 5.1. In the first step, the researcher constructs an estimator for $\hat{h}_s$ for the graph function $h_s$. In our Online Appendix Section D.3 we estimate the graph function in the dyadic regression setting  (the first example literature we provided in Section 3.2.2)  under a parametric restriction, as presented in Section 4 of \cite{graham2020network}. But parametric restrictions are not necessary here, see for instance the USVT estimator of \cite{chatterjee2015matrix}. We emphasize that since $h_s$ is only identified up to a measure preserving transformation, the goal is not to approximate the function $h_s$ exactly, but only to approximate an element of the identified set. \looseness=-1

In the second step, the researcher computes the eigenvalues of the function $\hat{h}_s$ and uses them as a substitute for those of $h_s$ in (\ref{SFH}) and (\ref{SM}). Intuitively, since eigenvalues are continuous (see Lemma B3 in Online Appendix Section B) and invariant to measure preserving transformations, the estimation error for these bounds should be small in large samples if the estimator $\hat{h}_s$ is consistent in the above sense. We formalize this logic in the dyadic regression setting in Online Appendix Section D.4. We also give a procedure to construct confidence intervals for the DPO and provide sufficient conditions for asymptotic validity. \looseness=-1 

We focus on the dyadic regression models because some version of this model is used in nearly all of the motivating examples we reference in our introduction. We believe that it is straightforward to apply our arguments to the exchangeable graph and strategic link formation settings (the other two example literatures described in Section 3.2.2), following the estimation and inference arguments made in these literatures, but we leave this to future work. \looseness=-1

\section{Empirical illustration}
In this section, we apply our framework to the setting of  \cite{banerjee2021changes}. Section 6.1 reviews background information, Section 6.2 applies the framework of Section 3, and Section 6.3 applies the estimation and inference results of Online Appendix Section D.3. The results are in Section 6.4. \looseness=-1 

\subsection{Background}
 \cite{banerjee2021changes} study the effect of a microfinance program on informal risk sharing in Karnataka, India. The authors find, among other things, that participating villages have one percentage point (1pp) less informal risk sharing links between households. We find disruptive effects that are sixteen to twenty-three times larger using our methodology. \looseness=-1 

The Karnataka study is centered around the planned introduction of microfinance in 75 villages by Bharatha Swamukti Samsthe (BSS). BSS selected 43 of these villages in 2006 and implemented the program between 2007 and 2010. They originally planned to implement the program in all of the  villages, but ultimately did not because of an external crisis. \cite{banerjee2021changes} argue that the villages are comparable after controlling for village size. The authors collected data on informal risk sharing connections between households at two time periods: before and after BSS implemented the program in the selected villages.  \looseness=-1 

We reanalyze the \cite{banerjee2021changes} data using our framework.\footnote{The data can be found at \url{https://zenodo.org/record/7706650\#.ZD9Tti-B2gQ}.} We focus on two villages: village 57, which participated in the microfinance program, and village 44, which did not. We chose these villages because they are the most similar in terms of pre-treatment covariates across all pairs of villages.\footnote{Specifically, we normalized the pre-treatment village-level covariates to have a standard deviation of 1. We picked the treatment-control pair whose pre-treatment covariates have the smallest Euclidean distance.}  \looseness=-1 

\subsection{Application of our framework}
We extend our Section 3 framework to the panel setting of \cite{banerjee2021changes}. Let $Y_{ij,t}(s)$ indicate whether there would be a risk sharing connection between households $i$ and $j$ when a village participates $(s = 1)$ or does not participate $(s = 0)$ in the microfinance program before $(t = 0)$ or after $(t = 1)$ the program is implemented in the participating villages. Following \cite{banerjee2021changes}, we take as the outcome of interest the change in network connections for a pair of households over time $\Delta Y_{ij}(s) := Y_{ij,1}(s) - Y_{ij,0}(s)$. These network connections are ternary, taking values in $\{-1,0,1\}$. \looseness=-1 


Our proposed measures of disruption are the expected fraction of risk sharing links created by implementing the microfinance program in a village,
\begin{align*}
C = \mathbbm{P}\left(\Delta Y_{ij}(1) > \Delta Y_{ij}(0)\right),
\end{align*}
and the expected fraction of risk sharing links destroyed by implementing the microfinance program in the village, 
\begin{align*}
D = \mathbbm{P}\left(\Delta Y_{ij}(1) < \Delta Y_{ij}(0)\right)
\end{align*}
where $\mathbbm{P}$ refers to the distribution of network connections as given by the model (\ref{model2}) in Section 3.2. Following \cite{banerjee2021changes}, we specify a dyadic regression model for our estimation and inference results in Section 6.3.\looseness=-1 

To bound $C$ and $D$, we first bound a discrete analog of the DPO, the joint density function\looseness=-1 
\begin{align*}
f(y_1,y_0) &= \mathbbm{P}\left( \Delta Y_{ij}(1) = y_1, \Delta Y_{ij}(0) = y_0\right) \\
&= \mathbbm{E}\left[ \mathbbm{P}\left(\Delta Y_{ij}(1) = y_1 | w_{i}, w_{j}\right) \mathbbm{P}\left(\Delta Y_{ij}(0) = y_0 | w_{i}, w_{j}\right)\right] \\
&= \int\int h_{1}(u,v)h_{0}(u,v)dudv
\end{align*}
where $y_1, y_0 \in \{-1,0,1\}$ is arbitrary and $h_{s}(u,v) = \mathbbm{P}\left(\Delta Y_{ij}(s) = y_s | w_i = u, w_j = v \right)$. We then use the bounds on $f(y_1,y_0)$ to bound $C$ and $D$ via the formulas
\begin{align*}
C = f(0,-1)+f(1,-1)+f(1,0) \text{ and }  
D = f(-1,0)+f(-1,1)+f(0,1). 
\end{align*}

We use a variation on the bounds in Proposition 2 of Section 4.1.2 that corrects for row and column heterogeneity, see Online Appendix Section D.2. Specifically, we define $\alpha_{s}(u) = \mathbbm{P}\left(\Delta Y_{ij}(s) = y_s | w_i = u\right) - \frac{1}{2}\mathbbm{P}\left(\Delta Y_{ij}(s) = y_s\right)$ and $\epsilon_s(u,v) = \mathbbm{P}\left(\Delta Y_{ij}(s) = y_s| w_i = u, w_j = v\right) - \alpha_{s}(u) - \alpha_{s}(v)$. Our lower bound on $f(y_1,y_0)$ is $$2\int\alpha_{1}^{+}(u)\alpha_{0}^{+}(1-u)du + 2\bar{\alpha}_{1}\bar{\alpha}_{0} + \max\left(\sum_{r}\left(\lambda_{r1}^{2} + \lambda_{r0}^{2}\right) - 1,\sum_{r}\lambda_{r1}\lambda_{\rho(r)0},0\right)$$ and our upper bound on $f(y_1,y_0)$ is $$2\int\alpha_{1}^{+}(u)\alpha_{0}^{+}(u)du + 2\bar{\alpha}_{1}\bar{\alpha}_{0} + \min\left(\sum_{r}\lambda_{r1}^{2},\sum_{r}\lambda_{r0}^{2},\sum_{r}\lambda_{r1}\lambda_{r0}\right)$$
where $\alpha_{s}^{+}$ is the quantile function of $\alpha_{s}$, $\bar{\alpha}_s = \int\alpha_s(u)du$,  $\lambda_{rs}$ is the $r$th eigenvalue of the function $\epsilon_s$, and the sums are as defined in Section 4.1.2. As we discuss in Online Appendix Section D.2, $\alpha^{+}_s(\cdot)$ and $\bar{\alpha}_s$ are identified from the distribution of $Y$ and $D$, while $\epsilon_s(\cdot,\cdot)$ is identified up to a measure-preserving transformation following the logic of Lemma 1 in Section 4.1.  \looseness=-1 

\subsection{Estimation and inference}
Our estimation and inference results use a parametric dyadic regression model along the lines specified by \cite{banerjee2021changes} in their Section 3.2. Specifically, we use \looseness=-1 
\begin{align}\label{parametric}
	h_{s}(w_i,w_j) := \mathbbm{P}\left(\Delta Y_{ij}(s) = y_s | w_{i}, w_{j}\right)  =  \Lambda\left(\tau_s + \sum_{k=1}^{K}(x_{ks}(w_{i}) - x_{ks}(w_{j}))^2\beta_{ks}\right)
\end{align}
where $\Lambda$ is the identity link function, $X_{i,s} = x_s(w_i)$ is a vector of $K$ observed household-specific characteristics represented as a function of $w_{i}$ following the logic of Example 1 in our Section 3.2.2, $\tau_s$ is an unknown intercept, $\beta_s$ is an unknown $K$-dimensional vector of coefficients, and $x_{ks}$ and $\beta_{ks}$ refer to the $k$th entry of $x_{s}$ and $\beta_{s}$.\footnote{While we follow \cite{banerjee2021changes} and take $\Lambda$ to be the identity link function, our theoretical results in Online Appendix Section D.3 allow for an arbitrary twice continuously differentiable function such as the logistic function or normal cdf. We use the same household covariates as  \cite{banerjee2021changes}, but exclude network statistics (functionals of $Y$ such as degree or eigenvector centrality) because of endogeneity concerns.} 

\begin{remark}
A benefit of using this dyadic regression model is that it allows us to directly compare the results we find with our proposed bounds to the results that those authors find in their regression analysis. In particular, when we find that our estimated bounds provide much larger disruptive effects in Section 6.4 below, we can conclude that this difference is because we focus on a different measure of disruption, and not because we model the effect of the microfinance policy on the network connections in a completely different way.\footnote{In a previous version of this paper, we estimated bounds for $C$ and $D$ nonparametrically using the USVT estimator proposed by Chatterjee (2015) and found disruptive effects of similar magnitudes.} \looseness=-1 
\end{remark}


\begin{remark}
We emphasize that imposing the dyadic regression model has no effect on the identification results of Section 4 or the validity of the bounds in Section 5. This is because, in those sections, we found that the DPO was determined by the graph functions $h_1$ and $h_0$ which were identified up to a measure-preserving transformation. Imposing the dyadic regression model restricts the shape of the function $h_s$, but still does not distinguish between graph functions  that are equivalent up to a measure preserving transformation. As a result, the identified set remains that described in Proposition 1. \looseness=-1 
\end{remark}

\subsection{Empirical findings}
\subsubsection{Estimated amount of social disruption using standard tools }
We first recreate the results of \cite{banerjee2021changes} . Using difference-in-differences, \cite{banerjee2021changes} find that the villages selected for the program experience a greater decline in social connections, with an ATE of approximately -1pp. They also estimate average treatment effects conditional on whether two households were linked in the pre-treatment period and whether households have a high (H) or low (L) propensity to borrow money from the microfinance program. For two households that were linked before treatment, those in the treated village were 6pp more likely to have their link destroyed. For two households that were not linked, those in the treated village were 2pp less likely to have a link form between them. The authors also find that the microfinance program has larger effects on connections between L households than H households. \looseness=-1  

We reproduce these results using villages 57 and 44 in Table 1. Specifically, Table 1 contains sample analogs of the average treatment effect (ATE), the average treatment effect conditional on being connected in the pre-treatment period (CATE(1)), and the average treatment effect conditional on not being connected in the pre-treatment period (CATE(0))\looseness=-1
\begin{align*}
		\text{ATE} &:= \mathbb{E}\left[\Delta Y_{ij}(1)\right] - \mathbb{E}\left[ \Delta Y_{ij}(0)\right] \\
		\text{CATE}(1) &:=  \mathbb{E}\left[\Delta Y_{ij}(1)| Y_{ij,0}(1) = 1\right]  - \mathbb{E}\left[\Delta Y_{ij}(0)| Y_{ij,0}(0) = 1\right] 	\\
		\text{CATE}(0) &:=   \mathbb{E}\left[\Delta Y_{ij}(1)| Y_{ij,0}(1) = 0\right]  - \mathbb{E}\left[\Delta Y_{ij}(0)| Y_{ij,0}(0) = 0\right] \end{align*} \looseness=-1
for the full sample of households, as well as subsamples of pairs of households that are both type $H$ ($HH$), both type $L$ ($LL$), or mixed ($HL$). 

\begin{table}[htbp]
  \centering
  \title{Table 1: Average and conditional average treatment effects} 
  \footnotesize
	\hspace*{-1.2cm}
    \begin{tabular}{clccccc}
	    \hline\hline
	          &       & ATE   & CATE(1) &  CATE(1)$\cdot P(1)$     & CATE(0) &  CATE(0)$\cdot P(0)$\\
	          \midrule
	    \multirow{2}[1]{*}{Full} & Est   & -0.0053 & -0.0652 & -0.0034 & 0.0028 & 0.0027 \\
	          & CI    & [-0.0107 , 0.0001] & [-0.0983 , -0.0321] & [-0.0052 , -0.0017] & [-0.0021 , 0.0078] & [-0.0020 , 0.0074] \\
	    \multirow{2}[0]{*}{HH} & Est   & -0.0022 & -0.0487 & -0.0043 & 0.0144 & 0.0132 \\
	          & CI    & [-0.0142 , 0.0098] & [-0.1035 , 0.0062] & [-0.0092 , 0.0005] & [0.0035 , 0.0254] & [0.0032 , 0.0231] \\
	    \multirow{2}[0]{*}{LL} & Est   & -0.0055 & -0.0591 & -0.0025 & 0.0088 & 0.0084 \\
	          & CI    & [-0.0152 , 0.0043] & [-0.1371 , 0.0190] & [-0.0058 , 0.0008] & [-0.0003 , 0.0178] & [-0.0003 , 0.0170] \\
	    \multirow{2}[0]{*}{HL} & Est   & -0.0090 & -0.1428 & -0.0070 & 0.0042 & 0.0040 \\
	          & CI    & [-0.0169 , -0.0012] & [-0.1948 , -0.0907] & [-0.0095 , -0.0044] & [-0.0032 , 0.0116] & [-0.0030 , 0.0111] \\
	    \hline\hline
    \end{tabular}%
	\begin{flushleft} \footnotesize \linespread{.75} Table 1 reports point estimates of various average treatment effects described in Section 6.4.1 and $95\%$ confidence intervals. \normalsize \end{flushleft}
\end{table}%

The average treatment effects in Table 1, particularly the ATEs, are on the order of $1$pp or smaller, consistent with the findings of \cite{banerjee2021changes}. The table does reveal nominally large negative effects for CATE(1), particularly for the HL subsample. However, the fraction of household pairs that are connected in the pre-treatment period is relatively small, so that total disruptive effect revealed by these estimators is still on the order of $1$pp. We demonstrate this in columns 3 and 5 of Table 1, where we scale CATE(1) by the (empirical analog of) the mass of links in the treatment group in the pre-treatment period,  $P(1) := \mathbbm{P}\left(Y_{ij,0}(1) = 1\right)$, and similarly scale CATE(0) by $P(0) := \mathbbm{P}\left(Y_{ij,0}(1) = 0\right)$.  We find substantially larger effects using our bounds below.     \looseness=-1

\subsubsection{Estimated amount of social disruption using our bounds}
Our main results are presented in Table 2. We report both our point estimates for the upper and lower bounds on $f(y_1,y_0)$, as well as confidence intervals, for $y_1,y_0 \in \{-1,0,1\}$. The first three rows report results using all of the households in villages 57 and 44. Recall that for link destruction, the mass of destroyed connections is given by $D = f(-1,0) + f(-1,1) + f(0,1)$. We find that, with high probability, \looseness=-1 $$f(-1,0) \in [0.0296, 0.0428] \quad , \quad f(0,1) \in [0.0517, 0.0725] $$
where $f(0,1)$ is the fraction of connections that would have been created if not for the microfinance program, $f(-1,0)$ is the fraction of connections that were destroyed because of the program, and $f(-1,1)$ is approximately $0$. The bounds together imply that the mass of destroyed connections $D \in [0.0830,0.1193]$ with high probability. Similarly for link creation, with $C = f(0,-1)+f(1,-1)+f(1,0)$, we find that, with high probability,\looseness=-1
$$f(0,-1) \in [0.0202, 0.0385] \quad ,\quad f(1,0) \in [0.0532, 0.0687]$$ where $f(1,-1)$ is also approximately $0$. These bounds together imply that the mass of created connections $C \in [0.0769,0.1103]$. The results for the HH, LL and HL subsamples are similar. 

Our bounds indicate that the microfinance program disrupted an approximately $D+C \in [0.1599, 0.2296]$ fraction of the connections between households. Using the -1pp ATE as a baseline, we conclude that the microfinance program is sixteen to twenty-three times more disruptive than what is indicated by the average treatment effects of Section 6.4.1. 
\looseness=-1

\begin{table}[htbp]
  \centering
  \title{Table 2: Bounds on the joint distribution of potential risk sharing links} \vspace{5mm}
    \begin{tabular}{cclccccc}\hline\hline
          &       &       & $\Delta Y_{ij,0} = -1$   &       & $\Delta Y_{ij,0} = 0$ &       & $\Delta Y_{ij,0} = 1$ \\
    \midrule
    \multirow{6}[6]{*}{Full} & \multirow{2}[2]{*}{$\Delta Y_{ij,1} = -1$} & Est   & [0.0013, 0.0016] &       & \textcolor[rgb]{ .753,  0,  0}{[0.0359, 0.0366]} &       & \textcolor[rgb]{ .753,  0,  0}{[0.0026, 0.0031]} \\
          &       & CI    & [0.0008, 0.0021] &       & \textcolor[rgb]{ .753,  0,  0}{[0.0296, 0.0428]} &       & \textcolor[rgb]{ .753,  0,  0}{[0.0017, 0.0040]} \\
\cmidrule{2-8}          & \multirow{2}[2]{*}{$\Delta Y_{ij,1} = 0$} & Est   & \textcolor[rgb]{ .188,  .329,  .588}{[0.0294, 0.0296]} &       & [0.7921, 0.7937] &       & \textcolor[rgb]{ .753,  0,  0}{[0.0619, 0.0626]} \\
          &       & CI    & \textcolor[rgb]{ .188,  .329,  .588}{[0.0202, 0.0385]} &       & [0.7753, 0.8106] &       & \textcolor[rgb]{ .753,  0,  0}{[0.0517, 0.0725]} \\
\cmidrule{2-8}          & \multirow{2}[2]{*}{$\Delta Y_{ij,1} = 1$} & Est   & \textcolor[rgb]{ .188,  .329,  .588}{[0.0021, 0.0024]} &       & \textcolor[rgb]{ .188,  .329,  .588}{[0.0607, 0.0615]} &       & [0.0044, 0.0051] \\
          &       & CI    & \textcolor[rgb]{ .188,  .329,  .588}{[0.0015, 0.0031]} &       & \textcolor[rgb]{ .188,  .329,  .588}{[0.0532, 0.0687]} &       & [0.0032, 0.0062] \\
    \midrule
    \multirow{6}[6]{*}{HH} & \multirow{2}[2]{*}{$\Delta Y_{ij,1} = -1$} & Est   & [0.0026, 0.0033] &       & \textcolor[rgb]{ .753,  0,  0}{[0.0589, 0.0606]} &       & \textcolor[rgb]{ .753,  0,  0}{[0.0037, 0.0045]} \\
          &       & CI    & [0.0003, 0.0055] &       & \textcolor[rgb]{ .753,  0,  0}{[0.0004, 0.0764]} &       & \textcolor[rgb]{ .753,  0,  0}{[0.0003, 0.0066]} \\
\cmidrule{2-8}          & \multirow{2}[2]{*}{$\Delta Y_{ij,1} = 0$} & Est   & \textcolor[rgb]{ .188,  .329,  .588}{[0.0372, 0.0382]} &       & [0.7460, 0.7489] &       & \textcolor[rgb]{ .753,  0,  0}{[0.0526, 0.0542]} \\
          &       & CI    & \textcolor[rgb]{ .188,  .329,  .588}{[0.0226, 0.0532]} &       & [0.6701, 0.8537] &       & \textcolor[rgb]{ .753,  0,  0}{[0.0354, 0.0723]} \\
\cmidrule{2-8}          & \multirow{2}[2]{*}{$\Delta Y_{ij,1} = 1$} & Est   & \textcolor[rgb]{ .188,  .329,  .588}{[0.0034, 0.0039]} &       & \textcolor[rgb]{ .188,  .329,  .588}{[0.0733, 0.0751]} &       & [0.0046, 0.0056] \\
          &       & CI    & \textcolor[rgb]{ .188,  .329,  .588}{[0.0011, 0.0063]} &       & \textcolor[rgb]{ .188,  .329,  .588}{[0.0134, 0.0922]} &       & [0.0007, 0.0080] \\
    \midrule
    \multirow{6}[6]{*}{LL} & \multirow{2}[2]{*}{$\Delta Y_{ij,1} = -1$} & Est   & [0.0008, 0.0010] &       & \textcolor[rgb]{ .753,  0,  0}{[0.0276, 0.0284]} &       & \textcolor[rgb]{ .753,  0,  0}{[0.0023, 0.0028]} \\
          &       & CI    & [0.0002, 0.0016] &       & \textcolor[rgb]{ .753,  0,  0}{[0.0183, 0.0368]} &       & \textcolor[rgb]{ .753,  0,  0}{[0.0011, 0.0040]} \\
\cmidrule{2-8}          & \multirow{2}[2]{*}{$\Delta Y_{ij,1} = 0$} & Est   & \textcolor[rgb]{ .188,  .329,  .588}{[0.0269, 0.0275]} &       & [0.8026, 0.8040] &       & \textcolor[rgb]{ .753,  0,  0}{[0.0727, 0.0735]} \\
          &       & CI    & \textcolor[rgb]{ .188,  .329,  .588}{[0.0109, 0.0428]} &       & [0.7706, 0.8358] &       & \textcolor[rgb]{ .753,  0,  0}{[0.0514, 0.0944]} \\
\cmidrule{2-8}          & \multirow{2}[2]{*}{$\Delta Y_{ij,1} = 1$} & Est   & \textcolor[rgb]{ .188,  .329,  .588}{[0.0015, 0.0017]} &       & \textcolor[rgb]{ .188,  .329,  .588}{[0.0514, 0.0524]} &       & [0.0043, 0.0049] \\
          &       & CI    & \textcolor[rgb]{ .188,  .329,  .588}{[0.0003, 0.0028]} &       & \textcolor[rgb]{ .188,  .329,  .588}{[0.0396, 0.0635]} &       & [0.0024, 0.0069] \\
    \midrule
    \multirow{6}[6]{*}{HL} & \multirow{2}[2]{*}{$\Delta Y_{ij,1} = -1$} & Est   & [0.0008, 0.0013] &       & \textcolor[rgb]{ .753,  0,  0}{[0.0344, 0.0352]} &       & \textcolor[rgb]{ .753,  0,  0}{[0.0026, 0.0032]} \\
          &       & CI    & [-0.0001, 0.0022] &       & \textcolor[rgb]{ .753,  0,  0}{[0.0203, 0.0493]} &       & \textcolor[rgb]{ .753,  0,  0}{[0.0010, 0.0048]} \\
\cmidrule{2-8}          & \multirow{2}[2]{*}{$\Delta Y_{ij,1} = 0$} & Est   & \textcolor[rgb]{ .188,  .329,  .588}{[0.0212, 0.0213]} &       & [0.7999, 0.8011] &       & \textcolor[rgb]{ .753,  0,  0}{[0.0650, 0.0656]} \\
          &       & CI    & \textcolor[rgb]{ .188,  .329,  .588}{[0.0078, 0.0347]} &       & [0.7677, 0.8334] &       & \textcolor[rgb]{ .753,  0,  0}{[0.0487, 0.0818]} \\
\cmidrule{2-8}          & \multirow{2}[2]{*}{$\Delta Y_{ij,1} = 1$} & Est   & \textcolor[rgb]{ .188,  .329,  .588}{[0.0016, 0.0019]} &       & \textcolor[rgb]{ .188,  .329,  .588}{[0.0644, 0.0652]} &       & [0.0048, 0.0056] \\
          &       & CI    & \textcolor[rgb]{ .188,  .329,  .588}{[0.0002, 0.0032]} &       & \textcolor[rgb]{ .188,  .329,  .588}{[0.0473, 0.0826]} &       & [0.0024, 0.0080] \\
    \hline\hline
    \end{tabular}%
    \vspace{-5mm}
	\begin{flushleft} \footnotesize \linespread{.75} Table 2 reports bounds on the joint density function of potential risk sharing connections $f(y_1,y_0) = \mathbbm{P}\left( \Delta Y_{ij}(1) = y_1, \Delta Y_{ij}(0) = y_0\right)$ and $95\%$ confidence intervals. \textcolor[rgb]{ .753,  0,  0}{Red} describes connections destroyed by the microfinance program and \textcolor[rgb]{ 0,  .439,  .753}{blue} describes connections created.\normalsize \end{flushleft}
\end{table}%

\subsubsection{Our results rule out monotonic treatment effects}
The reason why our bounds in Section 6.4.2 reveal a much larger amount of disruption than the average treatment effects of Section 6.4.1 is because the microfinance program both creates and destroys many connections between households. If, in contrast, the program only created or only destroyed connections, then we would expect them to be roughly equivalent. To see this, suppose that $\Delta Y_{ij,1} \geq \Delta Y_{ij,0}$  with probability one. In this case, $f(-1,0) = f(-1,1) = f(0,1) = 0$ so that $D = 0$ and  \looseness=-1 
\begin{align*}
ATE &=  \mathbb{E}\left[\Delta Y_{ij}(1)\right] -  \mathbb{E}\left[\Delta Y_{ij}(0)\right] \\
&= \left(\mathbbm{P}\left(\Delta Y_{ij}(1) = 1\right) - \mathbbm{P}\left(\Delta Y_{ij}(1) = -1\right) \right) 
-  \left(\mathbbm{P}\left(\Delta Y_{ij}(0) = 1\right) - \mathbbm{P}\left(\Delta Y_{ij}(0) = -1\right)\right) \\
& = 2f(1,-1) + f(1,0) + f(0,-1) = C + f(1,-1) \approx C
\end{align*}
where the last line is because $f(1,-1) \approx 0$ (see Table 2). Similarly, $C = 0$ and $ATE \approx D$ whenever $\Delta Y_{ij,1} \leq \Delta Y_{ij,0}$ with probability one.  \looseness=-1 

Since our confidence intervals for $C$ and $D$ both exclude $0$, we reject both monotonicity assumptions $\Delta Y_{ij}(1) \geq \Delta Y_{ij}(0)$ and $\Delta Y_{ij}(1) \leq \Delta Y_{ij}(1)$. As a result, the average treatment effects, even conditional on whether the households are linked in the pre-treatment period or have a high or low propensity to borrow money, necessarily understate the program's disruptive effect. \looseness=-1

\section{Conclusion}
This paper is about identifying social disruption: the amount of network connections created or destroyed by a policy. It focuses on a research design that is popular in the literature. We first formalize the informational content of the random assignment of agents to communities. We then show that the sharp identified set is given by an intractable quadratic assignment problem and propose outer bounds constructed by rearranging the eigenvalues of two graph function parameters that are identified from the experiment. Our empirical illustration demonstrates that our methodology is effective at identifying social disruption in practice. Alternative methods used in the literature can substantially understate the disruptive impact of the policy.    \looseness=-1 

How should researchers use measurements of social disruption to evaluate and design policy? One way to do this would be to specify a welfare function that values both conventional economic benefits like the health or wealth of agents as well as the amount of social disruption. Researchers could then choose a policy that maximizes the economic benefits subject to the constraint that it does not alter too many social connections (i.e. that the estimate for the upper bound on the amount of social disruption is sufficiently small). The idea here is that policies that destroy only a few connections between agents are unlikely to have the kinds of negative unintended consequences reported in the empirical literature \citep[see][for a recent review]{jackson2021inequality}.\looseness=-1 

Another way to use measurements of social disruption for policy is to actually assign a specific value to the number of created and destroyed connections, and directly weigh them against the other economic benefits or harms of the policy. This requires the researcher to take stance on the economic cost of social disruption. To our knowledge, there is currently little work explicitly on pricing disruption, and so we highlight this as an important area for future research.\looseness=-1

\renewcommand{\baselinestretch}{1}
\bibliographystyle{aer}
\bibliography{literature}
\renewcommand{\baselinestretch}{1.5}

\appendix

\section{Appendix: proof of Propositions 1-3}
\subsection{Definitions and lemmas}

\subsubsection{Hilbert-Schmidt integral operators and function embeddings}
In Section 4 of the main text we use graph function to describe the conditional distribution of the network connections between pairs of agents in the community. These functions are bounded, measurable, and symmetric. Any bounded symmetric measurable function $f: [0,1]^{2}\to \mathbb{R}$ defines a compact symmetric Hilbert-Schmidt integral operator $T_{f}: L_{2}([0,1]) \to L_{2}([0,1])$ where $(T_{f} g)(u) = \int f(u,\tau)g(\tau)d\tau$. It has a bounded countable multiset of real eigenvalues $\{\lambda_{r}\}_{r \in \mathbb{N}}$ with $0$ as the only limit point. It also admits the spectral decomposition $\sum_{r}\lambda_{r}\phi_{r}(u)\phi_{r}(v)$ where $\phi_{r}: [0,1] \to \mathbb{R}$ is the eigenfunction associated with eigenvalue $\lambda_{r}$, i.e. $\int f(u,\tau)\phi_{r}(\tau)d\tau = \lambda_{r}\phi_{r}(u)$. The functions $\{\phi_{r}\}_{r\in \mathbb{N}}$ are chosen to be orthogonal, i.e. $\int \phi_{r}(u)^{2}du = 1$ and $\int(\phi_{r}(u)- \phi_{s}(u))^{2}du = 2$ if $r \neq s$, and form a basis of $L_{2}([0,1])$. It follows that $\sum_{r}\lambda_{r}^{2} = \int\int f(u,v)^{2}dudv <  \infty$. See Chapter 9.2 of \cite{birman2012spectral} for a textbook reference.  \looseness=-1 

Any square symmetric matrix can be represented by a bounded symmetric measurable function sometimes called its function embedding. Let $F$ be an arbitrary $n\times n$ square symmetric matrix with $ij$th entry $F_{ij}$. The function embedding $f: [0,1]^{2}\to\mathbb{R}$ of $F$  is $f(u,v) = F_{\lceil nu\rceil \lceil nv\rceil}$ for $u,v \in [0,1]$. Intuitively, $f$ assigns the mass of agents in the region $S_{i}^{n} := \left(\frac{i-1}{n},\frac{i}{n}\right]$ to observation $i$.  Similarly, any $n \times n$ permutation matrix $\Pi$ can be represented as a measure preserving transformation $\varphi(u) = \lceil nu \rceil - nu + \Pi(\lceil nu \rceil)$ where $\Pi(k) = \{l \in [n]: \Pi_{kl} = 1\}$. Intuitively, if $\Pi_{kl} = 1$, $\varphi_{t}$ maps the interval $\left(\frac{k-1}{n},\frac{k}{n}\right]$ monotonically to $\left(\frac{l-1}{n},\frac{l}{n}\right]$. See Section 7.1 of \cite{lovasz2012large} for a textbook reference.  \looseness=-1 
 
The eigenvalues of matrices and their function embeddings are scaled differently. Specifically, if $(\lambda^{F}_{r},\phi^{F}_{r})$ is an eigenvalue and eigenvector pair of $F$ then $(\lambda^{F}_{r}/n,\sqrt{n}\phi^{F}_{r}(\lceil n\cdot\rceil))$ is an eigenvalue and eigenfunction pair of its function embedding where $\phi^{F}_{r}(i)$ is the $i$th entry of the vector $\phi^{F}_{r}$. As described in Section 5.2, we take inner products of eigenvalues in a specific way. That is, if $\{\lambda_{r1}\}$ and $\{\lambda_{r0}\}$ are the eigenvalues of $f_{1}$ and $f_{0}$, then $\sum_{r}\lambda_{r1}\lambda_{r0}$ refers to $\lim_{R \to \infty}\sum_{r \in [R]}\lambda_{r1}\lambda_{r0}$ where $\{\lambda_{r1}\}_{r \in [R]}$ and $\{\lambda_{r0}\}_{r \in [R]}$ are the $R$ largest (in absolute value) elements of $\{\lambda_{r1}\}$ and $\{\lambda_{r0}\}$ respectively (counting multiplicities) ordered to be decreasing.  These limits exist: the relevant series are absolutely summable because of the Cauchy-Schwarz inequality and since $\sum_{r}\lambda_{rs}^2 = \int\int h_s(u,v)^2dudv \leq 1$ for $s \in \{0,1\}$.

\subsubsection{Sets}
We use $\mathbb{N}$ for the positive integers, $\mathbb{R}$ for the real numbers, $[n]$ for $\{1,2,...,n\}$, $\mathcal{P}_{n}$ for $n\times n$ permutation matrices (square matrices with $\{0,1\}$ valued entries whose rows and columns sum to $1$), $\mathcal{D}_{n}^{+}$ for $n\times n$ doubly stochastic matrices (square matrices with nonnegative real entries whose rows and columns sum to $1$), $\mathcal{O}_{n}$ for $n\times n$ orthogonal matrices (square matrices with real entries where any two rows or any two columns have inner product $1$ if they are the same and $0$ if they are not), and $\mathcal{M} := \{\psi: [0,1] \to [0,1]\text{ with } |\psi^{-1}(A)| = |A| \text{ for any measurable } A \subseteq [0,1]\}$ for measure preserving transformations on $[0,1]$ where $|A|$ is the Lebesgue measure of $A$. \looseness=-1

\subsection{Proposition 1}
Proposition 1 follows directly from Lemma 1 in Section 4.1 of the main text, which we prove here. As discussed in Section 4.1, Lemma 1 makes two claims. We call these claims Lemmas 1A and 1B respectively. Lemma 1A says that the distribution of the data can distinguish between graph functions that are not equivalent up to a measure preserving transformation. That is, \looseness=-1 
\begin{flushleft}
\textbf{Lemma 1A:} Let $(g_{s},\varphi_{s})$ and $(g'_{s},\varphi'_{s})$ parametrize two different distributions of network connections and community assignments such that $\{Y,D\}$ are determined by $(g_{s},\varphi_{s})$ and $\{Y',D'\}$ are determined by $(g'_{s},\varphi'_{s})$. Let $h_{s}(a,b) =  \int \mathbbm{1}\{g_{s}(\varphi_s(a),\varphi_s(b),w) \leq y_s\}dw$ and $h'_{s}(a,b) =  \int \mathbbm{1}\{g'_{s}(\varphi'_s(a),\varphi'_s(b),w) \leq y_s\}dw$ be the two graph functions associated with $(g_{s},\varphi_{s})$ and $(g'_{s},\varphi'_{s})$.  Then $h_{s} \not\sim h'_{s}$ for some $s \in \{0,1\}$ implies that $\{Y,D\}$ and $\{Y',D'\}$ do not have the same distribution. 
\end{flushleft}

\begin{flushleft}
\textbf{Proof of Lemma 1A:} To demonstrate the result, we rely on a construction called a graph homomorphism. We define the construction here, but see also the first paragraph of Section 7.2 of Lovasz (2012) for a textbook exposition. Let $V$ be an arbitrary set of agents with $M = |V|$ and $E \subseteq V \times V$ be an arbitrary set of binary connections between pairs of agents in $V$. Then the graph homomorphism associated with $(V,E)$, policy $s \in \{0,1\}$, and $y_s \in \mathbbm{R}$ is $\mathbbm{P}\left( \{Y_{ij}(s) \leq y_s\}_{ij \in E}\right)$ where $Y_{ij}(s)$ is the potential outcome for agents $i$ and $j$ under the model $(g_{s},\varphi_{s})$ with $N \geq M$ as described in Section 3.2. The graph homomorphism can be represented as 
\begin{align*}
\mathbbm{P}\left( \{Y_{ij}(s) \leq y_s\}_{ij \in E}\right) 
&= \int_{u_1,\ldots,u_{M}}\prod_{ij \in E}\left[\int \mathbbm{1}\{g_{s}(\varphi_s(u_i),\varphi_s(u_j),w) \leq y_s\}dw\right]du_1\ldots u_{M} \\
&=  \int_{u_1,\ldots,u_{M}}\prod_{ij \in E}h_{s}(u_i,u_j)du_1\ldots u_{M}
\end{align*}
where the first equality follows from the model (\ref{model2}) and the second equality follows from the definition of the graph function: $h_{s}(a,b) := \int \mathbbm{1}\{g_{s}(a,b,w) \leq y_s\}dw$. \newline \\


We now demonstrate Lemma 1A. Specifically, we show its contrapositive. Let $\{Y,D\}$ and $\{Y',D'\}$ be data generated by the models $\{g_{s},\varphi_s\}$ and $\{g'_{s},\varphi'_s\}$ according to Section 3.2 of the main text. Assume that the models are such that $\{Y,D\}$ and $\{Y',D'\}$ have the same distribution. We first show that this implies that the graph homomorphisms for the two models must be the same, i.e. for any  $(V,E)$, $s \in \{0,1\}$, and $y_s \in \mathbb{R}$
\begin{align*}
\mathbbm{P}\left( \{Y_{ij}(s) \leq y_s\}_{ij \in E}\right) &= \mathbbm{P}\left( \{Y'_{ij}(s) \leq y_s\}_{ij \in E}\right).
\end{align*}
We then show that Lemma 1A follows from Corollary 10.35(a) of \cite{lovasz2012large}. \newline \\

To demonstrate the first part, that $\{g_{s},\varphi_s\}$ and $\{g'_{s},\varphi'_s\}$ must have the same graph homomorphisms, we start by decomposing 
\begin{align*}
\mathbbm{P}\left(\left\{Y_{ij} \leq y_{ij}, D_{i} = d_i\right\}_{i,j=1}^{2N}\right) 
&= \mathbbm{P}\left(\left\{Y_{ij} \leq y_{ij}\right\}_{i,j=1}^{2N}|\{D_i = d_i\}_{i=1}^{2N}\right)\mathbbm{P}\left(\{D_i = d_i\}_{i=1}^{2N}\right)  
\end{align*}
and
\begin{align*}
\mathbbm{P}\left(\left\{Y'_{ij} \leq y_{ij}, D'_{i} = d_i\right\}_{i,j=1}^{2N}\right) 
&= \mathbbm{P}\left(\left\{Y'_{ij} \leq y_{ij}\right\}_{i,j=1}^{2N}|\{D'_i = d_i\}_{i=1}^{2N}\right)\mathbbm{P}\left(\{D'_i = d_i\}_{i=1}^{2N}\right). 
\end{align*}
Since both $D$ and $D'$ are drawn uniformly from the set $\{d \in \{0,1\}^{2N}: \sum_{i=1}^{2N}d_i = N\}$, it follows that $\mathbbm{P}\left(\{D_i = d_i\}_{i=1}^{2N}\right) = \mathbbm{P}\left(\{D'_i = d_i\}_{i=1}^{2N}\right)$ for any $d \in \{0,1\}^{2N}$. And so the assumption that the entries of $\{Y,D\}$ and $\{Y',D'\}$ have the same distribution implies that 
\begin{align}\label{conditional}
\mathbbm{P}\left(\left\{Y_{ij} \leq y_{ij}\right\}_{i,j=1}^{2N}|\{D_i = d_i\}_{i=1}^{2N}\right) = \mathbbm{P}\left(\left\{Y'_{ij} \leq y_{ij}\right\}_{i,j=1}^{2N}|\{D'_i = d_i\}_{i=1}^{2N}\right)
\end{align} 
for any $y \in \mathbbm{R}^{2N\times 2N}$ and $d \in \{0,1\}^{2N}$ such that $\sum_{i=1}^{2N}d_i = N$. In particular, for any $(V,E)$, $s \in \{0,1\}$, $y_s \in \mathbb{R}$, and $i,j \in V$, we can choose $y_{ij} = y_s$ if $ij \in E$, $y_{ij} = \infty$ if $ij \not\in E$, and $d_i = d_j = s$ so that
\begin{align*}
\mathbbm{P}\left(\left\{Y_{ij} \leq y_{ij}\right\}_{i,j=1}^{2N}|\{D_i = d_i\}_{i=1}^{2N}\right) 
&= \mathbbm{P}\left(\left\{Y_{ij} \leq y_{s}\right\}_{ij \in E}|\{D_i = d_i\}_{i=1}^{2N}\right)  \\
&=  \mathbbm{P}\left(\left\{Y_{ij}(s) \leq y_{s}\right\}_{ij \in E}|\{D_i = d_i\}_{i=1}^{2N}\right)  \\
&= \mathbbm{P}\left(\left\{Y_{ij}(s) \leq y_{s}\right\}_{ij \in E}\right)
\end{align*} 
where the first equality follows from the choice of $y_{ij}$ fixed above, the second equality follows from the choice of $d_i$ fixed above, and the third equality follows from Assumption 1, that the community assignments and the potential outcomes are independent. The same logic applied to the distribution of $\{Y',D'\}$ gives $\mathbbm{P}\left(\left\{Y'_{ij} \leq y_{ij}\right\}_{i,j=1}^{2N}|\{D'_i = d_i\}_{i=1}^{2N}\right)$ $= \mathbbm{P}\left(\left\{Y'_{ij}(s) \leq y_{s}\right\}_{ij \in E}\right)$.\newline

It follows from these derivations and (\ref{conditional}) that $\mathbbm{P}\left( \{Y_{ij}(s) \leq y_s\}_{ij \in E}\right) = \mathbbm{P}\left( \{Y'_{ij}(s) \leq y_s\}_{ij \in E}\right)$ which, by the algebra in the first paragraph of this proof, is equivalent to 
\begin{align}\label{equaldensity}
  \int_{u_1,\ldots,u_{M}}\prod_{ij \in E}h_{s}(u_i,u_j)du_1\ldots u_{M} =   \int_{u_1,\ldots,u_{M}}\prod_{ij \in E}h'_{s}(u_i,u_j)du_1\ldots u_{M}.
  \end{align}
Finally, since the choice of $(V,E)$ was arbitrary, we can apply Corollary 10.35(a) of \cite{lovasz2012large} (Lemma B7 in Online Appendix Section B.1) which implies that $h_{s}$ and $h'_{s}$ are equivalent up to a measure preserving transformation. Since the choice of $s \in \{0,1\}$ and $y_s \in \mathbb{R}$ was arbitrary, this implies that $h_{1} \sim h'_{1}$ and $h_{0} \sim h'_{0}$, which demonstrates the (contrapositive of) Lemma 1A.  $\square$
\end{flushleft}

Lemma 1B says that the distribution of the data cannot distinguish between graph functions that are equivalent up to a measure preserving transformation. That is,
\begin{flushleft}
\textbf{Lemma 1B:} Let $(g_{s},\varphi_{s})$ and $(g'_{s},\varphi'_{s})$ be such that $h_{s} \sim h'_{s}$ where  $h_{s}(a,b)$ $=  \int \mathbbm{1}\{g_{s}(\varphi_s(a),\varphi_s(b),w) \leq y_s\}dw$ and $h'_{s}(a,b)$ $=  \int \mathbbm{1}\{g'_{s}(\varphi'_s(a),\varphi'_s(b),w) \leq y_s\}dw$. Then $\{Y,D\}$ and $\{Y',D'\}$ have the same distribution, where $\{Y,D\}$ is drawn from $(g_{s},\varphi_{s})$ and $\{Y',D'\}$ is drawn from $(g'_{s},\varphi'_{s})$. \end{flushleft}

\begin{flushleft}
\textbf{Proof of Lemma 1B:} 
Fix $s \in \{0,1\}$ and suppose that $h_{s} \sim h'_{s}$. Then, by definition, there exist $\psi,\psi' \in \mathcal{M}$ such that $h_{s}(\psi(a),\psi(b);y_s) = h'_{s}(\psi'(a),\psi'(b);y_s)$ almost everywhere for every $y_s \in \mathbb{R}$. Let $y \in \mathbbm{R}^{M\times M}$ and $d \in \{0,1\}^{M}$ be arbitrary constants where $M = 2N$. The distribution function for $\{Y,D\}$ can then be represented 
\begin{align*}
\mathbbm{P}\left(\left\{Y_{ij} \leq y_{ij}, D_{i} = d_i\right\}_{i,j=1}^{M}\right) 
&= \mathbbm{P}\left(\left\{Y_{ij} \leq y_{ij}\right\}_{i,j=1}^{M}|\{D_i = d_i\}_{i=1}^{M}\right)\mathbbm{P}\left(\{D_i = d_i\}_{i=1}^{M}\right)  \\
&= \mathbbm{P}\left(\left\{Y_{ij}(1)d_i d_j + Y_{ij}(0)(1-d_i)(1-d_j) \leq y_{ij}\right\}_{i,j=1}^{M}\right)\mathbbm{P}\left(\{D_i = d_i\}_{i=1}^{M}\right) \\
&= \mathbbm{P}\left(\left\{Y_{ij}(1)d_i d_j + Y_{ij}(0)(1-d_i)(1-d_j) \leq y_{ij}\right\}_{i,j=1}^{M}\right)\mathbbm{P}\left(\{D_i = d_i\}_{i=1}^{M}\right).
\end{align*}
The first part of the product can be rewritten
\begin{align*}
&\mathbbm{P}\left(\left\{Y_{ij}(1)d_i d_j + Y_{ij}(0)(1-d_i)(1-d_j) \leq y_{ij}\right\}_{i,j=1}^{M}\right) \\
&= \int_{u_1,\ldots,u_{M}}\prod_{i,j=1}^{M}\left[ \int_w\mathbbm{1}\left\{  \sum_{s \in \{0,1\}}g_{s}\left(\varphi_s(u_i),\varphi_s(u_j),w\right)\delta_{ij}(s) \leq y_{ij}\right\}dw\right]du_1\ldots du_{M} \\
&= \int_{u_1,\ldots,u_{M}}\prod_{i,j=1}^{M}\left[ \int_w\mathbbm{1}\left\{  \sum_{s \in \{0,1\}}g_{s}\left(\varphi_s(\psi(u_i)),\varphi_s(\psi(u_j)),w\right)\delta_{ij}(s)  \leq y_{ij}\right\}dw\right]du_1\ldots du_{M} \\ 
&= \int_{u_1,\ldots,u_{M}}\prod_{i,j=1}^{M} \sum_{s \in \{0,1\}}\left[ \int_w\mathbbm{1}\left\{ g_{s}\left(\varphi_s(\psi(u_i)),\varphi_s(\psi(u_j)),w\right)\leq y_{ij}\right\}dw\right]\delta_{ij}(s)  du_1\ldots du_{M} \\
&= \int_{u_1,\ldots,u_{M}}\prod_{i,j=1}^{M} \sum_{s \in \{0,1\}}h_{s}(\psi(u_i),\psi(u_j); y_{ij})\delta_{ij}(s) du_1\ldots du_{M} 
\end{align*}
where $\delta_{ij}(s) = d_i^sd_j^s(1-d_i)^{1-s}(1-d_j)^{1-s}$ and the second equality follows because $\psi$ is measure preserving. \newline

Similarly, the distribution function for $\{Y',D'\}$ is given by the product of $\mathbbm{P}\left(\left\{Y'_{ij}(1)d_i d_j + Y'_{ij}(0)(1-d_i)(1-d_j) \leq y_{ij}\right\}_{i,j=1}^{M}\right)$ and $\mathbbm{P}\left(\{D'_i = d_i\}_{i=1}^{M}\right)$. The second term  $\mathbbm{P}\left(\{D'_i = d_i\}_{i=1}^{M}\right)$ is equivalent to $\mathbbm{P}\left(\{D_i = d_i\}_{i=1}^{M}\right)$ because, under the model of Section 3.2, the community assignments are drawn uniformly at random from the set $\{d \in \{0,1\}^{M}: \sum_{i=1}^{M}d_i = N\}$. Following the logic of the previous paragraph, the first term $\mathbbm{P}\left(\left\{Y'_{ij}(1)d_i d_j + Y'_{ij}(0)(1-d_i)(1-d_j) \leq y_{ij}\right\}_{i,j=1}^{M}\right)$ can be written as 
\begin{align*}
 \int_{u_1,\ldots,u_{M}}\prod_{i,j=1}^{M} \sum_{s \in \{0,1\}}h'_{s}(\psi'(u_i),\psi'(u_j); y_{ij})\delta_{ij}(s) du_1\ldots du_{M} \\
 =  \int_{u_1,\ldots,u_{M}}\prod_{i,j=1}^{M} \sum_{s \in \{0,1\}}h_{s}(\psi(u_i),\psi(u_j); y_{ij})\delta_{ij}(s)du_1\ldots du_{M}
 \end{align*}
where the second equations is because $h_{s} \sim h'_{s}$. It follows that $\mathbbm{P}\left(\left\{Y'_{ij}(1)d_i d_j + Y'_{ij}(0)(1-d_i)(1-d_j) \leq y_{ij}\right\}_{i,j=1}^{M}\right)$ is equivalent to $\mathbbm{P}\left(\left\{Y_{ij}(1)d_i d_j + Y_{ij}(0)(1-d_i)(1-d_j) \leq y_{ij}\right\}_{i,j=1}^{M}\right)$ and so $\{Y,D\}$ and $\{Y',D'\}$ have the same distribution. $\square$

\end{flushleft}

\begin{remark}
Our framework considers a completely randomized balanced experiment where $N$ agents are chosen uniformly at random and assigned to community 1, however, the proofs of our Lemmas 1A and 1B do not use this assumption at all. Specifically, the arguments only require that the conditional distribution $\mathbbm{P}\left(\{Y_{ij} \leq y_{ij} \}_{i,j=1}^{2N}|\{D_i = d_i\}_{i=1}^{2N}\right)$ is identified for every $\{ d \in \{0,1\}^{2N} :  \sum_{i=1}^{2N}d_i = N\}$, which is the case if $D$ is distributed so that every element of $\{ d \in \{0,1\}^{2N} :  \sum_{i=1}^{2N}d_i = N\}$ occurs with positive probability. It is also straightforward to extend the proofs to allow for differently sized communities, but because this complicates the notation with little additional analytical insight, we leave it to future work.  \looseness=-1 
\end{remark}

\subsection{Lemmas}
The following are used to demonstrate Propositions 2 and 3. For these results, let $h_{s}(u,v)$ equal $\int\mathbbm{1}\{g_{s}(\varphi_{s}(u),\varphi_{s}(v),w) \leq y_{s}\}dw$ for an arbitrary measurable $g_{s} : [0,1]^{3} \to \mathbb{R}$, $y_{s} \in \mathbb{R}$, $\varphi_{s} \in \mathcal{M}$, and $s \in \{0,1\}$. For any $n\in \mathbb{N}$ let $S_{i}^{n} := \left(\frac{i-1}{n},\frac{i}{n}\right]$, $H_{s}^{n}$ be an $n\times n$ matrix with $H_{ij,s}^{n} \in \mathbb{R}$ as its $ij$th entry, and $h_{s}^{n}(u,v) = \sum_{ij}H_{ij,s}^{n}\mathbbm{1}\{u \in S_{i}^{n}, v \in S_{j}^{n}\}$ such that $\int\int\left(h_{s}(u,v)-h_{s}^{n}(u,v)\right)^{2}dudv \to 0$ as $n\to\infty$. In words, $H_{s}^{n}$ is an $n\times n$ matrix approximation of $h_{s}$ and $h_{s}^{n}$ is its function embedding. Intuitively, $h^{n}_{s}$ is a histogram approximation to the function $h_{s}$. We show that a sequence of matrices $H_{s}^{n}$ satisfying these properties exists in Lemma 1 below. Let $\{\lambda_{rs}\}$ denote the eigenvalues of  $h_{s}$ and $\{\lambda^{n}_{rs}\}$ the eigenvalues of  $h_{s}^{n}$.  \looseness=-1 

\begin{flushleft}
\textbf{Lemma 1:} For every bounded measurable $g: [0,1]^{2} \to \mathbb{R}$ there exists sequences $\{G^{n}\}_{n \in \mathbb{N}}$ and $\{g^{n}\}_{n \in \mathbb{N}}$ where $G^{n}$ is an $n\times n$ matrix with $ij$th entry $G^{n}_{ij}$ and $g^{n}:[0,1]^{2}\to\mathbb{R}$ with $g^{n}(u,v) = \sum_{ij}G_{ij}^{n}\mathbbm{1}\{u \in S_{i}^{n}, v \in S_{j}^{n}\}$ and $S_{i}^{n} := \left(\frac{i-1}{n},\frac{i}{n}\right]$ such that for every $\varepsilon > 0$ there exists an $m \in \mathbb{N}$ such that $\int\int\left(g(u,v)-g^{n}(u,v)\right)^{2}dudv \leq \varepsilon$ for every $n > m$.  
\end{flushleft}

\begin{flushleft}
\textbf{Proof of Lemma 1:} Fix an arbitrary $\varepsilon > 0$. Lusin's Theorem (Lemma B1 in Online Appendix Section B.1) implies that for any measurable $g: [0,1]^{2} \to \mathbb{R}$ and $\epsilon > 0$, there exists a compact $E^{\epsilon}_{g} \subseteq [0,1]^{2}$ of measure at least $1-\epsilon$ such that $g$ is continuous when restricted to $E^{\epsilon}_{g}$.  \newline

For any $n' \in \mathbb{N}$, define the $n'\times n'$ matrix  $G^{n'\epsilon}$ with $ij$th entry $G_{ij}^{n'\epsilon} = \frac{\int\int_{(u,v) \in E^{\epsilon}_{g}} g_{t}(u,v)\mathbbm{1}\{u \in S_{i}^{n'}, v \in S_{j}^{n'}\}dudv}{\int\int_{(u,v) \in E^{\epsilon}_{g}}\mathbbm{1}\{u \in S_{i}^{n'}, v \in S_{j}^{n'}\}dudv}$  if $\int\int_{(u,v) \in E^{\epsilon}_{g}}\mathbbm{1}\{u \in S_{i}^{n'}, v \in S_{j}^{n'}\}dudv > 0$ and $G_{ij}^{n'\epsilon} = 0$ otherwise. Let $g^{n'\epsilon}$ be the function embedding of $G^{n'\epsilon}$ so that for $u,v \in [0,1]$, $g^{n'\epsilon}(u,v) = \sum_{ij}G_{ij}^{n'\epsilon}\mathbbm{1}\{u \in S_{i}^{n'}, v \in S_{j}^{n'}\}$. Also let  $\bar{g} := \sup_{(u,v) \in [0,1]^{2}}|g(u,v)|^{2} < \infty$. \newline 

Since $g$ is continuous when restricted to $E^{\epsilon}_{g}$ there exists an $m(\epsilon) \in \mathbb{N}$ such that $\int\int_{(u,v) \in E^{ \epsilon}_{g}}\left(g(u,v) - g^{n'\epsilon}(u,v)\right)^{2}dudv \leq \epsilon$ for every $n' > m(\epsilon)$. In addition, $\int\int_{(u,v) \not\in E^{\epsilon}_{g}}\left(g(u,v) - g^{n'\epsilon}(u,v)\right)^{2}dudv \leq 4\bar{g}\epsilon$  for every $n'$. It follows that $\int\int_{(u,v) \in [0,1]^{2}}\left(g(u,v) - g^{n'\epsilon}(u,v)\right)^{2}dudv \leq \left(1+4\bar{g}\right)\epsilon$  for every $n' > m(\epsilon)$.  \newline

Let $e^{\dagger}(n') := \inf\{e > 0: m(e) \leq n'\}$ where $e^{\dagger}(n') \to 0$ as $n'\to \infty$ because $m(\epsilon) \in \mathbb{N}$ for every $\epsilon > 0$. For every $n \in \mathbb{N}$, define $G^{n} = G^{n e^{\dagger}(n)}$ and  $g^{n} = g^{n e^{\dagger}(n)}$. Then $\int\int_{(u,v) \in [0,1]^{2}}\left(g(u,v) - g^{n}(u,v)\right)^{2}dudv \leq \left(1+4\bar{g}\right)e^{\dagger}(n')$ for all $n > m(e^{\dagger}(n'))$ and $n' \in \mathbb{N}$. The claim follows by taking $n'$ sufficiently large so that $\left(1+4\bar{g}\right)e^{\dagger}(n') < \varepsilon$.  $\square$
\end{flushleft}

\begin{flushleft}
\textbf{Lemma 2:} $ \sum_{r \in [n]}\lambda^{n}_{\rho_{n}(r)0}\lambda^{n}_{r1} \leq \int\int h^{n}_{0}(u,v)h^{n}_{1}(u,v)dudv \leq \sum_{r \in [n]}\lambda^{n}_{r0}\lambda^{n}_{r1}$ where $\rho_{n}(r) = n-r+1$. 
\end{flushleft}

\begin{flushleft}
\textbf{Proof of Lemma 2:} This lemma follows the logic of \cite{finke1987quadratic}, Theorem 3. By construction $\int\int h^{n}_{1}(u,v)h^{n}_{0}(u,v)dudv = \frac{1}{n^{2}}\sum_{ij}H_{ij,1}^{n}H_{ij,0}^{n}$ so it is sufficient to show that $n^2\sum_{r \in [n]}\lambda^{n}_{\rho_{n}(r)0}\lambda^{n}_{r1} \leq \sum_{ij} H^{n}_{ij,1}H^{n}_{ij,0} \leq n^2\sum_{r \in [n]}\lambda^{n}_{r0}\lambda^{n}_{r1}$. Also if  $\{\lambda_{rs}^{n}\}_{r \in [n]}$ are the $n$ largest (in absolute value) eigenvalues of $h_{s}^{n}$  then $\{n\lambda_{rs}^{n}\}_{r \in [n]}$ are the eigenvalues of $H_{s}^{n}$.  \newline

Since $H^{n}_{s}$ is square and symmetric, the spectral theorem (Lemma B2 in Online Appendix Section B.1) implies that $H_{ij,s}^{n} = n\sum_{r \in [n]}\lambda_{rs}^{n}\phi_{ir,s}^{n}\phi_{jr,s}^{n}$ where $\phi_{ir,s}^{n}$ is the eigenvector of $H_{ij,s}^{n}$ associated with eigenvalue $n\lambda_{rs}^{n}$. As a result $\sum_{ij}H_{ij,1}^{n}H_{ij,0}^{n} = n^2\sum_{r,s \in [n]}\lambda_{r1}^{n}\lambda_{s0}^{n}\left[\sum_{i}\phi_{ir,1}^{n}\phi_{is,0}^{n}\right]^{2}$. \newline

The matrix $\left[\sum_{i}\phi_{ir,1}^{n}\phi_{is,0}^{n}\right]^{2}$ is doubly stochastic and so Birkhoff's Theorem (Lemma B4 in Online Appendix Section B.1)  implies that 
\begin{align*}
\sum_{r,s \in [n]}\lambda_{r1}^{n}\lambda_{s0}^{n}\left[\sum_{i}\phi_{ir,1}^{n}\phi_{is,0}^{n}\right]^{2} 
 =  \sum_{r,s \in [n]}\lambda_{r1}^{n}\lambda_{s0}^{n}\sum_{t\in[m]}\alpha_{t}P_{ij,t} 
 =  \sum_{t\in[m]}\alpha_{t}\sum_{r,s \in [n]}\lambda_{r1}^{n}\lambda_{s0}^{n}P_{ij,t}
 \end{align*}
 for some $m \in \mathbb{N}$, $\alpha_{1},...,\alpha_{m} > 0$ with $\sum_{t\in[m]}\alpha_{t} = 1$, and $P_{1},...,P_{m} \in \mathcal{P}_{n}$.  \newline
 
 Hardy-Littlewood-Polya's Theorem 368 (Lemma B5 in Online Appendix Section B.1) implies that
$  \sum_{r \in [n]}\lambda_{r1}^{n}\lambda_{\rho_{n}(r)0}^{n} 
  \leq \sum_{r,s \in [n]}\lambda_{r1}^{n}\lambda_{s0}^{n}P_{ij} 
  \leq \sum_{r \in [n]}\lambda_{r1}^{n}\lambda_{r0}^{n}$
   for any $P \in \mathcal{P}_{n}$ and so 

$  \sum_{r \in [n]}\lambda_{r1}^{n}\lambda_{\rho_{n}(r)0}^{n} 
  \leq \sum_{t\in[m]}\alpha_{t}\sum_{r,s \in [n]}\lambda_{r1}^{n}\lambda_{s0}^{n}P_{ij,t}
  \leq \sum_{r \in [n]}\lambda_{r1}^{n}\lambda_{r0}^{n}$
  because $\sum_{t \in [m]}\alpha_{t} = 1$. The claim follows. $\square$
\end{flushleft}

\begin{flushleft}
\textbf{Lemma 3:} For every $\varepsilon > 0$ there exists an $m \in \mathbb{N}$ such that  
\begin{enumerate}
\item[i.] $\left|\int\int h^{n}_{1}(u,v)h^{n}_{0}(u,v)dudv -  \int\int h_{1}(u,v)h_{0}(u,v)dudv\right| \leq \varepsilon$ and
\item[ii.] $\left|\sum_{r \in [n]}\lambda^{n}_{\sigma_{n}(r)0}\lambda^{n}_{r1} - \sum_{r}\lambda_{\sigma(r)0}\lambda_{r1}\right| \leq \varepsilon$, 
\end{enumerate}
for every $n > m$ where $\sum_{r}\lambda_{\sigma(r)0}\lambda_{r1}$ refers to $\lim_{R \to \infty}\sum_{r\in [R]}\lambda_{\sigma_{R}(r)0}\lambda_{r1}$, $\{\lambda_{rt}\}_{r \in [R]}$ is ordered to be decreasing, and $\sigma_{R}(r)$ refers to either $R$ or $\rho_{R}(r) := R-r+1$. 
\end{flushleft}

\begin{flushleft}
\textbf{Proof of Lemma 3:} Fix an arbitrary $\varepsilon > 0$. Part i. follows from
\begin{align*}
&\left|\int\int h^{n}_{1}(u,v)h^{n}_{0}(u,v)dudv -  \int\int h_{1}(u,v)h_{0}(u,v)dudv\right| \\
&= \left|\int\int \left(h^{n}_{1}(u,v)-h_{1}(u,v)\right)h^{n}_{0}(u,v)dudv + \int\int \left(h^{n}_{0}(u,v)-h_{0}(u,v)\right)h_{1}(u,v)dudv\right| \\
&\leq \left(\int\int \left(h^{n}_{1}(u,v)-h_{1}(u,v)\right)^{2}dudv\right)^{1/2}\bar{h}^{n}_{0} + \left(\int\int \left(h^{n}_{0}(u,v)-h_{0}(u,v)\right)^{2}dudv\right)^{1/2}\bar{h}_{1} \\
  &\leq \epsilon \left(\bar{h}^{n}_{0} + \bar{h}_{1}\right) \text{for $n > m(\epsilon)$ where $m(\epsilon)$ is from the hypothesis of Lemma 1}\\
 &\leq \varepsilon \text{ for any $n > m\left(\varepsilon\right)$ where $\varepsilon = \epsilon(\bar{h}^{n}_{0} + \bar{h}_{1})$}
\end{align*}
where $\bar{h}^{n}_{0} = \left(\int\int h^{n}_{0}(u,v)^{2}dudv\right)^{1/2}$ and $\bar{h}_{1} = \left(\int\int h_{1}(u,v)^{2}dudv\right)^{1/2}$, the first inequality is due to Cauchy-Schwarz and the triangle inequality, and the second is due to Lemma 1. \newline

To demonstrate Part ii, we bound $\left|\sum_{r \in [n]}\lambda^{n}_{\sigma_{n}(r)0}\lambda^{n}_{r1} - \sum_{r\in[n]}\lambda_{\sigma_{n}(r)0}\lambda_{r1}\right|$ where the sum $\sum_{r\in[n]}\lambda_{\sigma_{n}(r)0}\lambda_{r1}$ is a function of the $n$ largest eigenvalues of $f_{0}$ and $f_{1}$ in absolute value. The remainder $\left|\sum_{r\in[n]}\lambda_{\sigma_{n}(r)0}\lambda_{r1} - \sum_{r}\lambda_{\sigma(r)0}\lambda_{r1}\right|$ can be made arbitrarily small since $\sum_{r}\lambda_{\sigma(r)0}\lambda_{r1} := \lim_{n \to \infty}\sum_{r\in[n]}\lambda_{\sigma_{n}(r)0}\lambda_{r1}$. We write
\begin{align*}
&\left|\sum_{r \in [n]}\lambda^{n}_{\sigma_{n}(r)0}\lambda^{n}_{r1} - \sum_{r\in[n]}\lambda_{\sigma_{n}(r)0}\lambda_{r1}\right|
= \left|\sum_{r \in [n]}\left(\lambda^{n}_{\sigma_{n}(r)0}\lambda^{n}_{r1} - \lambda_{\sigma_{n}(r)0}\lambda_{r1}\right)\right| \\
&= \left|\sum_{r \in [n]}\left(\lambda^{n}_{\sigma_{n}(r)0}-\lambda_{\sigma_{n}(r)0}\right)\lambda^{n}_{r1} + \sum_{r \in [n]}\left(\lambda^{n}_{r1}-\lambda_{r1}\right)\lambda_{\sigma_{n}(r)0}\right| \\
&\leq  \left(\sum_{r \in [n]}\left(\lambda^{n}_{r0}-\lambda_{r0}\right)^{2}\right)^{1/2}\left(\sum_{r \in [n]}\left(\lambda^{n}_{r1}\right)^{2}\right)^{1/2} + \left(\sum_{r \in [n]}\left(\lambda^{n}_{r1}-\lambda_{r1}\right)^{2}\right)^{1/2}\left(\sum_{r \in [n]}\left(\lambda_{r0}\right)^{2}\right)^{1/2} \\
&=  \left(\sum_{r \in [n]}\left(\lambda^{n}_{r0}-\lambda_{r0}\right)^{2}\right)^{1/2}\bar{h}^{n}_{1} + \left(\sum_{r \in [n]}\left(\lambda^{n}_{r1}-\lambda_{r1}\right)^{2}\right)^{1/2}\bar{h}_{0}
\end{align*}
where the first inequality is due to Cauchy-Schwarz and the triangle inequality. Since $h_{s}^{n}$ and $h_{s}$ are bounded functions then for every $\epsilon > 0$ there exists a $R,m' \in \mathbb{N}$ such that $\sum_{r \in [n]-[R]}\left(\lambda^{n}_{rs}\right)^{2} < \epsilon$ and $\sum_{r \in [n]-[R]}\left(\lambda_{rs}\right)^{2} < \epsilon$ for every $n > m'$ and $s \in \{0,1\}$. As a result, 
\begin{align*}
&\left(\sum_{r \in [n]}\left(\lambda^{n}_{r0}-\lambda_{r0}\right)^{2}\right)^{1/2}\bar{h}^{n}_{1} + \left(\sum_{r \in [n]}\left(\lambda^{n}_{r1}-\lambda_{r1}\right)^{2}\right)^{1/2}\bar{h}_{0}\\
&\leq \left(\sum_{r \in [R]}\left(\lambda^{n}_{r0}-\lambda_{r0}\right)^{2}\right)^{1/2}\bar{h}^{n}_{1} + \left(\sum_{r \in [R]}\left(\lambda^{n}_{r1}-\lambda_{r1}\right)^{2}\right)^{1/2}\bar{h}_{0} + 2\sqrt{\epsilon}(\bar{h}^{n}_{1}+\bar{h}_{0}) \text{ for $n > m'(\epsilon)$}  \\
&\leq \sqrt{R} \left(\int\int\left(h^{n}_{0}(u,v) - h_{0}(u,v)\right)^{2}dudv \right)^{1/2}\bar{h}^{n}_{1} + \sqrt{R} \left(\int\int\left(h^{n}_{1}(u,v) - h_{1}(u,v)\right)^{2}dudv \right)^{1/2}\bar{h}_{0}  \\
&\hspace{20mm}+ 2\sqrt{\epsilon}(\bar{h}^{n}_{1}+\bar{h}_{0}) \text{ for $n > m'(\epsilon)$} \\
&\leq (\sqrt{R}\tilde{\epsilon} + 2\sqrt{\epsilon}) (\bar{h}^{n}_{1} + \bar{h}_{0}) \text{ for $n > \max(m'(\epsilon),m(\tilde{\epsilon}))$ where $m(\tilde{\epsilon})$ is from the hypothesis of Lemma 1}\\
&\leq  \varepsilon/2 \text{ for $n > \max(m'(\varepsilon^{2}/(8\bar{h}^{n}_{1} + 8\bar{h}_{0})^{2}),m( \varepsilon /(4\sqrt{R}\bar{h}^{n}_{1} + 4\sqrt{R}\bar{h}_{0}) ))$}
\end{align*}
where the third inequality follows because the eigenvalues of compact Hermitian operators are Lipschitz continuous (see the paragraph after Lemma B3 in Online Appendix Section B.1) and the last inequality follows if $\epsilon$, $R$, and $m'$ are chosen so that $\epsilon = \varepsilon^{2}/(8\bar{h}^{n}_{1} + 8\bar{h}_{0})^{2}$ and $\tilde{\epsilon}$ and $m$ are chosen so that $\tilde{\epsilon} = \varepsilon /(4\sqrt{R}\bar{h}^{n}_{1} + 4\sqrt{R}\bar{h}_{0}) $. The claim follows.  $\square$
\end{flushleft}

\begin{flushleft}
\textbf{Lemma 4:} $\max\left(\sum_{r \in [n]}\left((\lambda_{r0}^{n})^{2}+(\lambda_{r1}^{n})^{2}\right) - 1,0\right) \leq  \int\int h^{n}_{1}(u,v)h^{n}_{0}(u,v)dudv \leq \min\left(\sum_{r \in [n]}(\lambda_{r0}^{n})^{2},\sum_{r \in [n]}(\lambda_{r1}^{n})^{2}\right)$.
\end{flushleft}

\begin{flushleft}
\textbf{Proof of Lemma 4:} This lemma follows the logic of \cite{whitt1976bivariate}, Theorem 2.1. Since  $h_{0}^{n}$ and $h_{1}^{n}$ take values in $[0,1]$, the upper bound follows  
\begin{align*}
\int\int h^{n}_{1}(u,v)h^{n}_{0}(u,v)dudv 
\leq \min_{s \in \{0,1\}}\int\int (h^{n}_{s}(u,v))^{2}dudv 
= \min_{s \in \{0,1\}}\sum_{r \in [n]}(\lambda_{rs}^{n})^{2}.
\end{align*}
The lower bound follows
 \begin{align*}
 \int\int h^{n}_{1}(u,v)&h^{n}_{0}(u,v)dudv 
 = \int\int h^{n}_{1}(u,v)\left( 1 - \left(1 - h^{n}_{0}(u,v)\right)\right)dudv \\
 &\geq \int\int h^{n}_{1}(u,v)dudv - \min \left(\int\int h^{n}_{1}(u,v)dudv , \int\int\left(1 - h^{n}_{0}(u,v)\right)dudv\right)\\
  &= \max\left(0,\int\int (h^{n}_{1}(u,v))^{2}dudv + \int\int (h^{n}_{0}(u,v))^{2}dudv - 1\right) \\
 &= \max\left(\sum_{r \in [n]}\left((\lambda_{r0}^{n})^{2}+(\lambda_{r1}^{n})^{2}\right) - 1,0\right). 
 \end{align*}
 The claim follows. $\square$
\end{flushleft}

\subsection{Proposition 2}
Let $h_{s}(u,v) = \int\mathbbm{1}\{g_{s}(\varphi_s(u),\varphi_s(v),w) \leq y_{s}\}dw$. For any $n\in \mathbb{N}$ let $S_{i} := \left(\frac{i-1}{n},\frac{i}{n}\right]$, $H_{s}^{n}$ be an $n\times n$ matrix with $H_{ij,s}^{n} \in \mathbb{R}$ as its $ij$th entry, and $h_{s}^{n}(u,v) = \sum_{ij}H_{ij,s}^{n}\mathbbm{1}\{u \in S_{i}, v \in S_{j}\}$ such that $\int\int\left(h_{s}(u,v)-h_{s}^{n}(u,v)\right)^{2}dudv \to 0$ as $n\to\infty$ as per Lemma 1. Let $\{\lambda_{rs}\}$ denote the eigenvalues of  $h_{s}$ and $\{\lambda^{n}_{rs}\}$ the eigenvalues of  $h_{s}^{n}$. \newline

For any $\epsilon > 0$ there exists an $m \in \mathbb{N}$ such that for every $n > m$
\begin{align*}
 &\int\int h_{1}(u,v)h_{0}(u,v)dudv
 < \int h_{1}^{n}(u,v)h_{0}^{n}(u,v)dudv + \epsilon \\
& \leq  \min\left(\sum_{r}\lambda^{n}_{r1}\lambda^{n}_{r0},\sum_{r}(\lambda^{n}_{r1})^{2},\sum_{r}(\lambda^{n}_{r0})^{2}\right) + \epsilon  
< \min\left(\sum_{r}\lambda_{r1}\lambda_{r0},\sum_{r}\lambda^{2}_{r1},\sum_{r}\lambda^{2}_{r0}\right) + 2\epsilon
\end{align*}
where the first inequality is due to Part i of Lemma 3, the second inequality is the intersections of the upper bounds in Lemmas 2 and 4, and the third inequality is due to Part ii of Lemma 3. Similarly, 
\begin{align*}
 &\int\int h_{1}(u,v)h_{0}(u,v)dudv 
 > \int h_{1}^{n}(u,v)h_{0}^{n}(u,v)dudv - \epsilon \\
& \geq \max\left(\sum_{r}\lambda^{n}_{r1}\lambda^{n}_{\rho(r)0},\sum_{r }\left((\lambda_{r0}^{n})^{2}+(\lambda_{r1}^{n})^{2}\right) - 1,0\right) - \epsilon \\
 &> \max\left(\sum_{r}\lambda_{r1}\lambda_{\rho(r)0},\sum_{r}\left(\lambda_{r0}^{2}+\lambda_{r1}^{2}\right) - 1,0\right) - 2\epsilon.
\end{align*}
Since $\epsilon > 0$ is arbitrary, the claim follows. $\square$

\subsection{Proposition 3}
We use the same notation and definitions as in the proof of Proposition 2 above. For any $y_{1},y_{0} \in \mathbb{R}$ such that $y_{1}-y_{0} = y$ we have
\begin{align*}
\mathbbm{P}\left(Y_{ij}(1) - Y_{ij}(0) \leq y\right) &\geq \mathbbm{P}\left(Y_{ij}(1) \leq y_1, -Y_{ij}(0) < -y_0\right)\\
&=  \mathbbm{P}\left(Y_{ij}(1) \leq y_1\right) -  \mathbbm{P}\left(Y_{ij}(1) \leq y_1, Y_{ij}(0) \leq  y_0\right)\\
&= \int\int h_{1}(u,v) dudv - \int\int h_{1}(u,v)h_{0}(u,v) dudv \\
&\geq \int\int \left(h_{1}(u,v)\right)^2 dudv - \int\int h_{1}(u,v)h_{0}(u,v) dudv \\
&\geq \sum_{r}\lambda_{r1}^{2} - \min\left(\sum_{r}\lambda_{r1}^{2},\sum_{r}\lambda_{r0}^{2},\sum_{r}\lambda_{r1}\lambda_{r0}\right) \\
&= \max\left(\sum_{r}(\lambda_{r1}^{2}-\lambda_{r0}^{2}),\sum_{r}(\lambda_{r1}^{2} - \lambda_{r1}\lambda_{r0}),0\right)
\end{align*}
and
\begin{align*}
\mathbbm{P}\left(Y_{ij}(1) - Y_{ij}(0) \leq y\right) 
 &\leq  \mathbbm{P}\left(Y_{ij}(1) \leq y_1 \cup -Y_{ij}(0) < -y_0\right)\\
 &= \mathbbm{P}\left( -Y_{ij}(0) < -y_0 \right) +  \mathbbm{P}\left(Y_{ij}(1) \leq y_1, Y_{ij}(0) \leq y_0\right)\\
&= 1 -  \mathbbm{P}\left( Y_{ij}(0) < y_0 \right)  +   \mathbbm{P}\left(Y_{ij}(1) \leq y_1, Y_{ij}(0) \leq y_0\right)\\
&= 1 - \int\int h_{0}(u,v) dudv + \int\int h_{1}(u,v)h_{0}(u,v) dudv \\
&\leq 1 -  \int\int \left(h_{0}(u,v)\right)^2 dudv + \int\int h_{1}(u,v)h_{0}(u,v) dudv \\
&\leq 1 - \sum_{r}\lambda_{r0}^{2} + \min\left(\sum_{r}\lambda_{r1}^{2},\sum_{r}\lambda_{r0}^{2},\sum_{r}\lambda_{r1}\lambda_{r0}\right)  \\
&= 1 + \min\left(\sum_{r}(\lambda_{r1}^{2}-\lambda_{r0}^{2}),\sum_{r}(\lambda_{r1}\lambda_{r0} - \lambda_{r0}^{2}),0\right)
\end{align*}
where the the first inequality in both systems is because 
\begin{align*}
\mathbbm{1}\{Y_{ij}(1) \leq y_{1}\}\mathbbm{1}\{-Y_{ij}(0) < -y_{0}\} 
\leq \mathbbm{1}\{Y_{ij}(1)-Y_{ij}(0) \leq y\}\\
\leq \max\left(\mathbbm{1}\{ Y_{ij}(1) \leq y_{1} \},\mathbbm{1}\{-Y_{ij}(0) < -y_{0}\}\right),
\end{align*}
the second inequality is because $h_{s}$ takes values in $[0,1]$, and the third inequality is because the upper bound in Proposition 2. Since these inequalities hold for any $y_{1},y_{0} \in \mathbb{R}$ such that $y_{1} - y_{0} = y$, the claim follows. $\square$

\section{Auxiliary lemmas}
\begin{flushleft}
\textbf{Lemma B1 (Lusin):} For any measurable $f: [0,1]^{2} \to \mathbb{R}$ and $\epsilon > 0$ there exists a compact $E_{\epsilon} \subseteq [0,1]^{2}$ with Lebesgue measure at least $1-\epsilon$ such that $f$ is continuous when restricted to $E_{\epsilon}$. See \cite{dudley2002real} Theorem 7.5.2.    
\end{flushleft}

\begin{flushleft}
\textbf{Lemma B2 (Spectral):}  Let $f :[0,1]^{2}\to \mathbb{R}$ be a bounded symmetric measurable function and $T_{f}: L_{2}([0,1]) \to L_{2}([0,1])$ the associated integral operator $(T_{f}g)(u) = \int f(u,\tau)g(\tau)d\tau$.  $T_{f}$ admits the spectral decomposition $f(u,v) = \sum_{r=1}^{\infty}\lambda_{r}\phi_{r}(u),\phi_{r}(v)$ in the sense that $(T_{f}g)(u) = \int f(u,\tau)g(\tau)d\tau = \sum_{r=1}^{\infty}\lambda_{r}\phi_{r}(u) \int\phi_{r}(\tau)g(\tau)d\tau$ for any $g \in L_{2}([0,1])$. Each $(\lambda_{r},\phi_{r})$ pair satisfies $\int f(u,\tau)\phi_{r}(\tau)d\tau = \lambda_{r}\phi_{r}(u)$ where $\{\lambda_{r}\}_{r=1}^{\infty}$ is a multiset of bounded real numbers with $0$ as its only limit point and $\{\phi_{r}\}_{r=1}^{\infty}$ is an orthogonal basis of $L_{2}([0,1])$. See \cite{birman2012spectral} equation (5) preceding Theorem 4 in Chapter 9.2.  \looseness=-1  
\end{flushleft}

The spectral decomposition in Lemma B2 is related to the finite-dimensional version that is commonly used for matrices which is if $Y$ is an $N\times N$ dimensional symmetric real-valued matrix then $Y_{ij} = \sum_{r=1}^{N}\lambda_{r}\phi_{ir}\phi_{jr}$. Each $(\lambda_{r},\phi_{r})$ pair satisfies $\sum_{j=1}^{N}Y_{ij}\phi_{jr} = \lambda_{r}\phi_{ir}$ where $\{\lambda_{r}\}_{r=1}^{N}$ is a multiset of real numbers and $\{\phi_{ir}\}_{i,r=1}^{N}$ is an $N \times N$ orthogonal matrix with $r$th column denoted by $\phi_{r}$.  \looseness=-1

\begin{flushleft}
\textbf{Lemma B3 (Continuity):} Let $f,g :[0,1]^{2}\to \mathbb{R}$ be bounded symmetric measurable functions with positive eigenvalues $\{\lambda^{+}_{r}(f),\lambda^{+}_{r}(g)\}_{r \in \mathbb{N}}$ and negative eigenvalues $\{\lambda^{-}_{r}(f),\lambda^{-}_{r}(g)\}_{r \in \mathbb{N}}$ both ordered to be decreasing in absolute value. Suppose $\left(\int\int \left(f(u,v)-g(u,v)\right)^{2}dudv\right)^{1/2} \leq \epsilon$. Then $|\lambda^{+}_{r}(f)-\lambda^{+}_{r}(g)| \leq \epsilon$ and $|\lambda_{r}^{-}(f)-\lambda_{r}^{-}(g)| \leq \epsilon$ for every ${r \in \mathbb{N}}$. See \cite{birman2012spectral}, equation (19) following Theorem 8 in Chapter 9.2.  \looseness=-1   
\end{flushleft}

In Lemma 3 of Appendix Section A.3 in the main text, we use an implication of Lemma B3 and Theorem 368 of \cite{hardy1952inequalities} (Lemma B5 below) that $\left(\sum_{r \in [R]}\left(\lambda_{r}(f)-\lambda_{r}(g)\right)^{2}\right)^{1/2} \leq \sqrt{R}\left(\int\int \left(f(u,v)-g(u,v)\right)^{2}dudv\right)^{1/2}$ where $\{\lambda_{r}(f),\lambda_{r}(g)\}_{r \in [R]}$ are the $R$ largest in absolute value eigenvalues of $f$ and $g$ ordered to be decreasing. This result is an analog of the Hoffman-Wielandt inequality for matrices (Lemma B6 below) which, in a previous version of our paper, we refined in Proposition 4 of the main text. \looseness=-1

\begin{flushleft}
\textbf{Lemma B4 (Birkhoff):} For every $M \in \mathcal{D}^{+}_{n}$ there exists an $m \in \mathbb{N}$, $\alpha_{1},...,\alpha_{m} > 0$, and $P_{1},...,P_{m} \in \mathcal{P}_{n}$ such that $\sum_{t=1}^{m}\alpha_{t} = 1$ and $M_{ij} = \sum_{t=1}^{m}\alpha_{t}P_{ij,t}$. See \cite{birkhoff1946three}.   \looseness=-1  
\end{flushleft}

\begin{flushleft}
\textbf{Lemma B5 (Hardy-Littlewood-Polya Theorem 368):} For any $m\in \mathbb{N}$ and $g,h \in \mathbb{R}^{m}$ we have $\sum_{r=1}^{m} g_{(r)}h_{(m-r+1)} \leq  \sum_{r=1}^{m} g_{r}h_{r}  \leq  \sum_{r=1}^{m} g_{(r)}h_{(r)}$ where $g_{(r)}$ is the $r$th order statistic of $g$. See \cite{hardy1952inequalities}, Section 10.2, Theorem 368.  \looseness=-1  
\end{flushleft}

A version of Lemma B5 also holds for elements of $L^{2}([0,1])$, see  \cite{hardy1952inequalities}, Section 10.13, Theorem 378. Specifically, for any $g,h \in L^{2}([0,1])$ we have $\int g^{+}(u)h^{+}(1-u)du \leq \int g(u)h(u)du \leq \int g^{+}(u)h^{+}(u)du$ where $g^+$ is the quantile function of $g$. This result is also used in the second proof of Theorems 2.1 and 2.5 in \cite{whitt1976bivariate}.   \looseness=-1

\begin{flushleft}
\textbf{Lemma B6 (Hoffman-Wielandt):} Let $\{\lambda_{r}(F)\}_{r \in [n]}$ and $\{\lambda_{r}(G)\}_{r \in [n]}$ be the eigenvalues of two $n\times n$ real symmetric matrices $F$ and $G$, ordered to be decreasing. Then $\sum_{r=1}^{n}\left( \lambda_{r}(F) - \lambda_{r}(G)\right)^{2} \leq \sum_{i=1}^{n}\sum_{j=1}^{n}\left(F_{ij}-G_{ij}\right)^{2}$. See \cite{hoffman1953hw}.  \looseness=-1  
\end{flushleft}

\begin{flushleft}
\textbf{Lemma B7 (Lov\'asz Corollary 10.35(a)):} Let $f,g: [0,1]^2 \to \mathbbm{R}$ be bounded symmetric measurable functions and $G = (V,E)$ be an arbitrary graph where $V$ is a finite set of vertices and $E \subseteq V \times V$ be an arbitrary subset of vertex-pairs. Define the graph homomorphism $t(G,f) = \int_{u_1,\ldots,u_{|V|}}\prod_{ij \in E}f(u_i,u_j)du_1,\ldots,du_{|V|}$. Then $t(G,f) = t(G,g)$ for every graph $G$ if and only if $f$ and $g$ are equivalent up to a measure preserving transformation (see Definition 2 in Section 4.1). See \cite{lovasz2012large}, Section 10.7, Corollary 10.35(a).\footnote{When \cite{lovasz2012large} states this corollary, he does not explicitly state the condition that $t(G,f) = t(G,g)$ for every graph $G$. He instead only writes that $f$ and $g$ are weakly isomorphic. The former condition is the definition of a weak isomorphism, which he gives after Proposition 7.10 in Section 7.3.} \looseness=-1  
\end{flushleft}

\begin{flushleft}
\textbf{Lemma B8 (Whitt Lemma 2.7):} For any cdf $H$ on $\mathbbm{R}^{n}$ and uniform random variable $U$, there exits a measurable $x: [0,1]\to\mathbbm{R}^{n}$ such that $x(U)$ has cdf $H$. See \cite{whitt1976bivariate}, Lemma 2.7.
\end{flushleft}

\section{Quadratic assignment problem}
The quadratic assignment problem described in Section 5.1 of the main text is neither analytically solvable nor directly computable in general. However, there are  special cases for which there is a simple analytical solution. In this section we give two examples.  In both cases, the outer bounds we propose in Proposition 2 of Section 5.2 agree with the analytical solutions, which means that, in these examples, our outer bounds are sharp.  \looseness=-1

\subsection{Diagonal graph functions}
In this example, the graph functions are diagonal. That is, $h_{s}(u,v) = \alpha_{s}(u)\delta_{u}(v)$ where $\delta$ refers to the Dirac delta function. Following Proposition 1, the lower bound on the identified set for the DPO is \looseness=-1
\begin{align*}
\min_{\psi_{0},\psi_{1} \in \mathcal{M}}\int\int \prod_{s \in \{0,1\}}h_s(\psi_s(u),\psi_s(v))dudv 
= \min_{\varphi_{0},\varphi_{1} \in \mathcal{M}}\int \prod_{s \in \{0,1\}}\alpha_{s}(\psi_s(u)) du 
= \int \alpha^{+}_{0}(u)\alpha^{+}_{1}(1-u)du
\end{align*}
where the first equality is due to the definition of the Dirac delta function and the second equality is due to the functional version of the Hardy-Littlewood-Polya Theorem 368 (Lemma B5 in Online Appendix Section B). By the same arguments, the upper bound is \looseness=-1
\begin{align*}
\max_{\psi_{0},\psi_{1} \in \mathcal{M}}\int\int \prod_{t \in \{0,1\}}h_s(\psi_s(u),\psi_s(v))dudv = \int \alpha^{+}_{0}(u) \alpha^{+}_{1}(u)du.
\end{align*}
These bounds are equivalent to the bounds we propose in Proposition 2. This is because, in this example, $h_s(u,v)$ is diagonal, and so the lower bound can be represented \looseness=-1
\begin{align*}
\int \alpha^{+}_{0}(u)\alpha^{+}_{1}(1-u)du = \sum_{r}\lambda_{r0}\lambda_{s(r)1}
\end{align*}
and the upper bound can be represented \looseness=-1
\begin{align*}
 \int \alpha^{+}_{0}(u) \alpha^{+}_{1}(u)du = \sum_{r}\lambda_{r0}\lambda_{r1}
\end{align*}
which are weakly outside the Proposition 2 bounds. But since these bounds are sharp, it follows that the Proposition 2 bounds must be sharp as well. \looseness=-1

\subsection{Block graph functions} 
In this example, the graph functions have a block structure. That is, $h_{s}(u,v) = f_{s}(u)f_{s}(v)$ where $f_{s}(u) = \mathbbm{1}\{u \in A_{s}\}$ for some Lebesgue measurable sets $A_{s} \subseteq [0,1]$ and $s \in \{0,1\}$. Following Proposition 1, the lower bond on the identified set for the DPO is, for $y_{s} \in [0,1)$\looseness=-1
\begin{align*}
\min_{\psi_{0},\psi_{1} \in \mathcal{M}}\int\int \prod_{s \in \{0,1\}}h_s(\psi_s(u),\psi_s(v))dudv  
= \min_{\psi_{0},\psi_{1} \in \mathcal{M}}\int\int \prod_{s \in \{0,1\}}f_{s}(\psi_{t}(u))f_{s}(\psi_{s}(v))dudv \\
= \int f_{0}^{+}(u)f_{1}^{-}(u)du
= \max(|A_{0}|+|A_{1}| - 1, 0)
\end{align*}
where $|A_{s}| = \int f_{s}(\tau)d\tau$ refers to the measure of $A_{s}$, the second equality is due to the functional version of the Hardy-Littlewood-Polya Theorem 368 (Lemma B5 in Online Appendix Section B). By the same arguments, the upper bound is 
\begin{align*}
\max_{\psi_{0},\psi_{1} \in \mathcal{M}}\int\int \prod_{t \in \{0,1\}}h_s(\psi_s(u),\psi_s(v))dudv
= \int f_{0}^{+}(u)f_{1}^{+}(u)du
= \min(|A_{0}|,|A_{1}|).
\end{align*}

These bounds are equivalent to the bounds we propose in Proposition 2. This is because $h_s$ has  one non-zero eigenvalue that is equal to $\sqrt{|A_{s}|}$. As a result, the above lower bound can be rewritten \looseness=-1
\begin{align*}
\max(|A_{0}|+|A_{1}| - 1, 0) = \max\left(\sum_{r}(\lambda_{r0}^{2} + \lambda_{r1}^{2}) - 1, 0\right) 
\end{align*}
and the above upper bound can be rewritten
\begin{align*}
\min(|A_{0}|,|A_{1}|) = \min\left(\sum_{r}\lambda_{r0}^{2} , \sum_r\lambda_{r1}^{2}\right) 
\end{align*}
which are weakly outside the Proposition 2 bounds. But since these bounds are sharp, it follows that the Proposition 2 bounds must be sharp as well. \looseness=-1

\section{Additional results and details}
\subsection{Asymmetric outcome matrices}
Our identification arguments in Section 4 and bounds in Section 5 of the main text apply to undirected unipartite networks. These networks are represented by symmetric adjacency matrices where the rows and columns of the matrix are indexed by the same community of agents. $Y_{ij}(s)$  and $Y_{ji}(s)$ describe the magnitude of a connection between agents $i$ and $j$ under policy $s$. We require symmetry here because, in our proof of Proposition 2, when we take a spectral decomposition of the histogram approximation to the graph function, we assume that the right and left eigenvectors are the same. This is only the case when the graph function is symmetric.\footnote{While other spectral decompositions may be applied when the graph function is asymmetric, these decompositions must, by definition, have different right and left eigenvectors when the graph function is asymmetric. Our current proof strategy does not directly apply to this decomposition.} \looseness=-1

Directed networks connecting two or more types of agents (sometimes called bipartite or multipartite networks in the literature) can be incorporated through symmetrization, which is commonly used in the literature on U-statistics, going back to at least \cite{hoeffding1948class} (see his equation 3.3), A textbook reference is Section 5.1.1 of \cite{serfling2009approximation}. Formally, we consider a community of $N$ agents with $K$ types where $T_{i}$ takes value $k \in 1,\ldots,K$ if agent $i$ is of type $k$. The types are assumed to be mutually exclusive and collectively exhaustive. The researcher conducts a completely randomized balanced experiment with $2N_k$ agents of type $k$ where $\sum_{k=1}^{K}N_k = N$. The agents are randomly assigned to two communities as given by the vector $D$ drawn uniformly at random from the set $\{d \in \{0,1\}^N: \sum_{i=1}^{2N}\mathbbm{1}\{T_i = 1\}d_i = N_1,\ldots, \sum_{i=1}^{N}\mathbbm{1}\{T_i = K\}d_i = N_K\}$. We model the potential connection from agent $i$ of type $T_i = k$ to agent $j$ of type $T_j = l$ with the model $Y_{ij}(s) = g_{s,kl}(\varphi_s(w_i),\varphi_s(w_j),\eta_{ij})$, where $w_i$ and $\eta_{ij}$ are defined as in Section 3.2. We define the graph function as in Section 4.1: $h_{s,kl}(a,b) = \int\mathbbm{1}\{g_{s,kl}(\varphi_s(a),\varphi_s(b),w) \leq y_s\}dw$. Relative to the link function specified in Section 3.2 and the graph function specified in Section 4.1, the link and graph functions here are allowed to vary with the agent types $k$ and $l$ and may be asymmetric. We focus on the DPO for the connections from agents of type $k$ to agents of type $l$. This is the parameter $\mathbbm{P}\left(Y_{ij}(1) \leq y_1, Y_{ij}(0) \leq y_0 | T_i = k, T_j = l\right)$ which, following the logic of Section 4.2, can be rewritten as $\int\int h_{1,kl}(u,v)h_{0,kl}(u,v)dudv$.  \looseness=-1

To apply the results of Section 4 and 5, the symmetrization strategy works by replacing the graph functions $h_{0,kl}$ and $h_{1,kl}$ with the functions $h^{\dagger}_{0,kl}$ and $h^{\dagger}_{1,kl}$ that are symmetric. Specifically, for $a,b \in [0,1]$ and $s \in \{0,1\}$, we define 
\begin{align*}
h^{\dagger}_{s,kl}(a,b) = \left[h_{s,kl}(2a,2b-1)\mathbbm{1}\{a \leq 1/2, b > 1/2\} + h_{s,kl}(2b,2a-1)\mathbbm{1}\{a > 1/2, b \leq 1/2\}\right]/2. 
\end{align*} 
The function $h^{\dagger}_{s,kl}$ is symmetric and satisfies $\mathbbm{P}\left(Y_{ij}(1) \leq y_1, Y_{ij}(0) \leq y_0 | T_i = k, T_j = l\right)$ $= \int\int h_{1,kl}(u,v)h_{0,kl}(u,v)dudv$ $= \int\int h^{\dagger}_{1,kl}(u,v)h^{\dagger}_{0,kl}(u,v)dudv$. Since $h^{\dagger}_{s,kl}$ is symmetric, Proposition 1 in Section 4.2 and Propositions 2 and 3 of Section 5.2 can be applied. Furthermore, since $h^{\dagger}_{s,kl}$ is a simple function of $h_{s,kl}$ the researcher can use any estimate of the latter to construct an estimate of the former. \looseness=-1

\subsection{Row and column heterogeneity}
In practice, the bounds in Propositions 2 and 3 may be wide when there is nontrivial heterogeneity in the  row and column means of the graph functions. In such cases, we propose an adjustment building on Section 5 of  \cite{finke1987quadratic} that can, in some cases, lead to tighter bounds. Specifically, we write $h_{s}(u,v) = \alpha_{s}(u) + \alpha_{s}(v) + \epsilon_{s}(u,v)$ where $\alpha_{s}(u) = \int  h_{s}(u,v) dv - \frac{1}{2}\iint h_{s}(u,v)dudv$ and $\epsilon_{s}(u,v) = h_{s}(u,v) - \alpha_s(u) - \alpha_s(v)$. Since $h_{s}$ is identified up to a measure preserving transformation by Proposition 1 in the main text, $\alpha_s(u)$, the projection of $h_{s}$ onto its first argument, is identified up to a measure preserving transformation. It follows that $\epsilon_s$, a linear function of $h_{s}$ and $\alpha_s$, is also identified up to a measure preserving transformation.     \looseness=-1 

The DPO is then  \looseness=-1 
\begin{align*}
F(y_{1},y_{0}) = \int\int \prod_{s \in \{0,1\}}\left(\alpha_{s}(u) + \alpha_{s}(v) + \epsilon_{s}(u,v)\right) dudv \\
= \int\int \prod_{s\in \{0,1\}}\left(\alpha_{s}(u) + \alpha_{s}(v)\right) dudv + \int\int \prod_{s \in \{0,1\}}\epsilon_{s}(u,v)dudv.  
\end{align*}

We bound the two summands separately. Specifically, the upper bound is \looseness=-1 
\begin{align*}
F(y_{1},y_{0}) \leq \max_{\psi_{1},\psi_{0} \in \mathcal{M}}\left[\int\int \prod_{s \in \{0,1\}}\left(\alpha_{s}(\psi_{s}(u)) + \alpha_{s}(\psi_{s}(v))\right) dudv + \int\int \prod_{s \in \{0,1\}}\epsilon_{s}(\psi_{s}(u),\psi_{s}(v))dudv\right] \\
\leq  \max_{\psi_{1},\psi_{0} \in \mathcal{M}}\left[\int\int \prod_{s \in \{0,1\}}\left(\alpha_{s}(\psi_{s}(u)) + \alpha_{s}(\psi_{s}(v))\right) dudv\right] + \max_{\psi_{1},\psi_{0} \in \mathcal{M}}\left[\int\int \prod_{s \in \{0,1\}}\epsilon_{s}(\psi_{s}(u),\psi_{s}(v))dudv\right]. 
\end{align*}
The first summand is bounded from above by \looseness=-1 
\begin{align*}
2\max_{\psi_{1},\psi_{0} \in \mathcal{M}}\left[\int \alpha_{1}(\psi_{1}(u))\alpha_{0}(\psi_{0}(u)) du\right]  
+ 2\bar{\alpha}_{1}\bar{\alpha}_{0}
\leq 2\int\alpha_{1}^{+}(u)\alpha_{0}^{+}(u)du  + 2\bar{\alpha}_{1}\bar{\alpha}_{0}
\end{align*}
where $\bar{\alpha}_{s} = \int \alpha_{s}(u)du$ and $\alpha_{s}^{+}$ is the quantile function of $\alpha_{s}$. See the functional version of the Hardy-Littlewood-Polya Theorem 368 (Lemma B5 in Online Appendix Section B). Following Proposition 2 of the main text, the second summand is bounded from above by $\min\left(\sum_{r}\lambda_{r1}^{2},\sum_{r}\lambda_{r0}^{2},\sum_{r}\lambda_{r1}\lambda_{r0}\right)$ where $\lambda_{rs}$ refers to the $r$th eigenvalue of $\epsilon_{s}$ and the sums are as defined as in Section 4.1.2 of the main text. Together, the bounds imply that \looseness=-1 
\begin{align*}
F(y_{1},y_{0}) \leq 2\int\alpha_{1}^{+}(u)\alpha_{0}^{+}(u)du + 2\bar{\alpha}_{1}\bar{\alpha}_{0} + \min\left(\sum_{r}\lambda_{r1}^{2},\sum_{r}\lambda_{r0}^{2},\sum_{r}\lambda_{r1}\lambda_{r0}\right).
\end{align*}

By the same logic, the lower bound on the DPO is\looseness=-1 
\begin{align*}
F(y_{1},y_{0}) \geq 2\int\alpha_{1}^{+}(u)\alpha_{0}^{+}(1-u)du + 2\bar{\alpha_{1}}\bar{\alpha}_{0} + \max\left(\sum_{r}\left(\lambda_{r1}^{2} + \lambda_{r0}^{2}\right) - 1,\sum_{r}\lambda_{r1}\lambda_{s(r)0},0\right).
\end{align*}

We emphasize that $\alpha^{+}_s$, $\bar{\alpha}_s$, and $\lambda_{rs}$ are all invariant to measure preserving transformations of $\alpha_s$ and $\epsilon_s$. Since these functions are identified up to a measure preserving transformation, is follows that $\alpha^{+}_s$, $\bar{\alpha}_s$, and $\lambda_{rs}$ are all point identified, and so the upper and lower bounds provided above are point identified. Bounds on the DTE can be constructed from those on the DPO following the logic of Proposition 3 of the main text. \looseness=-1

\subsection{Estimation and inference}
In this section, we provide details about how we estimated the bounds and constructed confidence intervals in Table 2 of Section 6 in the main text. As described in Section 5.3, we first estimate the graph functions $h_1$ and $h_0$, compute the eigenvalues of the estimated functions, and then plug the eigenvalues into the relevant bounds. A large econometrics and statistics literature considers the problem of estimating graph functions for dyadic data, under a wide variety of conditions, see broadly, the handbook chapter by \cite{graham2020network}.  It is impossible for us to cover this entire literature here so instead we focus on the class of dyadic regression models described by \cite{graham2020network} in his Section 4. We focus on this class of models for two reasons. First, dyadic regression models are popular in the network economics literature, being used in nearly all of the motivating examples we referenced in the introduction of the main text. Second, we can build on the estimation and inference results provided by \cite{graham2020network} in that section.  \looseness=-1 

\subsubsection{Dyadic regression model}
We start by summarizing the class of dyadic regression models in Section 4 of the handbook chapter by \cite{graham2020network}. In his setting, the researcher observes network connections between $N$ agents. For each agent $i = 1,\ldots, N$, they observe a vector of covariates $X_i \in \mathbbm{X}$ where $\mathbbm{X}$ is a compact subset of $\mathbbm{R}^{L}$. For each of the $N(N-1)$ pairs of agents $i,j = 1,\ldots, {N\choose 2}$ with $i \neq j$ they observe the network connection $Y_{ij}$ determined by the model \looseness=-1 
\begin{align*}
Y_{ij} = g\left(X_i,X_j,\varepsilon_{ij}\right)
\end{align*}
where $\varepsilon_{ij}$ is an unobserved error. The error $\varepsilon_{ij}$ is decomposed into $\varepsilon_{ij} = \{U_i,U_j,V_{ij}\}$ where $\{X_i,U_{i}\}_{i=1}^{N}$ and $\{V_{ij},V_{ji}\}_{i,j=1}^{N}$ are iid, have mutually independent entries, and the marginal distributions of $U_i$ and $V_{ij}$ are normalized to be standard uniform. \cite{graham2020network} further normalizes the errors $\{U_{i}\}_{i=1}^{N}$ so that the entries of  $\{X_i\}_{i=1}^{N}$ and $\{U_{i}\}_{i=1}^{N}$ are mutually independent. These normalizations are made at the bottom of his page 141 in Section 4.1. \looseness=-1 

Under this model, the outcomes $\{Y_{ij}\}_{i,j = 1}^{N}$ are jointly exchangeable (i.e. their distribution is invariant to relabelings of the agent indices) and the outcomes $Y_{ij}$ and $Y_{i'j'}$ are independent if $\{i,j\}$ and $\{i',j'\}$ do not share an index. In contrast to our general model given in Section 3.2.2, the link function $g$ is not indexed by $N$ in this setting, so the network is either dense or empty in the limit. \cite{graham2020network} assumes that the marginal density of $Y_{ij}$ conditional on $X_i$ and $X_j$ belongs to a parametric family $\mathcal{F} =\{f_{Y_{12}|X_1,X_2}(Y_{12}|X_1,X_2;\theta): \theta \in \Theta \subseteq \mathbbm{R}^{K})\}$ in his Section 4.2, using $\theta_0$ to denote the parameter that describes the data. We will use $f_0$ for the corresponding element of $\mathcal{F}$.   \looseness=-1 

A concrete example of a dyadic regression model is the one \cite{fafchamps2007risk} specify for informal risk sharing that we described in Example 1 of Section 3.2.2 in the main text. Additional examples applying the model to international trade, supply chain linkages across firms, R\&D collaborations and more can be found in Section 4 of  \cite{graham2020network}. \looseness=-1 

\subsubsection{Estimation of $\theta_0$}
We now review the estimation strategy proposed by \cite{graham2020network}. Setting  $l_{ij}(\theta) = \ln f_{Y_{12}|X_1,X_2}(y_{ij}|X_i,X_j;\theta)$, \cite{graham2020network} considers the estimator that maximizes the composite likelihood\looseness=-1 
\begin{align*}
\hat{\theta} = \argmax_{\theta \in \Theta}\frac{1}{N(N-1)}\sum_{i \neq j}l_{ij}(\theta).
\end{align*}
The objective function on the right-hand side of this problem is called the composite likelihood, and not the log-likelihood function, for the connections given the covariates because it ignores the fact that the connections are dependent across pairs of agents that share an index. Still, \cite{graham2020network} writes that, under certain regularity conditions, the resulting estimator $\hat{\theta}$ is consistent and asymptotically normal in his Section 4.3. Specifically, he finds that 
  \begin{align*}
 \sqrt{N}(\hat{\theta}-\theta_0) \to_d \mathcal{N}\left(0,V\right).
 \end{align*}
(his equation 33 in Section 4.3) where $V = 4\left(\Gamma_0 \Sigma_1^{-1}\Gamma_0\right)^{-1}$, $\Gamma_0$ is the probability limit of the Hessian evaluated at $\theta$, $H_{N}(\theta) = \frac{1}{N(N-1)}\sum_{i \neq j}\frac{\partial^2l_{ij}(\theta)}{\partial\theta \partial\theta'}$, and $\Sigma_1$ is the variance of the projections of the symmetrized score $\frac{s_{ij} + s_{ji}}{2}$ where $s_{ij} = \frac{\partial l_{ij}(\theta)}{\partial\theta}$. {\cite{graham2020network} defines these parameters after his equation 28 in Section 4.3.  \looseness=-1 

To estimate the variance $V$, \cite{graham2020network} proposes $\hat{V} = \left(\hat{\Gamma}\hat{\Omega}\hat{\Gamma}\right)^{-1}$ in equation 43 of his Section 4.4 where $\hat{\Gamma} = H_{N}(\hat{\theta})$, the Hessian evaluated at $\hat{\theta}$ (equation 44), and 
\begin{align*}
\hat{\Omega} = 4\hat{\Sigma}_1 + \frac{2}{N-1}\left(\widehat{\Sigma_2 + \Sigma_3} - 2\hat{\Sigma}_1\right)
\end{align*}
where 
\begin{align*}
\hat{\Sigma}_1 = {N \choose 3}^{-1}\sum_{i < j < k}\frac{1}{3}\left\{\left(\frac{\hat{s}_{ij}+\hat{s}_{ji}}{2}\right)\left(\frac{\hat{s}_{ik}+\hat{s}_{ki}}{2}\right)' + \left(\frac{\hat{s}_{ij}+\hat{s}_{ji}}{2}\right)\left(\frac{\hat{s}_{jk}+\hat{s}_{kj}}{2}\right)' \right. \\ \left.+ \left(\frac{\hat{s}_{ik}+\hat{s}_{ki}}{2}\right)\left(\frac{\hat{s}_{jk}+\hat{s}_{kj}}{2}\right)' \right\},
\end{align*}
\begin{align*}
\widehat{\Sigma_2 + \Sigma_3} = {N \choose 2}^{-1}\sum_{i < j}\left(\frac{\hat{s}_{ij}+\hat{s}_{ji}}{2}\right)\left(\frac{\hat{s}_{ij}+\hat{s}_{ji}}{2}\right)',
\end{align*}
 and $\hat{s}_{ij}$ is the score $s_{ij}$ evaluated at $\hat{\theta}$ (equation 45). \cite{graham2020network} also proposes a jackknife estimator for the variance and considers bootstrap inference in his Section 4.5 following \cite{menzel2017bootstrap}. \looseness=-1 

\subsubsection{Potential outcomes and the parameter of interest}
In this subsection, we incorporate the dyadic regression model into our framework, following the logic of our Section 3.2.3. Specifically, we suppose that the potential connections under each policy $s \in \{0,1\}$ are described by a dyadic regression model
\begin{align*}
Y_{ij}(s) = g_s(X_{i,s},X_{j,s},\varepsilon_{ij,s}).
\end{align*}
where $X_{i,s} \in \mathbbm{X}$ is the vector of covariates observed under policy $s \in \{0,1\}$. Importantly, we do not assume that $X_{i,1}$ and $X_{i,0}$ are equal, so that the covariate value for agent $i$ depends on the policy that was implemented. Following Section 3.2.3, we allow $g_1$ and $g_0$ to be any function (so long as the conditional density of $Y_{ij}(s)$ given $X_{i,s}$ and $X_{j,s}$ belongs to an element of $\mathcal{F}$ defined in Section D.3.1) and we model the dependence between $X_{i,1}$ and $X_{i,0}$ as $X_{i,s} = \varphi_s(w_i)$ where $\{w_i\}_{i=1}^{N}$ is an unobserved iid random variable with standard uniform marginals and the function $\varphi_s$ is an unknown measure preserving transformation, except $\varphi_s$ now takes values in $\mathbbm{X}$ instead of $[0,1]$. We also assume that $\{\varepsilon_{ij,1}\}_{i,j=1}^{N}$ and $\{\varepsilon_{ij,0}\}_{i,j=1}^{N}$ have mutually independent entries. For $s \in \{0,1\}$ use $f_s \in \mathcal{F}$ to describe the data under policy $s$, determined by $\theta_s \in \Theta$.  \looseness=-1 

As in Section 4.1, we define the graph function to be 
\begin{align*}
 h_s\left(a,b;\theta_s\right) = \mathbbm{P}\left(Y_{ij}(s) \leq y_s| w_i = a, w_j = b; \theta_s\right)
\end{align*}
where we continue to suppress the dependence of the parameter on the value of $y_s$, but have now added the parameter $\theta_s$ that determines $f_s$. In this dyadic regression setting, we note that the graph function can be written as a functional of $f_s$, i.e.
\begin{align*}
 h_s\left(a,b;\theta_s\right) = \int f_{Y_{12}|X_1,X_2}(\tau | \varphi_s(a),\varphi_s(b);\theta_s)\mathbbm{1}\{\tau \leq y_s\}d\tau.
\end{align*}
We assume that $\mathcal{F}$ is defined so that $h_s$ is twice continuously differentiable in $\theta_s$ with a uniformly bounded derivative.  \looseness=-1 

Following the logic of Section 4.2, the DPO is
\begin{align*}
F(y_1,y_0) &= \mathbbm{P}\left(Y_{ij}(1) \leq y_1, Y_{ij}(0) \leq y_0\right) 
= \iint h_1(u,v;\theta _1)h_0(u,v;\theta _0)dudv.
\end{align*}
By Lemma 1 in the main text, the functions $h_1$ and $h_0$ are identified up to a measure-preserving transformation. The identified set for the DPO is given by Proposition 1. \looseness=-1 

\begin{remark}
We emphasize that even though in the dyadic regression setting the researcher observes covariates and assumes that the marginal distribution of connections conditional on the covariates belongs to a parametric family, this additional information has no effect on the identification analysis of Section 4. This is because, in that analysis, the graph functions were identified up to a measure preserving transformation. While the covariates and parametric assumptions further restrict the shape of the graph functions, they cannot distinguish between graph functions that are equivalent up to a measure preserving transformation because there is no restriction on how the covariates across the two policies are related. And so the identified set is unchanged. \looseness=-1 

Intuitively, this means that, in terms of identification, the covariates and parametric restriction are completely unnecessary because they only reveal information about the DPO that, according to our Proposition 1 in the main text, is already known from the distribution of the data. The additional information may be relevant for estimation and inference, however, which is why we make use of it in this section. \looseness=-1 
\looseness=-1 
\end{remark}

\subsubsection{Outer bounds on the identified set}
We specify outer bounds on the identified set adjusting for row and column heterogeneity as described in Section D.2. Specifically, we decompose $h_{s}(u,v) = \alpha_{s}(u) + \alpha_{s}(v) + \epsilon_{s}(u,v)$ where $\alpha_{s}(u) = \int h_s(u,v)dv - \frac{1}{2}\iint h_s(u,v)dudv$ and $\epsilon_s(u,v) = h_s(u,v) - \int h_s(u,v)du - \int h_s(u,v)dv  + \iint h_s(u,v)dudv$. Following the logic of Online Appendix Section D.2, the upper bound on the identified set for the DPO is  \looseness=-1 
\begin{align*}
U = 2\int\alpha_{1}^{+}(u)\alpha_{0}^{+}(u)du + 2\int\alpha_{1}^{+}(u)du\int\alpha_{0}^{+}(u)du + \min\left(\sum_{r}\lambda_{r0}^{2}, \sum_{r}\lambda_{r1}^{2}, \sum_{r}\lambda_{r0}\lambda_{r1}\right)
\end{align*}
and the lower bound on the identified set is \looseness=-1 
\begin{align*}
L = 2\int\alpha_{1}^{+}(u)\alpha_{0}^{+}(1-u)du &+ 2\int\alpha_{1}^{+}(u)du\int\alpha_{0}^{+}(u)du + \max\left(\sum_{r}\left(\lambda_{r0}^{2} +\lambda_{r1}^{2}\right) - 1, \sum_{r}\lambda_{r0}\lambda_{\rho(r)1},0\right) 
\end{align*}
where $\alpha^{+}_s$ is the quantile function associated with $\alpha_s$, $\lambda_{rs}$ is the $r$th eigenvalue  (ordered to be decreasing) of $\epsilon_s$, and the infinite sums are as defined in Section 4.1.2 of the main text. As pointed out in that section, the DPO, $F(y_1,y_0) \in [L,U]$. In addition, $\alpha_s^+$ and $\lambda_{rs}$ are invariant to measure preserving transformations of $h_s$ and so are point identified because $h_s$ is identified up to a measure preserving transformation by Lemma 1 of the main text.  \looseness=-1

\subsubsection{Estimation of the outer bounds}
For each $s \in \{0,1\}$, we separately estimate $\theta_s$ by maximizing the composite likelihood of \cite{graham2020network} using the data on the network connections and covariates for the agents assigned to community $s$, and call this estimator $\hat{\theta}_s$. We then use $\hat{\theta}_s$ to construct a plug-in estimator for the graph function $h_s$, $\tilde{h}_s(a,b) = \int f_{Y_{12}|X_1,X_2}(\tau | a,b;\hat{\theta}_s)\mathbbm{1}\{\tau \leq y_s\}d\tau$. \looseness=-1 

The estimator $\tilde{h}_s$ differs from its estimand $h_s$ in two ways. First, it is defined using $\hat{\theta}_s$ instead of $\theta_s$. Second, it omits the measure preserving transformations $\varphi_s$. The first difference results in an estimation error which we account for in our confidence interval below. The second difference does not result in an estimation error. This is because our outer bounds proposed in Section D.3.2 only depend on the quantiles of $\alpha_s$ and the eigenvalues of $\epsilon_s$, both of which are invariant to measure preserving transformations of the graph function. It follows that using the estimated graph function $\tilde{h}_s$ leads to bounds that are equivalent to the infeasible estimate $\hat{h}_s(a,b) = \int f_{s}(\tau | \varphi_s(a),\varphi_s(b);\hat{\theta_s})\mathbbm{1}\{\tau \leq y_s\}d\tau$.   \looseness=-1 

We use $\hat{\alpha}_s(u) =  \int \tilde{h}_s(u,v)dv - \frac{1}{2} \int\int \tilde{h}_s(u,v) dudv$ as our estimator for $\alpha_{s}(u)$ and $\hat{\epsilon}_{s}(u,v) = \tilde{h}_s(u,v)  - \int \tilde{h}_s(u,v) du - \int \tilde{h}_s(u,v) dv + \int \int \tilde{h}_s(u,v) dudv$ as our estimator for $\epsilon_{s}(u,v)$. Our estimators of the bounds described in Online Appendix Section D.3.2 above are then \looseness=-1 
\begin{align*}
\hat{U} &= 2\int\hat{\alpha}_{1}^{+}(u)\hat{\alpha}_{0}^{+}(u)du + 2\int\hat{\alpha}_{1}^{+}(u)du\int\hat{\alpha}_{0}^{+}(u)du + \min\left(\sum_{r}\hat{\lambda}_{r0}^{2}, \sum_{r}\hat{\lambda}_{r1}^{2}, \sum_{r}\hat{\lambda}_{r0}\hat{\lambda}_{r1}\right), \text{ and }\\
\hat{L} &= 2\int\hat{\alpha}_{1}^{+}(u)\hat{\alpha}_{0}^{+}(1-u)du + 2\int\hat{\alpha}_{1}^{+}(u)du\int\hat{\alpha}_{0}^{+}(u)du + \max\left(\sum_{r}\left(\hat{\lambda}_{r0}^{2} + \hat{\lambda}_{r1}^{2}\right) - 1, \sum_{r}\hat{\lambda}_{r0}\hat{\lambda}_{\rho(r)1},0\right)  \looseness=-1 
\end{align*}
where $\hat{\alpha}^{+}_s$ is the quantile function of $\hat{\alpha}_s$ and $\hat{\lambda}_{rs}$ is the $r$th eigenvalue  (ordered to be decreasing) of $\hat{\epsilon}_s$.  \looseness=-1 

Since these bounds are equivalent to those bounds based on the infeasible estimator $\hat{h}_s$, it is without loss to conduct our statistical analysis below as though we had used $\hat{h}_s$ instead of $\tilde{h}_s$ in their construction. See Section D.4 below for a discussion.  \looseness=-1 

%
%

\subsubsection{Confidence interval}
For a fixed $\alpha > 0$, we propose the confidence interval 
\begin{align*}
S_{\alpha} = \left[\hat{L} - N^{-1/2}C^{L}_{\alpha}, \hat{U} + N^{-1/2}C^{U}_{\alpha}\right]
\end{align*}
where $C^{L}_{\alpha}$ and $C^{U}_{\alpha}$ are chosen such that 
\begin{align}\label{size}
\mathbb{P}\left(\sum_{j=1}^{6}\left(\xi ' \hat{\Omega}_j^{L} \xi\right)^{1/2} \leq C_{\alpha}^{L},\sum_{j=1}^{4}\left(\xi ' \hat{\Omega}_j^{U} \xi\right)^{1/2} \leq C_{\alpha}^{U} \right) \geq 1-\alpha,
\end{align}
 $\xi \sim \mathcal{N}\left(0,I_{K}\right)$, $\hat{\Omega}_j^L = \hat{V}^{1/2}W_{j}^L \hat{V}^{1/2}$, $\hat{\Omega}_j^U = \hat{V}^{1/2}W_{j}^U \hat{V}^{1/2}$, and $W_j^L$ and $W_j^U$ are defined in Online Appendix Section D.3.5 below. It follows from Proposition D3 below that if $C^{L}_{\alpha}$ and $C^{U}_{\alpha}$ satisfy condition (\ref{size}) then $\liminf_{N\to\infty}\mathbb{P}\left( [L,U] \subseteq S_{\alpha}\right) \geq 1-\alpha$. Since our confidence interval is valid for the identified set, it is potentially conservative for the DPO, see broadly Section 4.3.1 of \cite{molinari2020microeconometrics}. The arguments of \cite{imbens2004confidence, stoye2009more} could be applied here to, in some cases, produce a shorter interval. However, we do not formally describe such a refinement here.\footnote{This literature typically focuses on uniform validity results, which we do not provide here. We can show that our intervals are uniformly valid if the limiting results provided by \cite{graham2020network}  in Section D.3.2 hold uniformly. However, since  \cite{graham2020network} only states pointwise results in his handbook chapter, we do the same here. We conjecture that it is straightforward to, under certainty additional conditions, extend his results to be uniform, but we leave the details to future work.}  \looseness=-1 

There are typically multiple choices of $C_{\alpha}^{L}$ and $C_{\alpha}^{U}$ that satisfy condition (\ref{size}). One way to choose these parameters is to first take $R$ draws $\{\xi_{r}\}_{r=1}^{R}$ from $\mathcal{N}\left(0,I_{K}\right)$, where $R$ is a large positive integer like $10000$, compute the $1-\alpha$ isoquant of the empirical joint distribution of $\left\{\left(\sum_{j=1}^{6}\left(\xi'_{r} \hat{\Omega}_j^{L} \xi_{r}\right)^{1/2},\sum_{j=1}^{4}\left(\xi '_{r} \hat{\Omega}_j^{U} \xi_{r}\right)^{1/2}\right)\right\}_{r=1}^{R}$, and find the point on the isoquant such that the sum of the inputs are minimized. This strategy is designed to approximate an interval with the smallest possible length. For our empirical results in Section 6 of the main text we use a conservative choice of $C_{\alpha}^{L}$ and $C_{\alpha}^{U}$ that avoids computing the isoquant of a joint distribution function.  That is, we choose $C_{\alpha}^{L}$ to be the $1-\alpha/2$ quantile of  $\left\{\sum_{j=1}^{6}\left(\xi'_{r} \hat{\Omega}_j^{L} \xi_{r}\right)^{1/2}\right\}_{r=1}^{R}$ and $C_{\alpha}^{U}$ to be the $1 - \alpha/2$ quantile of $\left\{\sum_{j=1}^{4}\left(\xi'_{r} \hat{\Omega}_j^{U} \xi_{r}\right)^{1/2}\right\}_{r=1}^{R}$. This choice is conservative because $\mathbb{P}\left(\sum_{j=1}^{6}\left(\xi ' \hat{\Omega}_j^{L} \xi\right)^{1/2} \leq C_{\alpha}^{L},\sum_{j=1}^{4}\left(\xi ' \hat{\Omega}_j^{U} \xi\right)^{1/2} \leq C_{\alpha}^{U} \right) $
$\geq \mathbb{P}\left(\sum_{j=1}^{6}\left(\xi ' \hat{\Omega}_j^{L} \xi\right)^{1/2} \leq C_{\alpha}^{L}\right) $$+ \mathbb{P}\left(\sum_{j=1}^{4}\left(\xi ' \hat{\Omega}_j^{U} \xi\right)^{1/2} \leq C_{\alpha}^{U} \right) - 1 = 1- \alpha.$ \looseness=-1 

\subsubsection{List of weight matrices}
Our confidence interval described above depends on several weight matrices which we describe here. We use the following definitions: $\hat{h}'_{s k}(u,v) = \partial h_s(u,v;\theta)/\partial \theta_{sk}$ where $\theta_{sk}$ is the $k$th entry of $\theta_s$, $\bar{\hat{\alpha}}_{s} = \int\hat{\alpha}_s^{+}(u)du$, and $\bar{\alpha}_{s} = \int\alpha_s^{+}(u)du$. For the upper bound weights, $(\tau, \tau') \in \argmin_{s,s' \in \{0,1\}}\sum_r \lambda_{rs}\lambda_{rs'}$. The upper bound weights are \looseness=-1 
\begin{align*}
W^{U}_{kl,1} &= 4\int \left(\int \hat{h}'_{1k}(u,v)dv + \iint \hat{h}'_{1 k}(u,v)dudv\right) \\
&\hspace{10mm}\left(\int \hat{h}'_{1 l}(u,v)dv + \iint \hat{h}'_{1 l}(u,v)dudv\right)du \left(\int \hat{\alpha}_0(u)^{2}du\right) \\
W^{U}_{kl,2} &= 4\left( \int \left( \hat{\alpha}_1(u) + \bar{\hat{\alpha}}_1\right)^2du\right) \int \left(\int \hat{h}'_{0k}(u,v)dv - \frac{1}{2}\int\int \hat{h}'_{0k}(u,v)dudv\right) \\
&\hspace{10mm}\left(\int \hat{h}'_{0l}(u,v)dv - \frac{1}{2}\int\int \hat{h}'_{0l}(u,v)dudv\right)du \\
W^{U}_{kl,3} &= \int\int\left(\hat{h}'_{\tau k}(u,v) - \int\hat{h}'_{\tau k}(u,v)du  - \int\hat{h}'_{\tau k}(u,v)dv\right. \\
&\hspace{5mm}\left.+ \int\int\hat{h}'_{\tau k}(u,v)dudv\right)\left(\hat{h}'_{\tau l}(u,v) - \int\hat{h}'_{\tau l}(u,v)dv \right. \\
&\hspace{5mm}\left.- \int\hat{h}'_{\tau l}(u,v)du  + \int\int\hat{h}'_{\tau l}(u,v)dudv\right)dudv \left(\sum_{r} \hat{\lambda}_{r\tau'}^{2}\right) \\
W^{U}_{kl,4} &= \int\int\left(\hat{h}'_{\tau' k}(u,v) - \int\hat{h}'_{\tau' k}(u,v)du  - \int\hat{h}'_{\tau' k}(u,v)dv\right. \\
&\hspace{5mm}\left.+ \int\int\hat{h}'_{\tau' k}(u,v)dudv\right)\left(\hat{h}'_{\tau' l}(u,v) - \int\hat{h}'_{\tau' l}(u,v)dv \right. \\
&\hspace{5mm}\left.- \int\hat{h}'_{\tau' l}(u,v)du  + \int\int\hat{h}'_{\tau' l}(u,v)dudv\right)dudv \left(\sum_{r} \hat{\lambda}_{r\tau}^{2}\right). 
\end{align*}
For the lower bound weights,  \[(\tau, \tau') \in \argmax_{t,t' \in \{0,1\}}\left(\left(\sum_r {\lambda}_{rt}{\lambda}_{s(r)t'}\right)\mathbbm{1}\{t \neq t'\} + \left(\sum_r {\lambda}_{rt}{\lambda}_{rt'} + \sum_r {\lambda}_{r(1-t)}{\lambda}_{r(1-t')} - 1 \right)\mathbbm{1}\{t = t'\}\right)_{+}\] and $\sigma \in \{0,1\}$. The lower bound weights are \looseness=-1 
\begin{align*}
W^{L}_{kl,1} &= 4\int \left(\int \hat{h}'_{1 k}(u,v)dv + \iint \hat{h}'_{1 k}(u,v)dudv\right) \\
&\hspace{10mm}\left(\int \hat{h}'_{1l}(u,v)dv + \iint \hat{h}'_{1l}(u,v)dudv\right)du \left(\int \hat{\alpha}_0(u)^{2}du\right) \\
W^{L}_{kl,2} &= 4\left( \int \left( \hat{\alpha}_1(u) + \bar{\hat{\alpha}}_1\right)^2du\right) \int \left(\int \hat{h}'_{0k}(u,v)dv - \frac{1}{2}\int\int \hat{h}'_{0k}(u,v)dudv\right)\\
&\hspace{5mm} \left(\int \hat{h}'_{0l}(u,v)dv - \frac{1}{2}\int\int \hat{h}'_{0l}(u,v)dudv\right)du\\
W^{L}_{kl,3} &= \left(\int\int\left(\hat{h}'_{\tau'k}(u,v) - \int\hat{h}'_{\tau'k}(u,v)du - \int\hat{h}'_{\tau'k}(u,v)dv + \int\int\hat{h}'_{\tau' k}(u,v)dudv\right)\right. \\
&\hspace{10mm}\left.\left(\hat{h}'_{\tau' l}(u,v) - \int\hat{h}'_{\tau'l}(u,v)dv - \int\hat{h}'_{\tau' l}(u,v)du\right.\right.\\
&\hspace{10mm}\left.\left.+ \int\int\hat{h}'_{\tau' l}(u,v)dudv\right)dudv \left(\sum_{r} \hat{\lambda}_{r\tau}^{2}\right)\right)\mathbbm{1}\{\tau \neq \tau'\} \\
W^{L}_{kl,4} &= \left(\int\int\left(\hat{h}'_{\tau'k}(u,v) - \int\hat{h}'_{\tau'k}(u,v)du - \int\hat{h}'_{\tau'k}(u,v)dv + \int\int\hat{h}'_{\tau'k}(u,v)dudv\right)\right. \\
&\hspace{10mm}\left.\left(\hat{h}'_{\tau l}(u,v) - \int\hat{h}'_{\tau l}(u,v)dv - \int\hat{h}'_{\tau l}(u,v)du\right.\right.\\
&\hspace{10mm}\left.\left.+ \int\int\hat{h}'_{\tau l}(u,v)dudv\right)dudv \left(\sum_{r} \hat{\lambda}_{r\tau'}^{2}\right)\right)\mathbbm{1}\{\tau \neq \tau'\} \\
W^{L}_{kl,5+\sigma} &= 4\left(\int\int\left(\hat{h}'_{\sigma k}(u,v) - \int\hat{h}'_{\sigma k}(u,v)du  - \int\hat{h}'_{\sigma k}(u,v)dv + \int\int\hat{h}'_{\sigma k}(u,v)dudv\right)\right.\\
&\hspace{10mm}\left.\left(\hat{h}'_{\sigma l}(u,v) - \int\hat{h}'_{\sigma l}(u,v)dv - \int\hat{h}'_{\sigma l}(u,v)du \right.\right. \\
&\hspace{10mm}\left.\left.+ \int\int\hat{h}'_{\sigma l}(u,v)dudv\right)dudv \left(\sum_{r} \hat{\lambda}_{r\sigma}^{2}\right)\right)\mathbbm{1}\{\tau = \tau'\}.
\end{align*}

\subsection{Proof of the consistency claim}
Our main justification for the confidence interval $S_{\alpha}$ is Proposition D3 in Online Appendix Section D.4.2 below. Its proof relies on Lemma D2  which we state and demonstrate first. \looseness=-1 
\subsubsection{Consistency lemma}
\begin{flushleft}
\textbf{Lemma D2:} (i) $\sup_{u \in [0,1]}\left|\hat{\alpha}^{+}_{s}(u) - \alpha^{+}_{s}(u)\right| = O_{p}(N^{-1/2})$, (ii) $\sup_{i \in \mathbb{N}}\left| \hat{\lambda}_{is}^{+} - \lambda_{is}^{+}\right| = O_{p}(N^{-1/2})$, and (iii) $\sup_{i \in \mathbb{N}}\left| \hat{\lambda}_{is}^{-} - \lambda_{is}^{-}\right| = O_{p}(N^{-1/2})$ where $\hat{\lambda}_{is}^{+}$  and $\lambda_{is}^{+}$ are the $i$th positive eigenvalue of $\hat{\epsilon}_s$ and $\epsilon_{s}$, and $\hat{\lambda}_{is}^{-}$ and $\lambda_{is}^{-}$ are the $i$th negative eigenvalue of $\hat{\epsilon}_s$ and $\epsilon_{s}$. The eigenvalues are all ordered to be decreasing in magnitude. \looseness=-1 
\end{flushleft}

\begin{flushleft}
\textbf{Proof of Lemma D2:} Since $h_s$ is continuously differentiable with a uniformly bounded derivative and $X_{s}$ is bounded, it follows from Assumption D1 and the mean value theorem that $\sup_{u,v}\left|\left(\hat{h}_{s}(u,v) - h_{s}(u,v)\right)\right| = O_{p}\left(|\hat{\theta} - \theta|\right) = O_{p}(N^{-1/2})$. The claim then follows from the fact that $\hat{\alpha}^{+}_{s}(u)$, $\hat{\lambda}_{is}^{+}$, and $\hat{\lambda}_{is}^{-}$ are Lipschitz continuous functions of $\hat{h}_{s}$, and $\alpha^{+}_{s}(u)$, $\lambda_{is}^{+}$, and $\lambda_{is}^{-}$ are Lipschitz continuous functions of $h_{s}$ (see Lemma B3 in Section B above). $\square$ \looseness=-1 
\end{flushleft}

\subsubsection{Result}
\begin{flushleft}
\textbf{Proposition D3:} $\liminf_{N\to\infty}\inf_{\theta \in \left[L,U\right]}\mathbb{P}\left( [L,U] \subseteq S_{\alpha}\right) \geq 1-\alpha$ 
\end{flushleft}

\begin{flushleft}
\textbf{Proof of Proposition D3:}  We break up the proof of Proposition D3 into three parts. The first two parts derive bounds on the estimation error of $\hat{U}$ and $\hat{L}$. The third part combines the bounds from the first two parts to demonstrate the claim. \looseness=-1 
\end{flushleft}

\subsubsection{Step 1: Estimation error for the upper bound}
We first bound the estimation error of $\hat{U}$.
\begin{align*}
\hat{U} - U &= 2\int \left(\hat{\alpha}_1^{+}(u) + \bar{\hat{\alpha}}_{1}\right)\hat{\alpha}_0^+(u) du - 2\int \left(\alpha_1^{+}(u) + \bar{\alpha}_{1}\right)\alpha_0^{+}(u) du \\
&\hspace{20mm}+ \min_{t,t' \in \{0,1\}}\sum_r \hat{\lambda}_{rt}\hat{\lambda}_{rt'} - \min_{t,t' \in \{0,1\}}\sum_r \lambda_{rt}\lambda_{rt'} \\
&= 2\int\left(\hat{\alpha}_1^{+}(u)  + \bar{\hat{\alpha}}_{1} - \alpha_1^{+}(u) - \bar{\alpha}_1 \right)\hat{\alpha}_0^{+}(u)du  + 2\int\left(\alpha_1^{+}(u) + \bar{\alpha}_1 \right)\left( \hat{\alpha}_0^{+}(u) -  \alpha_0^{+}(u)\right)du \\
&\hspace{20mm}+ \sum_{r}\left(\hat{\lambda}_{r\tau} - \lambda_{r\tau}\right)\hat{\lambda}_{r\tau'} +  \sum_{r}\lambda_{r\tau}\left(\hat{\lambda}_{r\tau'} - \lambda_{r\tau'}\right) 
\end{align*}
where $\bar{\hat{\alpha}}_{t} = \int\hat{\alpha}_t^{+}(u)du$, $\bar{\alpha}_{t} = \int\alpha_t^{+}(u)du$, $(\tau, \tau') \in \argmin_{t,t' \in \{0,1\}}\sum_r \lambda_{rt}\lambda_{rt'}$, and the second equality holds eventually because $\sum_{r}\hat{\lambda}_{r\tau}\left(\hat{\lambda}_{r\tau'} - \lambda_{r\tau'}\right) = \min_{t,t' \in \{0,1\}}\sum_r \lambda_{rt}\lambda_{rt'}$ by definition of $(\tau,\tau')$ and $\sum_{r}\left(\hat{\lambda}_{r\tau} - \lambda_{r\tau}\right)\hat{\lambda}_{r\tau'} = \min_{t,t' \in \{0,1\}}\sum_r \hat{\lambda}_{rt}\hat{\lambda}_{rt'}$ eventually by Lemma D2 and the continuous mapping theorem. We analyze the four summands in the second line of the displayed equation separately. The first summand can be bounded  \looseness=-1 
\begin{align*}
&\left| 2\int\left(\hat{\alpha}_1^{+}(u)  + \bar{\hat{\alpha}}_{1} - \alpha_1^{+}(u) - \bar{\alpha}_1 \right)\hat{\alpha}_0^{+}(u)du\right|  \\
&\hspace{2mm}\leq 2\left(\int \left(\hat{\alpha}_1^{+}(u) - \alpha_1^{+}(u) + \bar{\hat{\alpha}}_1 - \bar{\alpha}_1 \right)^2\right)^{1/2}\left(\int \hat{\alpha}_0(u)^{2}du\right)^{1/2} \\
&\hspace{2mm} \leq 2\left(\int \left(\hat{\alpha}_1(u) - \alpha_1(u) + \bar{\hat{\alpha}}_1 - \bar{\alpha}_1 \right)^2\right)^{1/2}\left(\int \hat{\alpha}_0(u)^{2}du\right)^{1/2} \\
&\hspace{2mm} = 2\left( \int\left(\left(\int \hat{h}_1(u,v) - h_1(u,v)dv\right) + \left(\iint \hat{h}_1(u,v) - h_1(u,v)dudv\right)\right)^2du\right)^{1/2}\left(\int \hat{\alpha}_0(u)^{2}du\right)^{1/2} \\
&\hspace{2mm}  =  2\left(\sum_{k,l=1}^{K}\left(\hat{
\theta}_{k}-\theta_{k}\right)\left(\hat{\theta}_{l}-\theta_{l}\right)\int \left(\left(\int \hat{h}'_{1k}(u,v)dv + \iint \hat{h}'_{1k}(u,v)dudv\right)\right.\right. \\
&\hspace{5mm}\left.\left.\left( \int  \hat{h}'_{1l}(u,v)dv + \iint \hat{h}'_{1l}(u,v)dudv\right)\right)^2du\right)^{1/2} \left(\int \hat{\alpha}_0(u)^{2}du\right)^{1/2} + o_{p}\left( N^{-1/2}\right) \\
&\hspace{2mm}  = \left(\sum_{k,l=1}^{K} \left(\hat{\theta}_{k} - \theta_{k}\right)\left(\hat{\theta}_{l} - \theta_{l}\right)W^{U}_{kl,1}\right)^{1/2}  + o_{p}\left( N^{-1/2}\right)
\end{align*}
where $\hat{h}'_{s k}(u,v)= h'_{s k}(\hat{\theta})$, $h'_{sk}(\theta) = \partial h_{s}(u,v;\theta)/\partial \theta_{sk}$ refers to the partial derivative of $h_s$ with respect to the $k$th element of $\theta_{s}$ at the point $(u,v)$ and $\theta$, the first inequality is due to Cauchy-Schwarz, the second inequality is due to the functional version of Hardy-Littlewood-Polya Theorem 368 (see Lemma B5 in Online Appendix Section B.1), the first equality is due to plugging in the definition of $\hat{\alpha}_1$, $\alpha_1$, $\bar{\hat{\alpha}}_1$, and $\bar{\alpha}_1$, and algebra, and the second equality is due to the definition of $\hat{h}_1$ and $h_1$, a first-order Taylor expansion, Proposition D3, and more algebra. The same exact arguments can be applied to the second summand \looseness=-1 
\begin{align*}
&\left|2\int\left(\alpha_1^{+}(u) + \bar{\alpha}_1 \right)\left( \hat{\alpha}_0^{+}(u) -  \alpha_0^{+}(u)\right)du\right| \\
&\leq 2 \left( \int \left( \alpha_1(u) + \bar{\alpha}_1\right)^2du\right)^{1/2}\left(\int\left( \hat{\alpha}_0(u) - \alpha_0(u)\right)^{2}du\right)^{1/2}\\
&\hspace{2mm}= 2 \left( \int \left( \hat{\alpha}_1(u) + \bar{\hat{\alpha}}_1\right)^2du\right)^{1/2}\left(\int\left( \hat{\alpha}_0(u) - \alpha_0(u)\right)^{2}du\right)^{1/2} + o_{p}\left(N^{-\gamma}\right)\\
&\hspace{2mm}= 2 \left(\sum_{k,l=1}^{K}(\hat{\theta}_{k}- \theta_{k})(\hat{\theta}_{l}- \theta_{l})\left( \int \left( \hat{\alpha}_1(u) + \bar{\hat{\alpha}}_1\right)^2du\right)\int \left(\int \hat{h}'_{0k}(u,v)dv - \frac{1}{2}\int\int \hat{h}'_{0k}(u,v)dudv\right) \right.\\
&\hspace{5mm}\left.\left(\int \hat{h}'_{0l}(u,v)dv - \frac{1}{2}\int\int \hat{h}'_{0l}(u,v)dudv\right)du\right)^{1/2} + o_{p}\left( N^{-1/2}\right) \\
&\hspace{2mm} = \left(\sum_{k,l=1}^{K} \left(\hat{\theta}_{k} - \theta_{k}\right)\left(\hat{\theta}_{l} - \theta_{l}\right)W^{U}_{kl,2}\right)^{1/2}  + o_{p}\left( N^{-1/2}\right).
\end{align*}
The third summand can be bounded \looseness=-1 
\begin{align*}
&\left|\sum_{r}\left(\hat{\lambda}_{r\tau} - \lambda_{r\tau}\right)\hat{\lambda}_{r\tau'} \right| 
\leq \left(\sum_{r} \left(\hat{\lambda}_{r\tau} - \lambda_{r\tau}\right)^{2}\right)^{1/2}\left(\sum_{r} \hat{\lambda}_{r\tau'}^{2}\right)^{1/2} \\
&\hspace{2mm}\leq \left(\int\int \left(\hat{\epsilon}_{\tau}(u,v) - \epsilon_{\tau}(u,v)\right)^{2}dudv\right)^{1/2}\left(\sum_{r} \hat{\lambda}_{r\tau'}^{2}\right)^{1/2}  \\
&\hspace{2mm}= \left(\int\int \left(\left(\hat{h}_{\tau}(u,v) - h_{\tau}(u,v)\right) - \int\left(\hat{h}_{\tau}(u,v) - h_{\tau}(u,v)\right)dv - \int\left(\hat{h}_{\tau}(u,v) - h_{\tau}(u,v)\right)du \right.\right.\\
&\hspace{5mm}\left.\left.  + \iint \left(\hat{h}_{\tau}(u,v) - h_{\tau}(u,v)\right)dudv\right)^{2}dudv\right)^{1/2}\left(\sum_{r} \hat{\lambda}_{r\tau'}^{2}\right)^{1/2} 
\end{align*}
\begin{align*}
&\hspace{2mm} = \left(\sum_{k,l=1}^{K}(\hat{\theta}_{k} - \theta_{k})(\hat{\theta}_{l} - \theta_{l})\int\int\left(\hat{h}'_{\tau k}(u,v) - \int\hat{h}'_{\tau k}(u,v)du - \int\hat{h}'_{\tau k}(u,v)dv + \int\int\hat{h}'_{\tau k}(u,v)dudv\right)\right.\\
&\hspace{5mm}\left.\left(\hat{h}'_{\tau l}(u,v) - \int\hat{h}'_{\tau l}(u,v)dv - \int\hat{h}'_{\tau l}(u,v)du  + \int\int\hat{h}'_{\tau l}(u,v)dudv\right)dudv \left(\sum_{r} \hat{\lambda}_{r\tau'}^{2}\right)\right)^{1/2} + o_{p}\left(N^{-1/2}\right) \\
&\hspace{2mm} =  \left(\sum_{k,l=1}^{K} \left(\hat{\theta}_{k} - \theta_{k}\right)\left(\hat{\theta}_{l} - \theta_{l}\right)W^{U}_{kl,3}\right)^{1/2}  + o_{p}\left( N^{-1/2}\right)
\end{align*}
where the first inequality is due to Cauchy-Schwarz, the second inequality is due to our Proposition 4, the first equality is due to plugging in the definition of $\hat{\epsilon}_{\tau}$ and $\epsilon_{\tau}$, and the second equality is due to the definition of $\hat{\mu}_\tau$ and $\mu_\tau$, a first-order Taylor expansion, Proposition D2, and some algebra. The same exact arguments can be applied to bound the fourth summand\looseness=-1 
\begin{align*}
&\left|\sum_{r}\lambda_{r\tau}\left(\hat{\lambda}_{r\tau'} - \lambda_{r\tau'}\right) \right| \\
&\hspace{2mm}\leq \left(\sum_{k,l=1}^{K}(\hat{\theta}_{k} - \theta_{k})(\hat{\theta}_{l} - \theta_{l})\int\int\left(\hat{h}'_{\tau' k}(u,v) - \int\hat{h}'_{\tau' k}(u,v)du  - \int\hat{h}'_{\tau' k}(u,v)dv + \int\int\hat{h}'_{\tau' k}(u,v)dudv\right)\right.\\
&\hspace{5mm}\left.\left(\hat{h}'_{\tau' l}(u,v) - \int\hat{h}'_{\tau' l}(u,v)dv - \int\hat{h}'_{\tau' l}(u,v)du  + \int\int\hat{h}'_{\tau' l}(u,v)dudv\right)dudv \left(\sum_{r} \hat{\lambda}_{r\tau}^{2}\right)\right)^{1/2} + o_{p}\left(N^{-1/2}\right) \\
&\hspace{2mm} =  \left(\sum_{k,l=1}^{K} \left(\hat{\theta}_{k} - \theta_{k}\right)\left(\hat{\theta}_{l} - \theta_{l}\right)W^{U}_{kl,4}\right)^{1/2}  + o_{p}\left( N^{-1/2}\right).
\end{align*}

Combining the bounds for all four results, it follows that 
\begin{align*}
N^{1/2}|\hat{U} - U| \leq \sum_{j = 1}^{4} \left(\sum_{k,l=1}^{K} N^{1/2}\left(\hat{\theta}_{k} - \theta_{k}\right)N^{1/2}\left(\hat{\theta}_{l} - \theta_{l}\right)W^{U}_{kl,j}\right)^{1/2}  + o_{p}\left(1\right).
\end{align*}

\subsubsection{Step 2: Estimation error for the lower bound}
We now bound the estimation error of $\hat{L} $.  \looseness=-1 
\begin{align*}
\hat{L} - L &= 2\int \left(\hat{\alpha}_1^{+}(u) + \bar{\hat{\alpha}}_{1}\right)\hat{\alpha}_0^{+}(1-u) du - 2\int \left(\alpha_1^{+}(u) + \bar{\alpha}_{1}\right)\alpha_0^{+}(1-u) du \\
&\hspace{5mm}+ \max_{t,t' \in \{0,1\}}\left(\left(\sum_r \hat{\lambda}_{rt}\hat{\lambda}_{s(r)t'}\right)\mathbbm{1}\{t \neq t'\} + 
\left(\sum_r \hat{\lambda}_{rt}\hat{\lambda}_{rt'} + \sum_r \hat{\lambda}_{r(1-t)}\hat{\lambda}_{r(1-t')} - 1 \right)\mathbbm{1}\{t = t'\}\right)_{+} \\
&\hspace{5mm}- \max_{t,t' \in \{0,1\}}\left(\left(\sum_r {\lambda}_{rt}{\lambda}_{s(r)t'}\right)\mathbbm{1}\{t \neq t'\} + \left(\sum_r {\lambda}_{rt}{\lambda}_{rt'} + \sum_r {\lambda}_{r(1-t)}{\lambda}_{r(1-t')} - 1 \right)\mathbbm{1}\{t = t'\}\right)_{+} \\
&= 2\int\left(\hat{\alpha}_1^{+}(u)  + \bar{\hat{\alpha}}_{1} - \alpha_1^{+}(u) - \bar{\alpha}_1 \right)\hat{\alpha}_0^{+}(1-u)du  \\
&\hspace{5mm}+ 2\int\left(\alpha_1^{+}(u) + \bar{\alpha}_1 \right)\left( \hat{\alpha}_0^{+}(1-u) -  \alpha_0^{+}(1-u)\right)du \\
&\hspace{5mm}+ \left(\left(\sum_r \hat{\lambda}_{r\tau}\hat{\lambda}_{s(r)\tau'}\right)_{+} - \left(\sum_r {\lambda}_{rt}{\lambda}_{s(r)t'}\right)_{+}\right)\mathbbm{1}\{\tau \neq \tau'\}  \\
&\hspace{5mm}+  \left(\left(\sum_r \hat{\lambda}_{r\tau}\hat{\lambda}_{r\tau'} + \sum_r \hat{\lambda}_{r(1-\tau)}\hat{\lambda}_{r(1-\tau')} - 1 \right)_{+} \right. \\
&\hspace{15mm}- \left. \left(\sum_r {\lambda}_{r\tau}{\lambda}_{r\tau'} + \sum_r {\lambda}_{r(1-\tau)}{\lambda}_{r(1-\tau')} - 1 \right)_{+}\right)\mathbbm{1}\{\tau = \tau'\} 
\end{align*}
where $\bar{\hat{\alpha}}_{t} = \int\hat{\alpha}_t^{+}(u)du$, $\bar{\alpha}_{t} = \int\alpha_t^{+}(u)du$, \[(\tau, \tau') \in \argmax_{t,t' \in \{0,1\}}\left(\left(\sum_r {\lambda}_{rt}{\lambda}_{s(r)t'}\right)\mathbbm{1}\{t \neq t'\} + \left(\sum_r {\lambda}_{rt}{\lambda}_{rt'} + \sum_r {\lambda}_{r(1-t)}{\lambda}_{r(1-t')} - 1 \right)\mathbbm{1}\{t = t'\}\right)_{+}\] and the second equality holds eventually by Lemma D2 and the continuous mapping theorem. We analyze the four summands in the second line of the displayed equation separately. The first two summands can be bounded uniformly over $\mathcal{P}$ \looseness=-1 
\begin{align*}
&\left|2\int\left(\hat{\alpha}_1^{+}(u)  + \bar{\hat{\alpha}}_{1} - \alpha_1^{+}(u) - \bar{\alpha}_1 \right)\hat{\alpha}_0^{+}(1-u)du\right| \\
&\hspace{2mm}\leq 2\left(\sum_{k,l=1}^{K}\left(\hat{\theta}_{k}-\theta_{k}\right)\left(\hat{\theta}_{l}-\theta_{l}\right)\int \left(\left(\int \hat{h}'_{1k}(u,v)dv + \iint \hat{h}'_{1k}(u,v)dudv\right)\right.\right. \\
&\hspace{5mm}\left.\left.\left( \int  \hat{h}'_{1l}(u,v)dv + \iint \hat{h}'_{1l}(u,v)dudv\right)\right)^2du\right)^{1/2} \left(\int \hat{\alpha}_0(u)^{2}du\right)^{1/2} + o_{p}\left( N^{-1/2}\right) \\
&\hspace{2mm}=  \left(\sum_{k,l=1}^{K} \left(\hat{\theta}_{k} - \theta_{k}\right)\left(\hat{\theta}_{l} - \theta_{l}\right)W^{L}_{kl,1}\right)^{1/2}  + o_{p}\left( N^{-1/2}\right) \text{ and }
\end{align*}
\begin{align*}
&\left|2\int\left(\alpha_1^{+}(u) + \bar{\alpha}_1 \right)\left( \hat{\alpha}_0^{+}(1-u) -  \alpha_0^{+}(1-u)\right)du\right| \\
&\hspace{2mm}\leq 2 \left(\sum_{k,l=1}^{K}(\hat{\theta}_{k}- \theta_{k})(\hat{\theta}_{l}- \theta_{l})\left( \int \left( \hat{\alpha}_1(u) + \bar{\hat{\alpha}}_1\right)^2du\right) \right. \\
&\hspace{5mm}\left.\int \left(\int \hat{h}'_{0k}(u,v) - \frac{1}{2}\int\int \hat{h}'_{0k}(u,v)dudv\right)\left(\int \hat{h}'_{0l}(u,v)dv - \frac{1}{2}\int\int \hat{h}'_{0l}(u,v)dudv\right)du\right)^{1/2} + o_{p}\left( N^{-1/2}\right) \\
&\hspace{2mm}=  \left(\sum_{k,l=1}^{K} \left(\hat{\theta}_{k} - \theta_{k}\right)\left(\hat{\theta}_{l} - \theta_{l}\right)W^{L}_{kl,2}\right)^{1/2}  + o_{p}\left( N^{-1/2}\right)
\end{align*}
where $h'$ is the first derivative of $h$, following the arguments of Step 1 from Section D.4.3. The second two summands can also be bounded \looseness=-1 
\begin{align*}
&\left|  \left(\left(\sum_r \hat{\lambda}_{r\tau}\hat{\lambda}_{s(r)\tau'}\right)_{+} - \left(\sum_r {\lambda}_{rt}{\lambda}_{s(r)t'}\right)_{+}\right)\mathbbm{1}\{\tau \neq \tau'\} \right| \\
 &\hspace{2mm}\leq \left|\left( \sum_{r}\left(\hat{\lambda}_{r\tau} - \lambda_{r\tau}\right)\hat{\lambda}_{s(r)\tau'} + \sum_{r}\lambda_{r\tau}\left(\hat{\lambda}_{s(r)\tau'} - \lambda_{s(r)\tau}\right) \right)\mathbbm{1}\{\tau \neq \tau'\}\right| \\
 \end{align*}
 \begin{align*}
 &\hspace{2mm}\leq \left(\left(\sum_{k,l=1}^{K}(\hat{\theta}_{k} - \theta_{k})(\hat{\theta}_{l} - \theta_{l})\int\int\left(\hat{h}'_{\tau k}(u,v) - \int\hat{h}'_{\tau k}(u,v)du \right.\right.\right. \\
&\hspace{5mm}\left.\left.\left. - \int\hat{h}'_{\tau k}(u,v)dv + \int\int\hat{h}'_{\tau k}(u,v)dudv\right)\right.\right.\\
&\hspace{5mm}\left.\left.\left(\hat{h}'_{\tau l}(u,v)- \int\hat{h}'_{\tau l}(u,v)dv - \int\hat{h}'_{\tau l}(u,v)du + \int\int\hat{h}'_{\tau l}(u,v)dudv\right)dudv \left(\sum_{r} \hat{\lambda}_{r\tau'}^{2}\right)\right)^{1/2}\right. \\
 &\hspace{2mm}+\left.\left(\sum_{k,l=1}^{K}(\hat{\theta}_{k} - \theta_{k})(\hat{\theta}_{l} - \theta_{l})\int\int\left(\hat{h}'_{\tau' k}(u,v) - \int\hat{h}'_{\tau' k}du - \int\hat{h}'_{\tau' k}(u,v)dv + \int\int\hat{h}'_{\tau' k}(u,v)dudv\right)\right.\right.\\
&\hspace{5mm}\left.\left.\left(\hat{h}'_{\tau' l}(u,v) - \int\hat{h}'_{\tau' l}(u,v)dv - \int\hat{h}'_{\tau' l}(u,v)du \right.\right.\right. \\
&\hspace{5mm}\left.\left.\left. + \int\int\hat{h}'_{\tau' l}(u,v)dudv\right)dudv \left(\sum_{r} \hat{\lambda}_{r\tau}^{2}\right)\right)^{1/2}+ o_{p}\left(N^{-\gamma}\right)\right)\mathbbm{1}\{\tau \neq \tau'\} \\
&\hspace{2mm}=  \left(\sum_{k,l=1}^{K} \left(\hat{\theta}_{k} - \theta_{k}\right)\left(\hat{\theta}_{l} - \theta_{l}\right)W^{L}_{kl,3}\right)^{1/2} + \left(\sum_{k,l=1}^{K} \left(\hat{\theta}_{k} - \theta_{k}\right)\left(\hat{\theta}_{l} - \theta_{l}\right)W^{L}_{kl,4}\right)^{1/2}  + o_{p}\left( N^{-1/2}\right)
\end{align*}
and 
\begin{align*}
&\left| \left(\left(\sum_r \hat{\lambda}_{r\tau}\hat{\lambda}_{r\tau'} + \sum_r \hat{\lambda}_{r(1-\tau)}\hat{\lambda}_{r(1-\tau')} - 1 \right)_{+} - \left(\sum_r {\lambda}_{r\tau}{\lambda}_{r\tau'} + \sum_r {\lambda}_{r(1-\tau)}{\lambda}_{r(1-\tau')} - 1 \right)_{+}\right)\mathbbm{1}\{\tau = \tau'\} \right|\\
 &\hspace{2mm}\leq \left|\sum_{\sigma\in \{0,1\}}\left( \sum_{r}\left(\hat{\lambda}_{r\sigma} - \lambda_{r\sigma}\right)\hat{\lambda}_{r\sigma} + \sum_{r}\lambda_{r\sigma}\left(\hat{\lambda}_{r\sigma} - \lambda_{r\sigma}\right) \right)\mathbbm{1}\{\tau = \tau'\}\right| \\
   &\hspace{2mm}\leq \left(\sum_{\sigma \in \{0,1\}}2\left(\sum_{k,l=1}^{K}(\hat{\theta}_{k} - \theta_{k})(\hat{\theta}_{l} - \theta_{l})\int\int\left(\hat{h}'_{\sigma k}(u,v)- \int\hat{h}'_{\sigma k}(u,v)du \right.\right.\right. \\
&\hspace{5mm}\left.\left.\left. - \int\hat{h}'_{\sigma k}(u,v)dv + \int\int\hat{h}'_{\sigma k}(u,v)dudv\right)\left(\hat{h}'_{\sigma l}(u,v) - \int\hat{h}'_{\sigma l}(u,v)dv - \int\hat{h}'_{\sigma l}(u,v)du \right.\right.\right. \\
&\hspace{5mm}\left.\left. \left. + \int\int\hat{h}'_{\sigma l}(u,v)dudv\right)dudv \left(\sum_{r} \hat{\lambda}_{r\sigma}^{2}\right)\right)^{1/2}+ o_{p}\left(N^{-1/2}\right)\right)\mathbbm{1}\{\tau = \tau'\} \\
&\hspace{2mm}= \left(\sum_{k,l=1}^{K} \left(\hat{\theta}_{k} - \theta_{k}\right)\left(\hat{\theta}_{l} - \theta_{l}\right)W^{L}_{kl,5}\right)^{1/2} + \left(\sum_{k,l=1}^{K} \left(\hat{\theta}_{k} - \theta_{k}\right)\left(\hat{\theta}_{l} - \theta_{l}\right)W^{L}_{kl,6}\right)^{1/2} + o_{p}\left( N^{-1/2}\right)
\end{align*}
following exactly the arguments of Step 1 from Section D.4.3. It follows that
\begin{align*}
N^{1/2}|\hat{L} - L| \leq\sum_{j = 1}^{6} \left(\sum_{k,l=1}^{K} N^{1/2}\left(\hat{\theta}_{k} - \theta_{k}\right)N^{1/2}\left(\hat{\theta}_{l} - \theta_{l}\right)W^{U}_{kl,j}\right)^{1/2}  + o_{p}\left( 1\right).
\end{align*}

\subsubsection{Step 3: Proof of Proposition D3}
From Steps 1 and 2, for any $\epsilon > 0$ there exists an $M \in \mathbb{N}$ such that for all $N > M$,   \looseness=-1 
\begin{align*}
 \mathbb{P}\left( [L,U] \subseteq S_{\alpha}\right) 
  &\geq \mathbbm{P}\left( L \geq \hat{L}-N^{-1/2}C_{\alpha}^{L} , U \leq \hat{U} + N^{-1/2}C_{\alpha}^{U} \right) \\
  &\geq \mathbb{P}\left( \left|\hat{L} - L\right| \leq N^{-1/2}C_{\alpha}^{L} , \left| \hat{U} - U\right|  \leq N^{-1/2}C_{\alpha}^{U} \right) \\
  &\geq \mathbb{P}\left(\sum_{j=1}^{6}\left(\xi ' \hat{\Omega}_j^{L} \xi\right)^{1/2} \leq C_{\alpha}^{L} + \epsilon,\sum_{j=1}^{4}\left(\xi ' \hat{\Omega}_j^{U} \xi\right)^{1/2} \leq C_{\alpha}^{U} + \epsilon \right) - \epsilon\\ &\geq 1-\alpha -\epsilon
 \end{align*}
 where the third inequality follows from the bounds from Steps 1 and 2, and the asymptotic normality result for $\hat{\theta}$ in Section D.3.2. Since the choice of $\epsilon > 0$ was arbitrary, the claim follows.

\bibliographystyle{aer}
\bibliography{literature}

\end{document}